\documentclass[11pt]{iopart}
\usepackage{amssymb}
\usepackage{bm}
\usepackage[dvips]{epsfig}
\newcommand{\ba}{\bm{a}}

\newcommand{\bfb}{\bm{b}}
\newcommand{\bd}{\bm{d}}
\newcommand{\be}{\bm{e}}
\newcommand{\bff}{\bm{f}}
\newcommand{\bF}{\bm{F}}
\newcommand{\bg}{\bm{g}}
\newcommand{\bh}{\bm{h}}
\newcommand{\bj}{\bm{j}}

\newcommand{\bq}{\bm{q}}

\newcommand{\bT}{\bm{T}}
\newcommand{\bx}{\bm{x}}
\newcommand{\bu}{\bm{u}}
\newcommand{\bv}{\bm{v}}
\newcommand{\bU}{\bm{U}}
\newcommand{\bw}{\bm{w}}

\newcommand{\al}{\alpha}
\newcommand{\del}{\delta}
\newcommand{\Del}{\Delta}
\newcommand{\eps}{\epsilon}
\newcommand{\gam}{\gamma}
\newcommand{\lam}{\lambda}
\newcommand{\Lam}{\Lambda}
\newcommand{\om}{\omega}
\newcommand{\Om}{\Omega}
\newcommand{\sg}{\sigma}
\newcommand{\bom}{\mbox{\boldmath $\omega$}}
\newcommand{\bOm}{\bm{\Omega}}
\newcommand{\vep}{\varepsilon}

\newcommand{\pl}{\partial}
\newcommand{\dd}{\mbox{d}}
\newcommand{\DD}{\mbox{D}}
\newcommand{\half}{\mbox{\small $\frac{1}{2}$} \,}
\newcommand{\Half}{\mbox{$\frac{1}{2}$} \,}
\newcommand{\third}{\mbox{\small $\frac{1}{3}$} \,}
\newcommand{\twothird}{\mbox{\small $\frac{2}{3}$} \,}
\newcommand{\A}{\mbox{$\mathcal{A}$}}
\newcommand{\Bb}{\mbox{$\mathcal{B}$}}
\newcommand{\D}{\mbox{D}}
\newcommand{\Dd}{\mbox{$\mathcal{D}$}}

\newcommand{\EE}{\mbox{$\mathcal{E}$}}
\newcommand{\F}{\mbox{$\mathcal{F}$}}

\newcommand{\Hm}{\mbox{$\mathcal{H}$}}

\newcommand{\V}{\mbox{$\mathcal{V}$}}

\newcommand{\ie}{{\it i.e.}\,}
\newcommand{\eg}{{\it e.g.}\ }
\newcommand{\etc}{{\it etc.}}
\renewcommand{\etal}{{\it et al.}\,}
\newcommand{\const}{{\it const}\,}
\newcommand{\curl}{\mbox{curl}}
\newcommand{\divg}{\mbox{div}\,}
\newcommand{\ine}{e_{\rm i}}

\begin{document}

\centerline{\small {\it Preprint} "Kambe 2017": \ \underline{\it New scenario of turbulence theory}\,\footnote{This 
was  originally prepared for submission to  "{\it Geophysical and Astrophysical Fluid Dynamics}".}
\hskip5mm (9 June 2017) } 
\vspace*{10mm}

\title[]{New scenario of turbulence theory and  wall-bounded turbulence:  Theoretical  significance}

\author{Tsutomu  Kambe\footnote{\copyright 2017 TKambe; \ Former Professor, University of Tokyo (Physics); 
\ {\it June} $9^{th}$, 2017}}

\address{Higashi-yama 2-11-3, Meguro-ku, Tokyo, Japan \ 153-0043}

\ead{kambe@ruby.dti.ne.jp}


\begin{abstract}

New general scenario of turbulence theory is proposed and applied to wall-bounded turbulence. This scenario 
introduces a new field of transverse waves. Significance of the theory rests on a mathematical theorem  associated 
with the fundamental conservation law of fluid current flux, expressed in a form of $4d$ physical space-time 
representation, which predicts a system of Maxwell-type equation and supports transverse waves traveling with 
a phase speed $c_t$.  In regard to the streaky wall flows, there exist both dynamical mechanism and energy channel 
which excite transverse waves and  exchange energy between flow field and transverse wave field. In developed 
state of the wave field, energy is supplied from the flow field to the  wave field if wavelengths are 
sufficiently large. The waves are accompanied with a new mechanism of energy dissipation, {\it i.e.} an internal 
friction analogous to the Ohm's effect. Energy is supplied from the main flow to the wave field, and some part of 
the energy is dissipated into heat. Thus, there exists a sustaining mechanism,  which implies that the streaky 
structure of wall-bounded turbulence is  a {\it dissipative structure}. 

The predictions are consistent with characteristic features of wall turbulence observed in experimental studies, 
with respect to three points. ($i$) {\it Existence of traveling waves} and their relation to streamwise streaks
and streamwise vortices: The traveling waves are characterized by two scales of wavelength $\lam$ and damping-length 
$d$.  ($ii$) {\it Existence of two large scales} (LSM and VLSM) observed in turbulent shear flows: Those  are 
interpreted by the waves amplified with the transient growth mechanism and maintained by  interaction with the new 
transverse wave field. The waves are robust since they have their own energy and momentum.  ($iii$) {\it Enhanced energy 
dissipation} in wavy turbulence.  Its bulk rate of energy dissipation takes a form resembling the models of 
eddy-viscosity, and its coefficient $\nu_{\mbox{\scriptsize D}}$ is estimated to be of the order of $c_t d$ and 
much larger than the molecular viscosity.

It must be emphasized as a  physical theory that no self-contradiction is incurred by the new field introduced. \\
\end{abstract}


\maketitle

\renewcommand{\thefootnote}{\alph{footnote}}
\baselineskip=4.5mm

\section{Introduction} \label{s1}

Large scale feature is one of the main subjects in the study of wall-bounded turbulence ({\it wall turbulence},
in short) such as boundary layer turbulence, channel turbulence or pipe turbulence.  Studies of pipe flows in 
the past decades  have revealed, at transition  to turbulence, that the flow supports transverse {\it traveling 
waves} of cross-stream modes.  It is now a hundred and thirty years since the great work of Osborn Reynolds (1883) 
reported it first historically.

Introducing a new field of transverse waves  into the turbulence field,  a new scenario of turbulence theory was 
presented by Kambe (2015, 2016), proposing that the turbulence field of wall-bounded parallel shear flows 
are composed of three components: {\it mean shear} field of velocity, {\it wavy} perturbation and {\it incoherent}
turbulent fluctuations.  Significance of this formulation rests on a mathematical {\it  theorem} and 
formulation by   Scofield \& Huq (2010, 2014), stating that {\it conservation law} of current flux implies existence of 
a field governed by Maxwell-like equations, which supports transverse traveling waves. Stating compactly, the 
current conservation implies existence of transverse waves in flow field structured with vorticity. Along with 
the wave field, a new dissipation mechanism  of energy is  introduced in this work as an enhanced effect, 
caused by a drift current in the turbulence field.  {\it Transverse  traveling waves} are well known in the linear 
stability theory of laminar shear flows. But apart from the stability problem, it is remarkable that such transverse   
waves were also  found in pipe-flows or channel flows by  numerical analyses 
(Brosa (1991), Waleffe (1998), Faisst \& Eckhardt (2003)).

\vskip2mm
\centerline{($a$)  {\it Wavy streaks in background turbulence and enhanced diffusivity} } \label{s1a}
\vskip1mm  \noindent
Near-wall turbulence is characterized by the structure of low-speed streaks which are wavy and non-uniform, and 
surrounded by a sea of turbulent fluctuations. It is noteworthy that Schoppa \& Hussain (2002) proposed a triple 
decomposition of the turbulent wall-flow $\bU$ consisting of ($i$)  mean streak field of velocity $\overline{\bU}$, 
($ii$) time-dependent wavy perturbation $\bu_w$,  and ($iii$)  incoherent turbulent fluctuations $\bu'$:
\begin{equation}
\bU = \overline{\bU} + \bu_w + \bu'.  		\label{triple-deco}	 
\end{equation}
They suggested  a sustaining mechanism called  the {\it streak transient growth} for generation of 
near-wall streamwise vortices and resulting array of streaks of low-speed and high-speed. 

Much earlier, a weak organized wave in a  channel turbulence had been detected already  by Hussain \& Reynolds (1970, 
1972), where  there is an important aspect worth being remarked: that is, an eddy-diffusivity representation had been 
found to work very well to describe the  weak organized wave observed in the background turbulent fluctuations
Reynolds \&  Hussain (1972).  In fact, the waves were obtained by solving the linearized (Orr--Sommerfeld) equation 
with additional Reynolds stress, modeled by an eddy-viscosity model in their work. In another stability analysis  
of channel  turbulence by  Del \'{A}lamo \& Jim\'{e}nez (2006) too, they  obtained  disturbance wave-modes 
computationally. Importantly  they used a variable turbulent eddy-viscosity instead of the constant molecular 
viscosity  to obtain such  waves.
 
These studies imply the importance of using the eddy-viscosity models  to describe coherent waves adequately 
in turbulent flows. We are going to consider these phenomena under the light of  present new scenario.

There exist ample  evidences of streamwise streaks and long meandering structures in wall turbulence,  
observed by a number of laboratory experiments. It is remarkable that, in atmospheric flows too, long meandering 
streaky structures were observed  by Hutchins \& Marusic (2007)  in their experimental study of atmospheric 
boundary-layer flows.  Eddy-viscosity models are necessary tools in numerical analyses and DNS of atmospheric 
turbulence.  In fact, the atmospheric flow is regarded as an important area of wall-bounded turbulence.

\vskip2mm
\centerline{($b$)  {\it Transient growth and bypass transition} } \label{s1b}
\vskip1mm  \noindent
In the standard stability theory of steady laminar viscous shear flows, the flow stability is examined in terms 
of {\it eigenvalues} by solving a linear perturbation equation (such as the Orr-Sommerfeld equation) to the 
Navier-Stokes equation under imposed boundary conditions.  Suppose that  the perturbation is expressed 
by the time factor $e^{-i\om t}= e^{\om_i t} e^{-i \om_rt}$ (where $t$ is the time, and $\om_r$ and  $\om_i $ 
are real and imaginary part of the eigenvalue $\om$). According as $\om_i <0$ or $>0$, the flow is said to 
be linearly stable or unstable, respectively. 

What makes  shear flows interesting and non-trivial is that the differential operator of the linear 
perturbation equation is characterized by the {\it non-normality} or {\it non-self-adjointness}
(Gustavsson 1991;  Butler \& Farrell 1992;  Henningson, Lundbladh \& Johansson 1993;  Reddy \& Henningson 1993;  
Trefethen,  Trefethen,  Reddy \& Driscoll 1993), mainly owing to the property of the mean base flow which is dependent 
on space coordinates and even three-dimensional such as in streaky flows.  In this case, infinitesimal 
perturbations are governed by the Orr-Sommerfeld equation supplemented by 
an equation (with Squire operator) for the vorticity component in the wall-normal direction including  an 
additional non-homogeneous term of forcing.  For purely two-dimensional waves, the forcing term is zero in the 
latter equation and the homogeneous part is associated with damped Squire modes. For oblique three-dimensional 
disturbances, a new element enters the problem and the non-zero forcing term appears for the normal vorticity. 
Although the modal analysis can be carried out to an eigenproblem mentioned above,  the eigen-functions resulting 
from the non-normal operator are not orthogonal to each other. An initial perturbation may consist of modes 
that are combined to have mutual interaction.  As a result, even in the case of a stable shear flow characterized 
with $\om_i <0$ of all eigenvalues,  we obtain  considerable growth of the perturbation in amplitude, with its 
energy growing by some orders of magnitude larger than the initial value,  before the decay due to the property 
$\om_i <0$ of all the modes. This is called {\it transient growth}.

This process implies a {\it bypass} transition avoiding the regular TS-wave transition process if there 
exists a mechanism sustaining the transiently amplified waves.  The present scenario aims to
provide a dynamical mechanism sustaining the transiently amplified waves robustly.

Gustavsson (1991) studied first this mechanism for plane Poiseuille flow and clarified transient growth of 
perturbation energy. The growth was found at sub-critical Reynolds numbers (\ie \ $\om_i <0$  for all 
eigenvalues), where both the normal vorticity and  normal velocity are expressed in terms of damped modes. 
Considerable amplitudes can be reached before decay by  the mechanism of vortex stretching acting on 
the normal vorticity to the boundary wall,

For the three-dimensional perturbations, growth by a factor of order $10^3$ can occur. 
Three-dimensionality plays a key role and allows for growth of the normal vorticity through the lift-up 
mechanism. This growth generates elongated structures in streamwise direction since the growth is largest 
at low streamwise wavenumbers. Thus the optimal perturbations which grow the most generate 
streaky structure.

\vskip2mm
\centerline{($c$) {\it Two large scales in wall turbulence} } \label{s1c}
\vskip1mm  \noindent
A wind-tunnel experiment of boundary layer flows by Fransson \etal (2006)  verified delay of 
transition to turbulence by  a worked-out design on the wall to enforce streaky flow in the wall layer. Recent studies 
of wall turbulence (Kim \& Adrian (1999); Guala \etal (2006); Adrian (2007); Monty \etal (2007); 
Hutchins \& Marusic (2007); Smits \etal (2011); Rosenberg \etal (2013))  recognize existence of two large scales
of the streaky  structures:  LSM (large-scale motions) and VLSM (very-large-scale motions)\footnote{Existence of 
{\it very-large-scale motions} was recognized by Kim \& Adrian (1999), proposing a newly coined term "VLSM".},  which 
characterize the streamwise streaks and long meandering structures.  The LSMs are considered to be created 
by the vortex packets consisting of hairpin-like structures.

For pipe turbulence, study of a streamwise (denoted by $x$-axis) energy spectrum $E_\parallel(k)$ with respect 
to the streamwise wave number $k_x$ clarified that the pre-multiplied streamwise spectrum $k_xE_{\footnotesize 
\parallel}(k_x)$ has two {\it peaks} at $k_{\rm lsm}= 2\pi/\lambda_{\rm lsm}$ corresponding to LSM of 
$\lambda_{\rm lsm}/R \approx 1\sim 3$ and at $k_{\rm vlsm}= 2\pi/\lambda_{\rm vlsm}$ corresponding to
VLSM of $\lambda_{\rm vlsm}/R \approx 15 \sim 20$ (where $R$ is the pipe radius and $\lambda_x=2\pi/k_x$ 
the streamwise wave length), and decays beyond VLSM. The energy spectrum $E_{\footnotesize \parallel}(k_x)$ 
takes a scaling form 		
\[  E_{\footnotesize \parallel} \  \propto   \ k_x^{\,-1} \quad  ( k_{\rm lsm} \, \gtrsim \ k_x \ 
		\gtrsim  \ k_{\rm vlsm} ); 	\qquad 	\quad	
E_{\footnotesize \parallel} \  \propto  \ k_x^{\,-5/3} \quad 
		 (k_x \, \gtrsim \ k_{\rm lsm}). 	 \]	
In the  formulation and analysis of the  study of wall-bounded turbulence, the following two observations 
are essential for application of the present scenario.  ($i$) Owing to the  turbulent motion under complex 
coupling of vorticity and current flux, the turbulence field supports propagation of transverse waves 
(\S\ref{s221}, \S\ref{s5}),  causing wavy nature of streaks  in background turbulence. In turbulent pipe 
flows, existence of such transverse traveling waves  is confirmed both experimentally and computationally 
(Hof \etal 2004; Faisst \&  Eckhardt 2003;   Brosa 1991).  \ ($ii$)  Such wavy streaks exist robustly in background 
turbulence, and the waves can be predicted much better by using enhanced diffusivity and dissipation which are 
modeled by formulae of eddy-viscosity. 

\vskip2mm
\centerline{($d$)  {\it A new approach to turbulence theory} } \label{s1d}
\vskip1mm  \noindent 
A new approach to the turbulence theory is proposed here, aiming that the present scenario would give a hint    
why the wavy streamwise streaks are maintained in a sea of turbulent fluctuations, and also why energy dissipation 
is enhanced in turbulence as evidenced by introducing free parameters of eddy viscosity  in turbulence modeling
(without being based on physical principle; see \eg Pope (2000)).

 The present approach is stimulated by the study proposed recently by  Scofield \& Huq (2014) who introduced a new 
field into the fluid-flow field, where the new field is governed by Maxwell-type equations. This is motivated by 
the {\it electromagnetic} pioneering work of Hehl \& Obukhov (2003).  Its fluid-version was formulated as the 
mathematical Theorem of Scofield \& Huq (2010),  stating that {\it conservation of fluid current flux} implies 
existence of fields of 4-vector potential governed by Maxwell equations. In the present paper, the field is called 
as {\it Transverse-Wave} field, or {\it in short} as {\it TW}-field.\footnote{ Scofield \& Huq (2010) called it 
{\it Vortex Field}. This may be called also as {\it SH}-field.}   In fluid turbulence, the current conservation 
is a basic property, hence it is reasonable to 
introduce a TW field in turbulence. Although the present study of  TW-field is motivated by the study of Scofield 
\& Huq (2014), the fluid-dynamic mechanisms formulated here are  the present author's own, which are  ($i$) the 
dynamical mechanism exciting the TW-waves, ($ii$) existence of a channel supplying energy to TW-field, and  
($iii$) a new dissipation mechanism by a drift current in turbulence field.\footnote{Scofield \& Huq (2014) 
mentioned also energy dissipation analogous to the Joule heating, without noting that the convection current 
$\bj_c =\rho \bv$ of Eq.(\ref{current-flux}) may have both signs of input and output of energy to the TW-field, 
nor mentioning significance of the drift current $\bj_d$ (a fluid version of the Ohm's law) in the turbulence theory.}  

The mechanism of energy dissipation of ($iii$)   is introduced by expressing the current flux $\bj$ in terms 
of two components: the convection current $\bj_c$ and  a new {\it drift} current $\bj_d$:
\begin{equation}
 \bj= \bj_c + \bj_d,  \qquad 		\label{current-flux}	 
\end{equation}
where $\bj_c = \rho\bv$ with $\bv$ the fluid velocity, and the drift current $\bj_d$ is explained briefly in the 
subsection ($f$) below. More detailed description of the drift current is given by  \S\ref{s232} and \S\ref{s7}.

It is remarkable that the TW-wave field  has its own momentum and energy like the electromagnetic waves, which 
explains the robustness of the wave, and accompanies its own mechanism of energy dissipation.  In fluid 
mechanics so far, the wave energy and momentum are considered only for longitudinal acoustic waves  
(Landau \& Lifshitz (1987), \S65; Lighthill (1978), Ch.1). 
Regarding the transversal waves in fluids, such consideration is not seen.  However, even 
the current theory of fluid mechanics (\S\ref{s21}) can describe transverse waves. The section \ref{a4} of {\it 
Appendix A} interprets this circumstance in the absence of TW-field. It is remarkable that the vorticity field 
$\bom(\bx, t)$ supports {\it transverse waves} under the constraint of the continuity equation. In fact, 
reflecting the hydrodynamic stability theory, disturbance waves obtained by solving the Orr-Sommerfeld equation 
are certainly such transverse traveling waves.

\vskip2mm
\centerline{($e$)  {\it Structure of the paper} } \label{s15}
\vskip1mm  \noindent 
It is proposed in the present study that the whole system under consideration consists of two fields: {\it Flow 
Field} (FF-field, in short) and  {\it Transverse-Wave} field (TW-field).  Fundamental equations governing the 
combined system are composed of conservation equations of {\it mass}, {\it energy} and {\it momentum}. Hence, 
the present formulation follows the fundamental principle of theoretical physics. This  paper is structured with 
three parts. 

($i$)\ In the first part, theoretical frame of the new scenario and its significance is presented in \S2\,({\it 
Governing equations}), \S3\,({\it Excitation of TW field}), and  \S4\,({\it Energy and momentum budgets of 
TW-field}).   State of the fluid is defined by the flow velocity, thermodynamic variables such as density, pressure,
entropy, \etc. The TW-wave field is described  in terms of a vector potential $\ba$ and a scalar potential $\phi_a$, 
from which  fluid $\be$-field and $\bfb$-field are defined by $\be= -\pl_t\ba -\nabla \phi_a$ and $\bfb = 
\nabla\times \ba$.  The  {\it entropy} equation accounts for the heat liberated by dissipative mechanisms 
including a new dissipation term. How  the new TW field is excited is considered in \S\ref{s3}.  General 
derivation in terms of the energy-momentum tensors are  also presented to support the validity of 
the present formulation.  

The new  formulation has more degrees of freedom than the current one. The new field has its own energy and 
momentum, which means robustness of the new field and in addition the new field is equipped with a new 
dissipation mechanism, of which scaling  estimate predicts enhanced dissipation  comparable with the  
eddy-viscosity models. 

($ii$)\ Second part is concerned with  application of the present formulation. One of the important  areas of 
possible application  would be the streaky shear-flow turbulence.  To begin with, wave-propagation traveling through  
turbulence is investigated in \S\ref{s5}\,({\it Traveling waves and wave dynamics (large scale motion)})
with having the two large scales LSM and VLSM in mind. Dynamical process of growth and decay of TW-waves is 
studied on the basis of wave equations equipped with terms of source and damping,  where the TW waves are 
characterized  with a wavelength $\lam$ and  a damping distance $d$. 

The section 6\,({\it Streaky wall turbulence}) presents an endeavor to clarify how the streaky structure is 
generated in wall turbulence and maintained robustly in turbulent environment, on the basis of the new scenario.
In these phenomena, two key ideas of transient growth mechanism and triple decomposition of velocity
play non-trivial roles,  and relevant studies are reviewed there.   After considering the wave dynamics of 
excitation and damping, it is proposed that the streaky structure in wall turbulence is a {\it dissipative 
structure} characterized with two large scales of wave length $\lam_{\rm lsm}$ and $\lam_{\rm vlsm}$. 

($iii$)\ The energy dissipation is caused  by an internal drift current $\bj_d$, which is driven by the $\be$ 
field and represented by a linear relation $\bj_d \propto \be$, which is called {\it turbulence}-Darcy effect or 
{\it D}-effect shortly, explained  briefly in the next subsection ($f$).  The section 7 studies the new mechanism 
of energy dissipation in detail. It is found that the bulk rate of {\it D}-effect dissipation takes a form analogous to 
the eddy-viscosity models, and its coefficient is  comparable in magnitude with the eddy-viscosities.  The 
dissipation formula is derived analytically from the basic governing equations, unlike usual eddy-viscosity models.

It is found in \S\ref{s5} that the damping distance $d$ owing to the {\it D}-effect gets reduced for larger 
waves ($\lam > \lam_{\rm lsm}$), meaning those to suffer stronger damping.  An interesting property is 
that the resistive drift current $\bj_d= \sg \be$ (with a positive constant) causes a phase shift between the 
flow perturbation $u_x(t)$ and the wave field  $a_x(t)$, enabling energy transfer from the flow field to the 
wave field.  If $\lam > \lam_{\rm lsm}$, the TW-wave  gains energy from the flow field.

\vskip5mm
\centerline{($f$)  {\it Dynamical and dissipative mechanisms of the new field} } \label{s16}
\vskip1mm  \noindent %
The dynamical mechanism  exciting the TW-waves is studied by a model equation derived in the present theory 
(\S\ref{s5}):
\[ \nabla^2 \be - c_t^{\,-2} \pl_t^{\,2} \be  = \mu \, \pl_t (\rho \bv) + \mu \sg \pl_t \be,	\]
(the first of (\ref{TW-eh-2})).  The left hand side (\underline{\it lhs}) is a wave equation describing waves 
traveling with the phase speed $c_t$ of a vector field $\be$ (a fluid-electric field defined in \S\ref{s22}). The 
second term on the \underline{\it rhs} (right hand side) is a damping term, while the first on \underline{\it rhs} 
is a term either exciting the wave and supplying energy to the TW-field by  extracting from the flow field $\bv$, 
or extracting energy from the TW-field and giving back to the flow field, depending on the sign of the source term 
$\bj \cdot \be$ (see \S\ref{s22}).

It is an important aspect of the present theory that the TW-field accompanies its own mechanism of energy 
dissipation caused by a drift current $\bj_d $ assumed to exist in turbulence field. It is proposed  that this 
current is generated  by an effect called  a {\it turbulence}-Darcy effect.
The well-known Darcy's law is a law to 
describe the current flux of a viscous fluid through a porous medium under an imposed pressure gradient.
In the present case of turbulent flow, the fluid in motion is acted on by an additional force from the TW-wave field, 
which is derived  as a force of $\bF \sim \rho \be$  (\S\ref{s23}).  In a turbulent state coexisting with the 
TW-field, this force would give rise to an internal drift current $\bj_d$ through a turbulent medium 
composed of a number of turbulent eddies. The turbulence-Darcy law is proposed to describe a proportional 
relationship between the current flux $\bj_d$ and an applied force $\rho\be$, given by $\bj_d=\sg \be$
where $\sg$ is a scalar constant. This resembles the Ohm's law in 
electromagnetism.\footnote{Although the Darcy's law is the discharge rate of  a viscous fluid through 
a porous medium under imposed pressure gradient, it is analogous to Ohm's law in the electromagnetism, or 
Fourier's law in the heat conduction.} 

Energy dissipation by the {\it turbulence}-Darcy effect, called  {\it D}-effect simply,\footnote{This effect 
is called so because it is analogous to the \underline{D}arcy effect, it drives \underline{d}rift current and 
it causes enhanced \underline{d}issipation and wave \underline{d}amping.}  is given by $Q_{\rm D} = 
\be \cdot \bj_d = \sg^{-1}|\bj_d|^2$.  As shown in \S\ref{s71}, the  dissipation  $Q_{\rm D}$ can be expressed 
in a form analogous to the viscous rate of dissipation. In fact, using $|\bj_d| =\rho v_d$  and 
$\sg \sim \rho d/c_t$, we have
\[ Q_{\mbox{\scriptsize D}}/\rho \sim\ \nu_{\mbox{\scriptsize D}}\,(v_d/d)^2 
	\hskip10mm 	\nu_{\mbox{\scriptsize D}} =c_t\,d.		\]
The coefficient $\nu_{\mbox{\scriptsize D}}= c_t\,\!d$ is analogous to the eddy-viscosity, 
with its magnitude  of the order of product of  a velocity $c_t$ (speed  of transverse 
waves in turbulence) and the damping distance $d$ of  the transverse wave. Its magnitude is estimated from the pipe 
turbulence data at $Re =2RU/\nu_m \approx 10^5$   (Kim \& Adrian, 1999) in \S\ref{s71}, and compared with the molecular 
kinematic viscosity of air $\nu_m$ at normal conditions. It is found that $\nu_{\mbox{\scriptsize D}}$ is much 
larger  than $\nu_m$ by some orders of magnitude.  It is in fact interesting to find that the turbulence-Darcy
dissipation in a volume takes a form analogous to  
eddy-viscosity models (see \S\ref{s7}).

\vskip2mm
Last but not least,  the system of equations  formulated in \S2 is not contradictory to the mathematical
framework of Theoretical Physics.  This formulation may shed a light to some unclear properties of 
turbulence or remove uncertain covering over them.  This is the aim of the present study.

\section{Governing equations }  \label{s2}
First of all, we consider the theoretical frame of the  scenario. Namely, 
the whole physical system  is composed of two fields: Fluid-Flow field  ({\it FF}-field) and Transverse-Wave field 
({\it TW}-field).  In  the beginning, the current theory of {\it FF}-field in the absence of the {\it TW}-field 
is reviewed first in \S\ref{s21}. Then in the next section \S\ref{s22}, the equations of {\it TW}-field  are  
presented.   The equations of combined field  (whole field) are  presented in \ref{s23}.


\vspace{-5mm}
\subsection{Equations of fluid flows (FF-field)  {\rm (}Review of the current theory{\rm )}}  \label{s21}

\subsubsection{Current system\,{\rm :}}   \label{s211}   
We consider flows of a viscous fluid of non-uniform density, which are 
governed by the system of equations of conservation  of {\it mass, momentum} and {\it energy}:
\begin{eqnarray} 
&& \pl_t\rho  + \divg \rho\bv = 0,						\label{FF-C}  \\
&& \pl_t (\rho \bv)_i + \pl_j \Pi_{ij} = 0,,					\label{FF-M0}  \\ 
&& \pl_t \big[ \rho(\frac{1}{2} v^2 + \ine ) \big] + \divg\,\bq_{\rm f}  = 0 ,  \label{FF-E}
\end{eqnarray} 
(Landau \& Lifshitz, 1987),  {\it Fluid Mechanics}), where $\pl_t=\pl/\pl t$,  $\pl_j=\pl/\pl x^j$ with $x^j$ being the 
Cartesian coordinates, $\bv$ is the 
fluid velocity, $\rho$ the fluid density, $\ine$ the specific internal energy (\ie per unit mass) of the fluid, 
$\bq_{\rm f}$ is the FF-energy flux defined by  
\begin{equation}
\bq_{\rm f} = \bq_{\rm f}^{(0)}  - \bv\cdot \tau^{(vis)} - k_T\,\nabla T, \qquad  
		\bq_{\rm f}^{(0)} \equiv  \rho \bv (\frac{1}{2} v^2+ h),			\label{q-FF}
\end{equation}
where  $h =\ine + p/\rho$ is the specific enthalpy, $T$ the temperature and $k_T$ the thermal diffusivity. The 
viscous stress tensor $\tau^{(vis)}= (\tau^{(vis)}_{ij})$ is  defined by (\ref{vis-str}) of Appendix B.  

The momentum flux density tensor $\Pi_{ij}$ of (\ref{FF-M0}) is defined by
\begin{equation}
\Pi_{ij} = \rho v_i v_j + p\,\del_{ij} - \tau^{(vis)}_{ij}.		\label{Pi-FF}
\end{equation}
Using the mass conservation equation (\ref{FF-C}) and a thermodynamic equation $(1/\rho) \dd p= \dd h -T \dd s$ 
($s$: specific entropy), the momentum equation (\ref{FF-M0}) is transformed to an equation of motion of 
a viscous incompressible fluid (by assuming $\rho=\const$):
\begin{equation}  
\pl_t \bv + \nabla(h +\half |\bv|^2) = - \bom \times \bv + T \nabla s + \nabla\cdot \tau^{(vis)},  \label{EqM-S21}  
\end{equation}
where $( \nabla \cdot \tau^{(vis)})_i = \pl_j \tau^{(vis)}_{ji}$.  Taking $\curl$, we obtain the vorticity equation:
\begin{equation}  \hspace*{-15mm} 
 \pl_t \bom = - \curl( \bom \times \bv  - T \nabla s - \nabla\cdot \tau^{(vis)}) 
        = - \curl( \bom \times \bv) + \nabla T \times \nabla s + \nu \nabla^2 \bom,	\label{Vort-S21}
\end{equation}
where $\nu=\eta/\rho$ is the kinematic viscosity, and (\ref{vis-for}) is used to obtain the last term.  

The entropy equation is given by
\begin{eqnarray}
\rho T \frac{\D}{\D t} s & = & Q_{\rm vis} + Q_T ,				\label{FF-entropy}  \\
   && \quad Q_{\rm vis} \equiv \mbox{$\sum_{i,j}$} \,\pl_jv_i\ \tau^{(vis)}_{ij}, 
   	\quad Q_T \equiv \divg (k_T \nabla T),  				\label{Q-vis}
\end{eqnarray}
where $\DD/\DD t \equiv \pl_t+ \bv\cdot\nabla$ is the convective derivative. The term $ Q_{\rm vis}$ is the 
energy dissipated into heat by the viscosity and can be shown to be non-negative (Landau \& Lifshitz (1987), \S49).
The term $Q_T$ is the heat conducted into the volume concerned. \\[-3mm]

\vspace{-5mm}
\subsubsection{General derivation in terms of the energy-momentum tensor of fluid flow\,{\rm :}}  \label{s212}
Having in mind later formulation of  the whole combined field of {\it FF}-field and {\it TW}-field (in \S\ref{s23}), 
general formalism of {\it theoretical physics} is applied to the present system of 
{\it FF}-field  on the basis of the Lagrangian density $\Lam_{\rm f}$ and hence the variational principle.  In this 
section, we  derive the same equations (\ref{FF-M0}) and (\ref{FF-E})   from the general principle.

Field equations are derived in accordance with the  general principle of least action in four-dimensional space-time \ 
$x^\mu =(t, x^1, x^2, x^3)$.  The Lagrangian density $\Lam_{\rm f}$ is a certain functional of the fields 
$q_\gam(x^\mu)$ describing the state of the system, where in the field $q_\gam(x^\mu)$ included are three  components 
of velocity field and two thermodynamic variables, \etc.  The action  $S_{\rm f}$ for the fluid flow is defined 
by the form, $S_{\rm f} = \int \Lam_{\rm f}(\,q_\gam(x^\mu)\,)\,\dd \Om$, where $\dd \Om= \dd x^0 x^1 x^2 x^3$. 
The governing equations of motion are derived as the Lagrange's equation in general with taking variation of 
the Lagrangian density $\Lam_{\rm f}$ by varying $q_\gam$. 

However, we are interested here in deriving the conservation equations of energy and momentum, which are 
represented by   {\small 
\begin{equation}	
	\pl_\al T^{\al\beta}_{\rm f} = 0,		\label{T-cons-s21}
\end{equation}
where $T^{\al\beta}_{\rm f} $ is the {\it energy-momentum} tensor (or stress tensor) of fluid flow, defined by
\begin{equation}	
T^{\al\beta}_{\rm f}  =   \left( \begin{array}{cccc}  
	T^{00}  	&  T^{01}  	& T^{02}  	& T^{03} 		\\	 
        T^{10}  	&  \Pi^{11}     & \Pi^{12}      & \Pi^{13}  		\\	
	T^{20}  	&  \Pi^{21}     & \Pi^{22}      & \Pi^{23}  		\\	
	T^{30}  	&  \Pi^{31}     & \Pi^{32}      & \Pi^{33}  		
        \end{array}  \right), 					
\qquad   \left.   \begin{array}{lcl}  
	  T^{00}    & = &  \half \rho v^2 + \rho \ine ,	     	\\
          T^{k0}    & = &  \rho v^k (\half v^2 + h ),  		\\
          T^{0k}    & = &  \rho v^k, \\
          \Pi^{ik}  & = &  \rho v^i v^k + p\,\del_{ik} = \Pi^{ki}
        \end{array}   \quad  \right\}    		      			\label{T-f_ab}  
\end{equation}} 
\noindent where $i,\, k = 1, 2, 3$ and $h =\ine + p/\rho$ is the enthalpy. General relativistic derivation of the tensor 
$T^{\al\beta}_{\rm f}$ is given concisely in Appendix \ref{b4} and Landau \& Lifshitz (1987 \S133).

The {\it time} component ($\beta=0$) of Eq.(\ref{T-cons-s21}) is  given by $\pl_\al T^{\al 0}_{\rm f} = 0$; 
namely, we have 
\begin{equation}
 \pl_t T^{00} + \pl_k T^{k0} =  \pl_t \big[ \rho ( \half  v^2 
 	+ \ine )\big]  + \divg \big[ \rho \bv (\half v^2 + h ) \big]  = 0,	 \label{e-cons-s21}
\end{equation}
where the terms of (\ref{T-f_ab}) are used. This  is equivalent to the energy  equation 
(\ref{FF-E}) with the energy flux vector $\bq_{\rm f}$ replaced by $\bq_{\rm f}^{(0)}$ of (\ref{q-FF}) without 
viscosity effect and thermal conduction effect.

The {\it space} component ($\beta= k$ with $k =1,,2,3$) is  given by $\pl_\al T^{\al k}_{\rm f}= 0$; namely, we have 
\begin{equation}
 \pl_t T^{0 k} + \pl_i \Pi^{ik} = \Big[ \pl_t \big( \rho v^k ) + \pl_i \big( \rho v^i v^k + p\,\del_{ik} \big) \,\Big]  
        = 0,  	\quad (k =1, 2, 3),	\label{mom-cons-s21}  
\end{equation}
 This  is equivalent to the momentum  equation (\ref{FF-M0}).
 
Thus, it is shown that the conservation equations of energy and momentum in the current theory are interpreted by 
the general formalism of theoretical physics.

\vspace{-3mm}
\subsection{Equations of a new field of transverse waves (TW-field)}  \label{s22}
\vspace{-2mm}
Conservation law of the current flux $\bj$ allows existence of a transverse wave field governed by Maxwell-type 
equations. Governing equations are presented concisely here, while details of its derivation are 
described in Appendix A. In order to present the equations of {\it Transverse Wave} (TW) field,  we take analogy 
with the electromagnetism (EM). Needless to say, the Maxwell equations of EM-theory support transverse waves. 
Even within the framework of the current theory of fluid mechanics (\S\ref{s21}), transverse waves are excited. 
The section \ref{a4} of {\it Appendix A} interprets this circumstance in the absence of TW-field. It is 
remarkable that the vorticity field $\bom(\bx, t)$ supports {\it transverse waves} under the constraint of 
the continuity equation. 

In the present new scenario, one can point out two  features characterizing difference from the current theory: 
($i$) there is an increased freedom expressed by the 4-vector-potential $a_\mu = (\phi_a, - a_x, - a_y, - a_z)$ of 
(\ref{4-vec-pot}) under the gauge condition of (\ref{Lorenz-c}), which are  introduced newly to represent 
the TW-field, and   ($ii$) the TW-field accompanies its own dissipation mechanism  termed 
as {\it turbulence}-Darcy effect, which leads to enhanced rate of energy dissipation.

\vspace{-5mm}
\subsubsection{A system of fluid-Maxwell equations} 		\label{s221} 
Conservation law of fluid current allows existence of transverse wave field governed by Maxwell-type 
equations. Details of derivation are given in Appendix A, and only the resulting equations are given 
here compactly. In order to represent the {\it Transverse Wave} (TW) field,\footnote{The {\it TW}-field is 
called by Scofield \& Huq (2014) as {\it Vortex Field}.} we take analogy with the electromagnetism (EM) faithfully. 

\setcounter{footnote}{0}
Firstly, suppose that we have a 4-component current  $j^\mu=(\rho, j_x, j_y, j_z)$ of fluid 
flow, satisfying the following mass conservation equation (see (\ref{A-C-cons1})):
\begin{equation}
 \pl_t \rho +  \divg \bj =0,   \hskip10mm \bj= (j_x,  j_y, j_z) . 		\label{C-cons1}
\end{equation}
where   $\bj$ is the space part of  $j^\mu$.  According to detailed mathematical analysis 
described in Appendix \ref{a2}  ({\it helped} by de Rahm Theorem),  existence of two excitation fields $\bd$ and $\bh$ 
are deduced, which are governed by a pair of equations (a pair of Maxwell-like equations, \S\ref{a2}):
\begin{equation} 
- \pl_t \bd + \curl \,\bh = \bj, 	\hskip15mm  \divg \bd = \rho. 		\label{TW-bf} 
\end{equation}
It is essential in the present formulation to realize that  the conservation law (\ref{C-cons1}) is satisfied 
identically by this set of equations.\footnote{ It is assumed that both of secondly-differentiable fields 
$\bd$ and $\bh$ satisfying (\ref{TW-bf}) exist.  Denoting the constant mean density by $\overline{\rho}$,
the second equation of (\ref{TW-bf}) may be replaced by $\divg \overline{\bd} =\overline{\rho}$ and 
$\divg \bd' = \rho'$, where $\bd' = \bd - \overline{\bd}$. }

Secondly, associated with the two potentials of vector  fields $\bd$ and $\bh$  in the 4-space-time $x^\mu=(t, x, y, z)$,  
one can introduce a field of 4-vector-potential by
\begin{equation} 
a_\mu = (\phi_a, - a_x, - a_y, - a_z) = (\phi_a, - \ba),		\label{4-vec-pot} 
\end{equation}
and define an 1-form $\A^1$  by\footnote{In the mathematics of exterior differential forms, a differential 1-form, 
$\A^1 = a_\mu\, \dd x^\mu$, is defined by a pair of a coordinate 4-{\it vector} $x^\mu = 
(x^0, x^1, x^2, x^3) \equiv (t, x,y,z)$ and a potential 4-{\it covector} $a_\mu = 
(a_0, a_1, a_2, a_3) \equiv (\phi_a, -a_x, -a_y, -a_z)$.}  
\begin{equation}
 \A^1 = a_\mu \dd x^\mu = \phi_a \,\dd t - a_x \dd x - a_y \dd y -  a_z \dd z.  	\label{1-formA}
\end{equation}
Assuming existence of the 1-form $\A^1$  and denoting the space part of $a_\mu$ by $- \ba = -(a_x, a_y, a_z)$,  
one can define a pair of {\it fluid}-electric field $\be$ and {\it fluid}-magnetic field  $\bfb$ by 
\begin{equation}
\be \equiv -\pl_t\ba -\nabla \phi_a, \hskip15mm  \bfb \equiv \nabla\times \ba,  \label{be-bfb}
\end{equation}
where $\ba$ plays the part of a vector potential analogous to the electromagnetism. It will be shown 
below (in \S\ref{s31}) that $\ba$ has the dimension of velocity.  From the definitions of $\be$ and $\bfb$ 
of (\ref{be-bfb}), it is evident that the  following set of  equations, \ie another pair of Maxwell-type equations, 
are satisfied identically (Appendix \ref{a1}):
\begin{equation}
	\pl_t \bfb + \curl \,\be =0, 	\hskip15mm  \divg \bfb=0. 		\label{TW-a}
\end{equation}
Thus, the dynamics of TW-field is defined by the system of Maxwell-type equations (\ref{TW-bf}) and 
(\ref{TW-a}) (Appendix \ref{a4}),  where the constitutive relations, 
\begin{equation} 
	\bd = \eps \be, \qquad \bh = \mu^{-1} \bfb,			\label{eps-mu} 
\end{equation}
are assumed with using field parameters $\eps$ and $\mu$  analogous to those of 
the EM theory.  Here, lower-case bold-letters are used for the TW-field vectors in order to show clear 
analogy to the Maxwell equations of electromagnetism (Jackson 1999) with corresponding upper-case letters. 

Transverse waves are naturally  supported by (\ref{TW-bf}) and (\ref{TW-a}). In fact, wave equations 
are derived  for  $\be$ and $\bh$ from the above four equations:
\begin{eqnarray} 
\Bigl[\nabla^2 - c_t^{-2} \pl_t^{\,2} \Bigr]\,\be & = & \mu\,\pl_t \bj +\eps^{-1}\nabla \rho,  \label{WE-be} \\
\Bigl[ \nabla^2 - c_t^{-2} \pl_t^{\,2} \Bigr] \,\bh & = & - \nabla \times \bj,      \label{WE-bfb}  	\\
 &&    c_t  = 1/ \sqrt{\eps \mu}.			\label{TW-vel} 
\end{eqnarray}
where $\eps$ and $\mu$ are assumed constant.

It is found that transverse waves of phase velocity $c_t$ are excited by sources 
characterized with unsteadiness and rotationality of the current $\bj$, expressed by $\pl_t \bj$ and  
$\nabla \times \bj$ respectively, and also with density non-uniformity ($\nabla \rho$).

Under these circumstances, the $\ba$-field and $\phi_a$-field are excited.   However there exists arbitrariness  
in the definitions $\ba$ and $\phi_a$ of (\ref{be-bfb}) in connection with the Maxwell equations. 
Usually the following Lorenz condition is imposed in order to resolve it:
\begin{equation}
\nabla \cdot \ba + c_t^{-2} \,\pl_t \phi_a = 0. 		\label{Lorenz-c}
\end{equation}
Thus, using (\ref{be-bfb}), the equations of $\ba$ and $\phi_a$  are given by
\begin{equation} 
\Bigl[ \nabla^2 - c_t^{-2} \pl_t^{\,2} \Bigr] \,\ba = - \mu\, \bj, \hskip8mm 
\Bigl[\nabla^2 - c_t^{-2} \pl_t^{\,2} \Bigr]\,\phi_a = - \eps^{-1} \rho', 		\label{WE-aphi}
\end{equation}
where  $\rho'=\rho - \overline{\rho}$ denotes fluctuating part of density with $\overline{\rho}$ the
mean density.

\vspace{-4mm}
\subsubsection{Equations of energy and momentum} \label{s222}
Equations of energy and momentum are  derived immediately from (\ref{TW-bf}) and (\ref{TW-a}) as 
\begin{eqnarray}
\hspace*{-5mm} \pl_t w_e  + \divg \bq_{\mbox{\tiny fP}} & = & - \bj \cdot\be,   	\label{TW-e}  \\
\pl_t \bg  + \nabla\cdot M  \ & = & - \bF_{\mbox{\tiny L}}[\ba],    \label{TW-m}
\end{eqnarray}
respectively (\ref{b3}; and Jackson (1999), Scofield \& Huq (2014)),	
({\footnotesize  See Jackson (1999, \S12.10), where $T_{ij}^{(M)} = - M_{ij}$ is used in stead of $M_{ij}$ 
for the Maxwell stress.} ),
where $w_e$ and $\bg$ are  the energy density  and  momentum density of TW-field, defined respectively by 
\begin{eqnarray}
w_e= \half (\be\cdot\bd + \bh\cdot\bfb),  & \qquad  &  \bg= \bd \times \bfb  \ ;   \label{en-mom}  \\
 &&  \hspace*{-40mm}  \mbox{and}  \hspace*{30mm}   \bq_{\mbox{\tiny fP}} = \be \times \bh = c^2 \bg,
 	\qquad  (\nabla\cdot M)_j = \pl_i M_{ij}.			 \label{Poynting}
\end{eqnarray}   
The vector $\bq_{\mbox{\tiny fP}}$  is a {\it fluid} Poynting vector (TW-energy flux). The tensor 
 $M=(M_{ij})$  is a {\it fluid} Maxwell-stress defined by (\ref{M-stress}). Fluid Lorentz force
$\bF_{\mbox{\tiny L}}[\ba]$ is defined by  
\begin{eqnarray}  
\bF_{\mbox{\tiny L}}[\ba] = \rho\,\bff_{\mbox{\tiny L}}[\ba],  & \qquad  & 
\bff_{\mbox{\tiny L}}[\ba] \equiv  \be + \rho^{-1} \bj\times \bfb.		\label{F_Lor}  
\end{eqnarray}  
(See (\ref{be-bfb})  for the definition of $\be$ and $\bfb$.)  
Note that $\bff_{\mbox{\tiny L}}[\ba]$ is a {\it fluid} Lorentz force per unit mass.  The term $- \bj\cdot \be$ 
on the right of (\ref{TW-e}) is an energy source (\,if $ \bj\cdot \be <0$) or loss (\,if $ \bj\cdot \be > 0$),  
and the term $- \bF_{\mbox{\tiny L}}$ on the right of (\ref{TW-m}) is Lorentz-force reaction. These are 
identical in form to those of the electromagnetic theory.

\vspace{-2mm}
\subsubsection{General formulation in terms of the energy-momentum tensor of TW wave} \label{s223}
\vspace{-1mm}
As done for the FF-field in \S\ref{s212},  general formalism is presented here for the TW wave too to derive the 
same equations of energy (\ref{TW-e}) and  momentum (\ref{TW-m}) of the {\it TW}-field on the basis of  Lagrangian 
density, aiming at general derivation of the  equations of the whole combined 
field of {\it FF}-field and {\it TW}-field in  \S\ref{s23}, 

Suppose that  we have a 4-vector potential $a^\mu =(\phi_a, a_1, a_2, a_3) =(\phi_a, \ba)$ (a {\it contra-variant} 
vector) and its {\it covariant} version $a_\mu  =( g_{\mu\nu} a^\nu) =(\phi_a, - \ba)$ in the space-time  
$\xi^\mu =(x^0, \bx)$ with $x^0 \equiv \tau =c_t t$ and $\bx=(x^1, x^2, x^3)$ ($a_\mu$ is already defined in 
\S\ref{s221}),\footnote{In the space-time representation, greek letters such as $\al,\,\beta,\, \mu,\, \nu, \,\lam$ 
denote $(0,1,2,3)$ and roman letters such as $i, \,k$ denote $(1,2,3)$, and  the metric tensor is defined by 
$ g_{\mu\nu} = g^{\mu\nu} ={\rm diag}(1, -1, -1, -1)$.}, where the coordinate variable $\xi^\mu$ is used instead 
of $x^\mu$ since $\xi^0$ is defined by $\tau =c_t t$ having the same dimension as $x^k$. and 
$c_t= 1/\sqrt{\eps\mu}$ of (\ref{TW-vel}) is the phase velocity of transverse wave under consideration.

Free-field Lagrangian density $\Lam_0$ and  {\it energy-momentum} tensor $T_{\rm w}^{\al\beta}$ (\ie the stress 
tensor) are defined in Appendix \S\ref{b2}. In the presence of external excitation represented by the current 
4-vector $j^\nu= (c_t \rho, j_1, j_2, j_3)$, the {\it Lagrangian density} $\Lam_{\rm w}$ of the present TW-field 
is given by 
\begin{eqnarray}
\Lam_{\rm w} & = & \Lam_0 - c_t^{-1} j_\lam a^\lam, \quad \Lam_0  = \frac{1}{2\mu} 
			[\,(\overline{\be},\, \overline{\be}) -  (\bfb, \bfb )] ,	  \label{LTW-s22}   \\
\overline{\be} & \equiv & \be/c_t , \qquad \overline{e}_{\lam} = - \pl_\tau a^\lam 
	- \pl_\lam \overline{\phi}, \qquad   j_\lam = (c_t \rho, -j_1, -j_2, -j_3),  \label{ec-s22}
\end{eqnarray}
where $\Lam_0$ is the free-field Lagrangian given by (\ref{LD-eb}),  and  $\overline{\phi} = \phi_a/c_t$.   
The Lagrangian $\Lam_{\rm w}$ of {\it TW}-wave has an interaction term $- c_t^{-1}j_\lam a^\lam$ to 
represent excitation by  the 4-current $j^\mu$.

Fundamental conservation laws are represented by the form $\pl_\al T_{\rm w}^{\al\beta} = f^\beta$ in general 
theoretical physics, where $T^{\al\beta}_{\rm w}$ is the energy momentum tensor and  $f^\al$ is an external 
forcing to excite the {\it TW}-field.  From the canonical stress tensor $T^{\al\beta}_{\rm w}$ defined  by 
(\ref{EM-tensor-b2}),  one can construct a symmetric stress tensor $\Theta^{\al\beta}$, which satisfies
\begin{equation}
\pl_\al T^{\al\beta}_{\rm w} = \pl_\al \Theta^{\al\beta}_{\rm w}  .  				\label{d-Tab-s22} 
\end{equation}
The symmetric stress tensor $\Theta^{\al\beta}$ is given by 
\begin{eqnarray}
        \frac{1}{\mu} \, g^{\al\nu}\overline{F}_{\nu\lam}\overline{F}^{\lam\beta}  + \frac{1}{4\mu} \,
        g^{\al\beta} \overline{F}_{\nu\lam} \overline{F}^{\nu\lam} = \Theta^{\beta\al}_{\rm w}, \label{Th-ab} 
\end{eqnarray} 
where $F^{\al\beta}= \pl^\al a^\beta - \pl^\beta a^\al$ is the field-strength tensor,  and  $\pl_{\al} \equiv 
(\pl_\tau, \nabla)$, \ $\pl^{\beta}=g^{\beta\al}\pl_\al = (\pl_\tau, -\nabla)$ \  (see  \S\ref{b2}, and 
Jackson (1999) \S12.10).   
Using (\ref{Th-ab}), we have 
\begin{eqnarray}
\pl_\al \Theta^{\al\beta}_{\rm w} 
& = & \frac{1}{\mu} \,(\pl^{\nu} \overline{F}_{\nu\lam}) \overline{F}^{\lam\beta}  
  	+    \frac{1}{\mu} \,\Big[  \overline{F}_{\nu\lam} (\pl^{\nu} \overline{F}^{\lam\beta} ) 
        +  \half \overline{F}_{\nu\lam} \,\pl^\beta \overline{F}^{\nu\lam} \Big]  .      \label{dT-ab}    
\end{eqnarray} 
The factor $\pl^{\nu} \overline{F}_{\nu\lam}$ of the first term is given by $\mu \,j_\lam= \mu (c_t\rho, -j_1, 
-j_2, -j_3)$. In fact, we have 
\begin{eqnarray*} 
\pl^{\nu} \overline{F}_{\nu 0}  & = & \pl^1 \overline{F}_{10} + \pl^2 \overline{F}_{20} + \pl^3 \overline{F}_{30} 
	= \pl_k \overline{e}_k	= \mu c_t \,\pl_k d_k = \mu\, c_t \rho = \mu\,j_0,		\label{F-n0}	\\
\pl^{\nu} \overline{F}_{\nu k}  & = & \pl^0  \overline{e}_k + \vep_{k\al\beta} \pl^\al b_\beta 
        = c_t^{-2} \pl_t e_k- (\nabla \times \bfb)_k = \mu\eps \,\pl_t e_k- \mu (\nabla \times \bh)_k 	\nonumber \\
               & = & - \mu \, j_k .			\label{F-nk}
\end{eqnarray*}
where the two equations of (\ref{TW-bf}) are used to obtain the last expression of the two. The second term 
$(1/\mu) [ \,\cdots\,]$ on the \underline{\it rhs} of (\ref{dT-ab}) can be shown to vanish (see  Jackson (1999) 
\S12.10 C).  Thus, the equation (\ref{dT-ab}) reduces to 
\begin{equation}
\pl_\al \Theta^{\al\beta}_{\rm w} =  j_\lam \,\overline{F}^{\lam\beta} . 		\label{cons-law-s223}
\end{equation}
This is the general form of conservation equations of energy and momentum under external forcing of
$j_\lam \,\overline{F}^{\lam\beta}$.  From  \ref{b2}, the symmetric stress tensor $\Theta^{\al\beta}$ is given by
 {\small 
\begin{equation}  \hspace*{-20mm}	
\Theta^{\al\beta}_{\rm w}  =   \left( \begin{array}{cccc}  
	\Theta^{00}  	&  \Theta^{01}  & \Theta^{02}  	& \Theta^{03} 		\\	 
        \Theta^{10}  	&  M^{11}     & M^{12}      & M^{13}  		\\	
	\Theta^{20}  	&  M^{21}     & M^{22}      & M^{23}  		\\	
	\Theta^{30}  	&  M^{31}     & M^{32}      & M^{33}  		
        \end{array}  \right), 					
\qquad   \left.   \begin{array}{lcl}  
	  \Theta^{00}    & = &  w_e  = (2\mu)^{-1} [\,(\overline{\be},\, \overline{\be}) + (\bfb, \bfb )], \\
          \Theta^{k0}    & = &  c_t^{-1} (\be \times \bh)_i = c_t^{-1} (\bq_{\mbox{\tiny fP}})_i,   		\\
          \Theta^{0k}    & = &   c_t (\bd \times \bfb ) = c_t \,\bg \ (= c_t^{-1} \,\bq_{\mbox{\tiny fP}}\,),   \\
          M^{ik}         & = &  - (\eps \,e_i e_k + \mu^{-1}\, b_ib_k) + w_e\, \del_{ik}
        \end{array}   \quad  \right\}    		      			\label{The-ab}
\end{equation}   }   
\vspace{-3mm}
\begin{equation}   
w_e = \frac{1}{2}[\,\eps (\be,\, \be) + \mu^{-1} (\bfb, \bfb )], \qquad \bg = (g_i) \equiv \bd \times \bfb. \label{E-M-s22} 
\end{equation}		 	
Setting $\beta=0$ in (\ref{cons-law-s223}), we have $\pl_\al \Theta^{\al 0}_{\rm w} = j_\lam \,\overline{F}^{\lam 0}$. 
From (\ref{The-ab}), this is written explicitly as 
\begin{equation}
	c_t^{-1} \pl_t w_e  + \divg c_t^{-1} \bq_{\mbox{\tiny fP}} = - \bj \cdot (\be/c_t). \label{b0-s223}
\end{equation}
where  $\bq_{\mbox{\tiny fP}}$ is defined by (\ref{Poynting}). Multiplying $c_t$ on both sides, we obtain 
the energy equation (\ref{TW-e}):
\begin{equation}
\pl_t w_e  + \divg \bq_{\mbox{\tiny fP}} = - \bj \cdot \be.  				\label{energy-s223}
\end{equation}
Likewise, for $\beta=1,2,3 \, (=k)$, the right hand side of (\ref{cons-law-s223}) is given by 
\begin{equation}
  j_\lam \,\overline{F}^{\lam k} = c_t \rho (- \overline{e}_k)  - (\bj \times \bfb )_k
		= - (\rho \be + \bj \times \bfb )_k = - \bF_{\mbox{\tiny L}}.			 \label{bk-s223}
\end{equation}
Using (\ref{The-ab}), the space components ($\beta= k= 1,2,3$) of (\ref{cons-law-s223}) reduce to
the momentum equation (\ref{TW-m}) with $\bg$ denoting the momentum density of TW-wave:
\begin{equation}
  \pl_t \bg  + \nabla\cdot M  \ = - \bF_{\mbox{\tiny L}}.    				\label{momentum-s223}
\end{equation}

\vspace{-5mm}
\subsection{Equations of the whole field (a combined field)}   \label{s23}
\subsubsection{Equations of energy and momentum} 	\label{s231}
Present system under investigation is a combined system consisting of {\it FF}-field and {\it TW}-field.  
According to general principle of theoretical physics\footnote{In particle physics, four fundamental interactions are  
known: weak interaction,  strong interaction, electromagnetism and gravitation.  The Lagrangian density of 
an elementary particle under consideration is represented in a standard theory by linear combination of several terms 
associated with the particle's wave-function field in free state and interaction terms.},  total Lagrangian 
density $\Lam_{\rm fw}$ of such a combined field is defined by linear combination of  $\Lam_{\rm f}$ and  
$\Lam_{\rm w}$ of each constituent field, hence $\Lam_{\rm fw} = \Lam_{\rm f} + \Lam_{\rm w}$, where $\Lam_{\rm f}$ 
is the Lagrangian density of fluid flow free from the {\it TW}-field and $\Lam_{\rm w}$ is that of  the  TW-field 
given by (\ref{LTW-s22}).  In view of the linearity of the energy-momentum tensor with 
respect to each constituent Lagrangian density, total  energy-momentum tensor $T^{\al\beta}_{\rm fw}$ is given 
by linear combination of $T^{\al\beta}_{\rm f}$ and  $T^{\al\beta}_{\rm w}$: namely $T^{\al\beta}_{\rm fw} 
= T^{\al\beta}_{\rm f} + T^{\al\beta}_{\rm w}$.  Assuming no external force acting on the whole system, the 
basic governing equation is described by $\pl_\al T^{\al\beta}_{\rm fw} =0$. Hence we have
\begin{equation}	
\pl_\al T^{\al\beta}_{\rm f} + \pl_\al T^{\al\beta}_{\rm w}  
			= \pl_\al T^{\al\beta}_{\rm f} + \pl_\al \Theta^{\al\beta}_{\rm w} = 0,    \label{T-cons-s23}
\end{equation}
(Scofield \& Huq (2014) \S3), where $T^{\al\beta}_{\rm f}$ and $\Theta^{\al\beta}_{\rm w}$ are given by (\ref{T-f_ab}) 
and (\ref{The-ab}) respectively.

Thus, we obtain the energy equation of the combined system by setting $\beta=0$ as 
\begin{equation}
\pl_t \big[ \rho(\frac{1}{2} v^2 + \ine) + w \big] + \divg(\bq_{\rm f}^{(0)} + \bq_{\mbox{\tiny fP}}) =0.   
\label{total E} 
\end{equation}
where $\bq_{\rm f}^{(0)}$ and  $\bq_{\mbox{\tiny fP}}$ are defined by (\ref{q-FF}) and (\ref{Poynting}) respectively.
Setting $\beta=k$, the momentum equation $\pl_\al T^{\al k}_{\rm f} + \pl_\al \Theta^{\al k}_{\rm w} = 0
\  (k=1,2,3)$ of the combined system  takes the following form:
\begin{equation}
\pl_t( \rho \bv +\bg) + \nabla \cdot (\Pi + M) = 0. 			\label{total-M}
\end{equation}
where  (\ref{T-f_ab}) is used for the fluid part, while (\ref{The-ab}) is used  for the wave part.

The equation of fluid flow under interaction with the wave field is obtained by substituting (\ref{cons-law-s223}) 
into (\ref{T-cons-s23}) as 
\begin{equation}	
 \pl_\al T^{\al\beta}_{\rm f} = - j_\lam \,\overline{F}^{\lam\beta},   \label{EMf-eq-s23}
\end{equation}
(note \ $ -\overline{F}^{\lam\beta}= \overline{F}^{\beta\lam}$). 
For the space part ($\beta=k$), the \underline{\it lhs} is given by (\ref{mom-cons-s21}), whereas the \underline{\it rhs} is  given by the \underline{\it rhs} of 
(\ref{bk-s223}) (for $\beta=k$) with $-1$ multiplied.  Thus, we obtain the  momentum equation for each of the fluid 
system and wave system as
\begin{eqnarray}
\pl_t(\rho \bv) +\nabla \cdot \Pi & = & \ \ \bF_{\mbox{\tiny L}},		\label{FF-m-s23}  \\
\pl_t \bg  + \nabla\cdot M  \     & = & - \bF_{\mbox{\tiny L}},    		\label{TW-m-s23}  \\  
\bF_{\mbox{\tiny L}}[\ba] \       & = &  \ \rho\, \be +  \bj\times \bfb, 	\label{FL-s23}
\end{eqnarray}
respectively,  where the second equation is the equation $\pl_\al \Theta^{\al\beta}_{\rm w} 
= j_\lam \,\overline{F}^{\lam\beta}$ of (\ref{cons-law-s223}) itself. The equation (\ref{FF-m-s23}) was given 
by Scofield \& Huq (2014). The term $\bF_{\mbox{\tiny L}}$ expresses the Lorentz-force-like interaction  between the 
two field components,  and $\Pi=(\Pi_{ij})$ is  given by  (\ref{Pi-FF}).  

The  energy equation for each of the fluid system and wave system is obtained by substituting  (\ref{e-cons-s21})
to the  \underline{\it lhs} of (\ref{EMf-eq-s23}), and \underline{\it rhs} of (\ref{b0-s223})  to its \underline{\it rhs} of (\ref{EMf-eq-s23}) 
with $-1$ multiplied. Thus, 
\begin{eqnarray}
\pl_t \big[ \rho(\frac{1}{2} v^2 + \ine) \big] + \divg(\bq_{\rm f}^{(0)}) & = & \ \ \bj \cdot \be, \label{FF-e-s23} \\
\pl_t w_e  + \divg \bq_{\mbox{\tiny fP}}  & = & - \bj \cdot \be,   		\label{TW-e-s23}
\end{eqnarray}
where $\ine$ is the  internal energy which varies {\it thermodynamically} by absorbing heat liberated 
by dynamical and dissipative  mechanisms occurring within the system.
The current flux $\bj$ consists of the convection current $\bj_c=\rho\bv$ and the drift current $\bj_d$
(see \S\ref{s232}):
\begin{equation}
 \bj= \bj_c + \bj_d,  		\label{current-flux-s23}	 
\end{equation}
The second term $\bj_d$ leads to the  energy dissipation   $\bj_d \cdot \be = Q_{\mbox{\scriptsize D}}$,  while 
the  term $\bj_c \cdot \be$ from the former $\bj_c $ denotes energy transfer between the FF-field and TW-field  
without energy loss.   

Owing to the dissipation terms $Q_{\rm vis}$ and $Q_{\mbox{\scriptsize D}}$, the entropy equation 
(\ref{FF-entropy}) must be modified, and a new entropy equation  is given by
\begin{equation}
\rho T \frac{\D}{\D t} s =  Q_{\rm vis} + Q_{\mbox{\scriptsize D}} +Q_T,		\label{Entropy}
\end{equation}
where  $\D/\D t=\pl_t+ \bv \cdot\nabla$, and $Q_{\rm vis}$ is defined by (\ref{Q-vis}).  For derivation of this
equation, see Appendix \ref{bb}.  In regard to the   heating  due to the turbulence Darcy effect,  the  sections
\S\ref{s232} and \S\ref{s7} consider more details.

\vspace{1mm} 
\subsubsection{Current flux $\bj$}  \label{s232}   \vspace{-1mm} 
The current flux $\bj$ of (\ref{current-flux-s23})  consists of two components. 
The second component $\bj_d$,  called   a {\it drift} current,  contributes to enhanced  energy dissipation in 
turbulent flows.  The drift current $\bj_d$ is considered to emerge in turbulent state agitated with intense 
vorticity causing  strong local acceleration, which is regarded as existing inherently in  turbulent medium. 
This current is a reaction of the fluid  in response to the local acceleration caused by the field $\be$.  

Regarding the TW-field, its energy source (or loss) is given by the right hand side of (\ref{TW-e-s23}): 
\begin{equation}
	S[\bj] = - \bj \cdot\be = - \rho\bv \cdot\be - \bj_d \cdot\be.   \label{S-bj-s23}
\end{equation}
In regard to the second term $\bj_d \cdot\be$,  it is proposed that the drift current $\bj_d$ 
is defined as a current component driven directly by the {\it fluid}-electric field $\be$.  Hence, the  current 
vector $\bj_d$  should be a function of $\be$ and vanishes when $\be=0$. Assuming statistical isotropy of 
fluctuating components of the turbulence field, the above implies that the vector $\bj_d$ (a polar vector) is 
related to $\be$ (a polar vector) by a linear relation:
\begin{equation}
\bj_d =  \sg \be, 			\label{drift-C}	
\end{equation}
where  $\sg$ is assumed  to be a constant or a scalar field depending on $\bx$ and $t$. This is analogous to  
the Darcy current through a porous medium when the pressure gradient acts on a viscous fluid existing in the 
porosity.  This is a {\it turbulence} Darcy effect and called '{\it D}-effect'.  The turbulent medium is 
composed of  a number of turbulent eddies behaving like a porous medium.  Also, this law resembles the Ohm's law  
in the electromagnetism.  This is called  {\it fluid} Ohm's law. 

It is essential to recognize that the contribution from this $\bj_d$ to the source $S$ is negative: 
\begin{equation}
S_d \equiv - \bj_d \cdot \be =  - \sg |\be|^2  <0, 			\label{J1-loss}	
\end{equation}
if $\sg>0$.  Thus,  the present theory is equipped with  a new mechanism of dissipation (if $\sg>0$),  called  
{\it D}-effect.  Moreover, it is remarkable,  shown in \S\ref{s71},  that this mechanism leads to the 
dissipation law analogous to that of the eddy-viscosity models.  Note that  the rate of heating 
$Q_{\mbox{\scriptsize D}} = \bj_d \cdot \be = \sg |\be|^2$  yields the entropy increase.

\section{Excitation of TW field}  \label{s3}

Transverse waves of phase velocity $c_t$ are excited when the current flux $\bj$ is time-dependent or
rotational, according to (\ref{WE-be}) and (\ref{WE-bfb}) of \S\ref{s221}:
\[	\Bigl[\nabla^2 - c_t^{-2} \pl_t^{\,2} \Bigr]\,\be = \mu\,\pl_t \bj +\eps^{-1}\nabla \rho,  \hskip15mm 
	\Bigl[ \nabla^2 - c_t^{-2} \pl_t^{\,2} \Bigr] \,\bh  = - \nabla \times \bj.  		\]
Also, density non-uniformity ($\nabla \rho$) excites the wave $\be$.  In addition, under the Lorenz condition
$\nabla \cdot \ba + c_t^{-2} \,\pl_t \phi_a = 0$, the fields of vector-potential $\ba$ and scalar-potential $\phi_a$
 are excited by $\bj$ and  $\rho'=\rho - \overline{\rho}$ respectively (see (\ref{WE-aphi})):
\[  \Bigl[ \nabla^2 - c_t^{-2} \pl_t^{\,2} \Bigr] \,\ba = - \mu\, \bj, \hskip10mm 
\Bigl[\nabla^2 - c_t^{-2} \pl_t^{\,2} \Bigr]\,\phi_a = - \eps^{-1} \rho'.	\]

\subsection{Governing equations modified in the presence of TW-field}  \label{s31}
Governing equations of an incompressible fluid of the current theory are presented in Appendix \ref{c1} in 
the absence of TW-field (for convenience of later formulation), where the operator ${\rm NS}[\bv]$ 
is defined by
\begin{eqnarray}
{\rm NS}[\bv]  & \equiv & \pl_t \bv + (\bv\cdot\nabla) \bv +\nabla P_v - \nu \nabla^2\bv,   \label{NS-a} 
\end{eqnarray}  
where $P_v = p_{\bv}/\rho$ with $p_{\bv}$ the pressure field associated with the $\bv$-field, and 
both of the density $\rho$ and kinematic viscosity $\nu$ are assumed  constant. The flow field is 
governed by ${\rm NS}[\bv]=0$ in the absence of TW-field.

Let us investigate how this is modified in the coupled system. Now the FF-momentum equation (\ref{FF-m-s23}) 
of the combined system reduces to  the following {\it modified} NS-type equation:
\begin{eqnarray}
{\rm NS}[\bv]  & = & \bff_{\rm L}[\ba] , \hskip10mm  
		\bff_{\mbox{\tiny L}}[\ba] = \be + \rho^{-1} \bj\times \bfb,   	\label{NS-fLa}		\\
&& \hspace*{10mm} \be = -\pl_t\ba -\nabla \phi_a, \qquad \bfb = \nabla\times \ba, 	\label{ea-ba-s3}
\end{eqnarray}
where  the mass conservation equation (\ref{FF-C}) is used, and the suffix $a$ is added to $\phi$ 
in order to make clear its definition as a potential of TW-wave.  From (\ref{NS-fLa}), the vector field 
$\be$ has obviously the same  [{\it acceleration}] dimension as ${\rm NS}[\bv]$, the physical dimension 
of $\ba$ is [{\it velocity}] and that of $\bfb$ is  [{\it vorticity}].\footnote{Denoting the physical dimension of 
{\it length, time} and {\it mass} by $L$, $T$ and $M$, the vector fields $\ba$, $\be$ and $\bfb$ have  the dimension: 
 $LT^{-1}$, $LT^{-2}$ and $T^{-1}$ respectively. In view of eqn. (\ref{TW-bf}), the dimensions of
the vectors $\bd$ and $\bh$ are $ML^{-2}$ and $ML^{-1}T^{-1}$ respectively.  From the relation (\ref{eps-mu}),
the parameters $\eps$ and $\mu$ have the dimensions $ML^{-3}T^2$ and $M^{-1}L^1$ respectively.  Hence the product 
$\eps \mu = c_t^{\,2}$ has the dimension $L^{-2}T^2$. Thus it is seen that $c_t$ has the dimension of velocity.}
  The total current flux is represented as
\begin{equation}
\bj = \bj_c + \bj_d = \rho(\bv + \bv_d),  \qquad \rho^{-1} \bj = \bv_t = \bv + \bv_d,  	\label{tot-j}
\end{equation}
where $\bv_t$ is the total velocity, and $\bj_d /\rho$ has a dimension of velocity denoted by $\bv_d$ for brevity.

We consider {\it channel} flows or {\it boundary-layer} flows in the cartesian frame of reference with 
$x, \,y$ and  $z$ denoting streamwise, normal and spanwise coordinates respectively. In addition. {\it pipe} 
flows with a circular cross-section are also  considered.  The flow fields to be studied here are 
streaky wall flows  perturbed by waves, or turbulent wall flows combined with  {\it TW}  field.

Two neighboring flow fields are denoted by $\bU$ and $\bv=\bU+\bw$ having similar vortex structures or 
similar coherent structures, where $\bU$ is a time-independent vector field representing a steady  flow or 
mean part of turbulent  flow. The second part $\bw$ represents a time-dependent component of either 
($a$) infinitesimal perturbation  or ($b$)  irregular turbulent field.

Let us consider how we can apply the present formulation of fluid flows coupled with a {\it TW} wave field to 
time-dependent wall-bounded flows, characterized with streaky structures which are turbulent,
We consider three specific  examples of $\bU=(U, 0, 0)$:  ($i$) a steady laminar channel flow or boundary-layer 
flow, for which   $U=U(y)$;  ($ii$) a steady streaky flow with streamwise vortices, for which  $U=U_s(y,z)$; 
and ($iii$) a turbulent channel flow expressed by the mean flow $U=U_{\rm turb}(y)$.

Let us rewrite the equation (\ref{NS-fLa}) in such a  way as the  equation (\ref{NS-U1}) was 
derived from ${\rm NS}[\bv]=0$. The equation (\ref{NS-fLa}) is transformed to the following  form: 
\begin{eqnarray}
{\rm NS}[\bU] + \pl_t \bw  & = &   \be  - \nabla (\phi_w) + \bv \times \bom_w + \bw\times \bOm 
		+ \bv_t \times \bfb + \nu  \nabla^2\bw.				\label{U-u-a}
\end{eqnarray}
where $\phi_w $ is defined below (\ref{bf_L}) with $\bu$ replaced with $\bw$, $\bom_w= \nabla\times \bw$, and 
$\bv_t =\bv + \bv_d = \bU + \bw + \bv_d$.  

\subsection{Equation of a combined field $\bw_a \equiv \bU+\bw+\ba$}   \label{s32} 
\subsubsection{Derivation}  \label{s321}  
Let us consider a dynamical mechanism  of generation of the TW-field.   We are interested in  a flow-dynamics and 
 an energy  channel, through which an energy is  extracted  from the  $\bv$-field  and the same amount 
of energy is transmitted to the $\ba$-field.  Because of simplicity of arguments,  the fluid is assumed 
incompressible in the undisturbed state.

Let us rewrite the modified  NS equation (\ref{U-u-a}) of the coupled system. By using the total velocity 
$\bv_t=\bv +\bv_d$ defined by (\ref{tot-j}) and $\bv= \bU +\bw$, the equation  (\ref{U-u-a}) is rewritten as
\begin{eqnarray}
{\rm NS}[\bU]  & = & \bff_{\rm L}[\bw+\ba] + \bw\times \bOm + \bv_d \times  \bfb + \nu  \nabla^2\bw,
			    	\label{NS-U-u+a} 	\\
&& \bff_{\rm L}[\bw+\ba] \equiv -\pl_t(\bw+\ba) - \nabla (\phi_w + \phi_a) + \bv \times (\bom_w + \bfb).
				\label{fL-u+a}
\end{eqnarray}
where the explicit expression of ${\rm NS}[\bU]$ is given by using the definition (\ref{NS-b-c}) of Appendix C:
\[ {\rm NS}[\bU] =\pl_t\bU +\bOm \times \bU +\nabla(P_U +\half |\bU|^2) -\nu \nabla^2\bU.  \]
Substituting (\ref{fL-u+a}) to the modified  NS equation   (\ref{NS-U-u+a}),  we obtain 
\begin{eqnarray}  
\hspace*{-20mm} 
{\rm NS}[\bU]  + \pl_t(\bw+\ba) + (\bom_w + \bom_a)\times \bv + \bom_U \times \bw 
+ \nabla(\phi_w +\phi_a)  & =  & \bv_d \times \bfb + \nu \nabla^2\bw,                \label{mNS-Uua}
\end{eqnarray}
where $\bom_a=\bfb = \nabla\times \ba$, and $\phi_w = P_v-P_U + \half (|\bv|^2 - |\bU|^2)$. 
In this equation, it is observed that first two terms of ${\rm NS}[\bU]$ and middle three terms (\ie \ 
$\pl_t(\bw+\ba),\ (\bom_w + \bom_a)\times \bv, \  \bom_U \times \bw$) of \underline{\it lhs} of 
(\ref{mNS-Uua}) are arranged as follows:
\[   \pl_t \bw_a + (\nabla \times \bw_a) \times \bw_a - (\nabla \times \bw_a) \times \ba, 	\]
where  $\bw_a = \bU+\bw+\ba= \bv+\ba$. Hence, the \underline{\it lhs} of (\ref{mNS-Uua}) is given by
\[  [\,\mbox{\underline{\it lhs} of (\ref{mNS-Uua})}\,] = \pl_t \bw_a + (\bw_a\cdot\nabla) \bw_a + \nabla P_{wa} 
	- (\nabla \times \bw_a) \times \ba - \nu \nabla^2\bU,  				\]
where $P_{wa} = \phi_a +P_v + \half (v^2  - w^2)$.  Therefore, the equation (\ref{mNS-Uua}) reduces to 
\begin{equation}
\pl_t \bw_a + (\bw_a \cdot\nabla) \bw_a + \nabla P_{wa}  = (\nabla \times \bw_a) \times \ba 
	+ \bv_d \times \bfb + \nu\nabla^2 \bv,					\label{mNS-W}
\end{equation}   
where $\bw_a= \bv+\ba= \bU+\bw+\ba$. Thus, the equation of coupled system, \ie {\it modified} NS-type equation of  
(\ref{NS-fLa}), is first transformed to (\ref{NS-U-u+a}), and finally reduced to (\ref{mNS-W}). 
From this, one can derive interesting equations for two particular cases.

($i$)  If the $\ba$-field is absent, the equation (\ref{mNS-W}) reduces to the usual NS equation 
for $\bw_a=\bv$:
\begin{equation}
 {\rm NS}[\bv] =0 \,: \quad  \mbox{equivalently,} \qquad 
		\pl_t \bv +(\bv \cdot\nabla) \bv +\nabla P_v = \nu\nabla^2 \bv.  \label{NS-v}
\end{equation}

($ii$) If all the terms on the right hand side are neglected, it reduces to an equation of Euler-type. 
In fact, last two terms of \underline{\it rhs} are those causing dissipation, while the first term  
vanishes if \ $\ba \parallel \nabla \times \bw_a$.  If we can  omit all of those \underline{\it rhs} terms, we obtain 
\begin{equation}
   \pl_t \bw_a + (\bw_a\cdot\nabla) \bw_a + \nabla P_{wa} = 0, \hskip9mm \bw_a =\bv + \ba. 	\label{mEul-W}
\end{equation}
This implies that the Euler equation can absorb the $\ba$-field  and  is transformed  to the equation of 
the combined field $\bw_a = \bv + \ba$  as far as $\ba \parallel \nabla \times \bw_a$.

\subsubsection{Infinitesimal  $\ba$-field (a possible case of excitation)}  \label{s322}  

Infinitesimal  $\ba$-field may be exited  without appreciable opposing reaction. In fact,  suppose that we 
have unperturbed steady flow field $\bU$ in the absence of TW-field, and consider excitation 
of  an infinitesimal $\ba$-field in the flow field. This results in excitation of two infinitesimal 
fields $\bw$ and $\ba$.  The second term of \underline{\it rhs} of (\ref{mNS-W}) is $O(|\ba|^2)$ (assuming 
$|\bv_d| =$O$(|\ba|)$), which is higher order and neglected.  By the same approximation, the first term is 
approximated as $ (\nabla \times \bU) \times \ba$. Thus we have 
\begin{equation}
\pl_t (\bv+\ba) + (\bv+\ba) \cdot\nabla (\bv+\ba) + \nabla P_{wa} =  (\nabla \times \bU) \times \ba
		+ \nu\nabla^2 \bv,	\label{mNS-W2}
\end{equation}
from (\ref{mNS-W}) neglecting terms of $O(|\ba|^2)$ and $O(|\ba| |\bw|)$. If the $\ba$-field is excited to the 
direction parallel to $\nabla\times \bU$, the first term on  \underline{\it rhs} can be neglected too. 
Resulting differential equation reduces to
\begin{equation}
 \pl_t \bw_a + (\bw_a\cdot\nabla) \bw_a = - \nabla P_{wa} + \nu \nabla^2 \bv,  \label{NS-va} 
\end{equation}  
where $\bw_a=\bU+\bu+\ba$. This is approximated by  ${\rm NS}[\bw_a] =0$ as far as $\nu \nabla^2 \ba$ is 
negligibly small, implying that the governing equation is insensitive whether an infinitesimal $\ba$-field 
exists or not. It appears that the $\ba$-field may have a high affinity with the FF-field and may be 
excited {\it without} appreciable opposing reaction from the flow as far as $\ba \parallel \nabla \times \bU$.

\subsection{Dynamical excitation of TW-field}  \label{s33}
From the energy consideration, dynamical excitation of $\ba$-field is investigated here.
For that purpose, let us take scalar product of $\bv$ with (\ref{NS-fLa}). Then we have
\begin{equation}
v_k\,{\rm NS}[\bv]_k  =  v_k\,(\bff_{\rm L}[\ba])_k,. 	\label{NSv-fLa}
\end{equation} 
Using (\ref{NS-b-c}), the left hand side  is transformed to
\begin{eqnarray} 
v_k\,{\rm NS}[\bv]_k  & = & \pl_t(\half v^2) + \nabla \cdot \bT_m + D_\nu,  \label{v-NSL} 		\\ 
  & & \bT_m \equiv (\half v^2 +P_v)\, \bv - \nu\nabla(\half v^2), \quad D_\nu = \nu(\pl_l v_k)^2 (\ge 0), \nonumber
\end{eqnarray}
where the solenoidal condition $\nabla \cdot \bv=0$ and the identity $v_k \,\pl_l^{\,2}v_k 
= \pl_l( v_k \,\pl_l v_k)- (\pl_l v_k)^2$ are used.  We integrate the equation (\ref{NSv-fLa})  over 
a simply connected 3-dimensional volume $V_3$ in the fluid, enclosed by 2-dimensional boundary 
surface $\pl V_3$ (chosen arbitrarily). Thus, we obtain
\begin{eqnarray}
\frac{\dd}{\dd t} K_v + D_{vis} & = & W_f[\bv, \ba] + \ I_{\pl V_3}, 		\label{dE-eq}	\\										
 K_v  & = & \int_{V_3} \half v^2\,\dd^3\bx, \hskip15mm      D_{vis} = \int_{V_3} D_\nu\,\dd^3\bx,   \label{dE-def}
\end{eqnarray}	  
(see \S\ref{c2} for the current theory without $\ba$-field), 
where the integral $K_v$ is the total kinetic energy of the flow field $\bv$ in $V_3$,  the term $W_f[\bv, \ba]$ 
is explained just below, and the term $D_{vis}$ denotes the bulk rate of viscous dissipation of energy $K_v$ in $V_3$.
The term $I_{\pl V_3}$ denotes integration  over the surface $\pl V_3$, obtained from integrating the term 
like $\pl_k(\cdot)$ over $V_3$, but those surface integrals  are neglected here and below. 

The integral $W_f[\bv,\ba]$ is defined by
\begin{equation}
W_f[\bv, \ba] = \int_{V_3}  v_k\,(\bff_{\rm L}[\ba])_k \,\dd^3\bx  ,  	\label{Wf-va}
\end{equation}
If $W_f[\bv, \ba] <0$, then the $\ba$-field would be produced. This is confirmed in the following way. 
By the definition, we have $\bff_{\mbox{\tiny L}}[\ba] = \be + (\bv + \bv_d) \times \bfb$  with using 
(\ref{tot-j}). Hence,
\begin{eqnarray}		
\bv\cdot \bff_{\mbox{\tiny L}}[\ba] & = & \bv \cdot\be + \bv \cdot(\bv_d \times \bfb)    
    \quad \approx \ \bv \cdot\be \quad (\mbox{to the order of $|\ba|$}),  		\label{v-fLa}  
\end{eqnarray}
since \ $\bv_d= (\sg/\rho) \be$ (see (\ref{drift-C})) and $|\bv_d \times \bfb|=$O$(|\ba|^2)$.

In regard to the TW-field $\ba$, its energy source  is given by the right hand side of (\ref{TW-e-s23}):
\begin{equation}
	S[\bj] \equiv - \bj \cdot\be = - \rho\, \bv \cdot \be - \rho\, \bv_d \cdot \be,
        \hskip8mm  \bj=\bj_c +\bj_d , 	\label{TW-S-34}
\end{equation}
where  $\bj_c=\rho \bv$ and $\bj_d =\rho \bv_d $.  Magnitude of the second term $|\rho\bv_d \cdot\be|$ is 
$O(|\ba|^2)$, and hence can be neglected. Up to the first order of $|\ba|$, the equation (\ref{Wf-va}) implies that,
if $W_f[\bv,\ba] <0$, the term $\bv \cdot\be $ is dominated by negative values, and hence that the energy 
source of TW-field $S[\bj] = - \rho\, \bv \cdot \be$ is dominated by positive values. Therefore, an infinitesimal 
$\ba$-field is excited according to (\ref{TW-e-s23}),  while the energy of  FF-field  is damped according 
to (\ref{dE-eq}).  Once the TW-field is sufficiently developed, the second term $- \rho\, \bv_d \cdot \be$ 
of (\ref{TW-S-34}) would  become significant, which is negative by (\ref{J1-loss}) (see  \S\ref{s42}), 
resisting growth of TW-field, \ie  a part of TW-energy is dissipated by the second term.
In  \S\ref{s53} and  \S\ref{s64},  a sustaining process of the TW-field is studied by solving a model equation 
in which there are mechanisms of energy pumping and dissipation.

Thus it is seen that there exist both of flow mechanism from momentum aspect (\S\ref{s322}) and  energy-feeding 
mechanism of this section for the {\it TW-field} by extracting energy from the {\it FF-field}. Once energy is 
transferred to the TW-field, some of the energy is transformed into heat and dissipated. However, total 
energy is conserved if all components of kinetic energy, internal energy (including entropy increase) and TW energy
are counted, which is expressed by (\ref{total E}) and (\ref{Entropy}) taking account of the thermodynamics. 

\subsection{An infinitesimal $\ba$-field and  energy source }  \label{s34}
Let us consider the above mechanism by taking example of  steady flows $\bU(\bx)=(U, 0, 0)$ in the $x$-direction,
satisfying ${\rm NS}[\bU] =0$. They are either  ($i$) a steady laminar  flow  with $U=U(y)$, or ($ii$) a steady 
streaky flow $\bU_s$  with $U=U_s(y,z)$ having nonzero streamwise vorticity ($(\nabla \times \bU_s)_x \ne 0$).  
Suppose that the flow  $\bU$  is perturbed with a small fluctuation  $\bu(\bx, t)$, and 
an infinitesimal $\ba$-field is growing by extracting energy from the basic flow $\bU$. 

In the equation   (\ref{mNS-Uua}) derived from ${\rm NS}[\bU+\bu] = \bff_{\rm L}[\ba]$ of (\ref{NS-fLa}), the first 
term ${\rm NS}[\bU]$  disappears by the assumption (where $\bu$ is used instead of $\bw$ with assuming to be 
infinitesimal).  Retaining only the terms linear with respect to $\bu$ and $\ba$ and neglecting the viscosity term, 
the equation (\ref{mNS-Uua}) reduces to 
\begin{equation} 
	\pl_t(\bu+\ba) + (\bom_u + \bom_a)\times \bU - \bu \times (\nabla \times \bU)  
		+ \nabla  (\phi_u +\phi_a) =0. 					\label{bu-ba}  
\end{equation}
This is solved by 
\begin{equation}
 \ba(\bx,t)  = - \bu(\bx,t) + \nabla \psi, \hskip10mm 
	\nabla(\pl_t\psi + \phi_u + \phi_a) = \bu \times (\nabla \times \bU). 		\label{sol-ua}
\end{equation}
It is seen that the $\ba$-field (given by $\ba= -\bu +\nabla \psi$) is excited by counteracting the 
perturbation field $\bu$. From the energy equation (\ref{TW-e-s23}), the convection current 
$\bj_c = \rho (\bU+\bu) \approx \rho \bU$ is regarded as a source  to excite the TW-field  by the term, 
\begin{equation} 
S_c[\bj_c]  =- \bj_c\cdot \be \approx - \rho\bU \cdot \be \approx \rho\,\bU \cdot \pl_t\ba , \label{Sw}  
\end{equation}	
(neglecting $\bU\cdot\nabla \phi_a$  and higher order terms).  Hence, if the field $\ba$ is excited to 
the direction of $\bU$, \ie if $\bU \cdot \pl_t\ba >0$, then we have $S_c>0$. This would induce positive 
value of $\pl_t w_e$ where $w_e$ is the TW-energy density.  
Thus, the wave energy $w_e$ grows according to (\ref{TW-e-s23}).  

Note that if the field $\ba$ is periodic with respect to time $t$ (considered in the next section),
time average of the $\rho\,\bU \cdot \pl_t\ba$ over a periodic time vanishes and  other terms come into play.

\subsection{Transient growth mechanism exciting  $\ba$-field }  \label{s35}
By the scenario of {\it transient growth} of disturbances  (Gustavsson 1991;  Butler \& Farrell 1992;  
Henningson, Lundbladh \& Johansson 1993;    Trefethen,  Trefethen,  Reddy \& Driscoll 1993),
perturbation modes can grow sufficiently  even for conditions under which disturbance modes  are stable with respect to
linearized dynamical systems associated with laminar shear flows such as plane Poiseuille flow or plane Couette 
flow. This growth  occurs in the absence of nonlinear effects and can be as large as $O(10^3)$ times of initial 
energy. Then afterward, nonlinear mechanisms are expected to work for the growing modes to develop into turbulent 
states.  This scenario is different from the normal stability theory in which the disturbance mode under 
investigation grows or decays exponentially with respect to time.  The transient growth mechanism is explained
by the non-normality of the governing linear operator and the non-orthogonality of the eigenfunctions of 
the linear problem.  The transient growth is sometimes said to be  a bypass transition.  

For the plane Poiseuille flow, eigenvalue analysis of the normal stability 
theory predicts a critical Reynolds number $R_c=5772$ at which exponential temporal instability should set in, 
while the plane  Couette flow is predicted to be stable for all values of Reynolds numbers. The optimal 
perturbations in the former transient growth scenario  are not of normal-mode form of the standard stability 
theory,  and those which grow the most divert the basic flow energy into the perturbation to grow by as much 
as three orders of magnitude (\ie \, about $10^3$ times the initial energy) (Butler \& Farrell 1992). 

In the laboratory experiments, it is known that  transition from  a steady laminar state to a state of 
disturbance waves is observed at Reynolds numbers  much less than the critical value for transition according 
to the linear stability theory of normal-mode analysis. The optimal perturbation of the transient-growth 
scenario provides us with examples of small disturbances that grow rapidly and robustly in shear flows
as much as two or three orders of magnitudes.

The analysis of the previous section \S\ref{s34} implies that there exists an excitation mechanism of the  
$\ba$-field by this transient growth scenario, because the streamwise velocity $u$ predicts a streamwise 
$\ba$-field by the first  of (\ref{sol-ua}).  The rate of change of the wave energy $w$ of $\ba$-field 
is given by (\ref{Sw}).  Its \underline{\it rhs} can be positive locally where the $\ba$-field is growing to the
direction of  $\bU$ (and reversed at the other phase).  This is in fact the case (see \S\ref{s621}). 
Thus, the {\it TW} field can be excited in shear flows  by the transient growth scenario.

\section{Energy and momentum budgets of TW-field}  \label{s4}
\subsection{Mechanical properties of TW-waves}					\label{s41}
Mechanical properties of the TW-field are described by the conservation equations of energy and momentum, 
(\ref{energy-s223}) and (\ref{momentum-s223}) respectively.  In particular, the right hand sides of 
both equations give the sources of energy and momentum of the TW-field.  Thus, the TW-field is characterized 
by the following important properties: ($i$) a TW-wave has its own energy and momentum. This reminds us of 
the  particle-like property of electromagnetic waves which yields the photon in the quantum physics. 
($ii$) the TW-wave acts on the fluid flow by the Lorentz force term $\rho \bff_{\mbox{\tiny L}}$, and
($iii$) the wave gains energy or loses it, depending on the sign of the source term $\bj \cdot \be$, 
possibly loses it by the term $\bj_d \cdot \be$  (which is positive definite, \S\ref{s42}).

The TW-wave of the property ($i$) explains its robustness and long-living within turbulent environment. 
It keeps long unless it loses (changes) momentum and energy by interaction with other components by the 
properties ($ii$) and ($iii$).  An experimental evidence is seen in Hussain \& Reynolds (1970, 1972)

Mechanical consequences of the Lorentz acceleration $\bff_{\mbox{\tiny L}}$ have been studied already in 
the previous section 3. It is found in \S\ref{s32} that there exists certainly a flow dynamics exciting the 
TW-field, and also in  \S\ref{s33} that there exists  a mechanism  of energy-feeding to the {\it TW-field} 
by extracting energy from the {\it FF-field}. Once energy is transferred to the TW-field, some of the energy 
is dissipated into heat. This is investigated next.

\subsection{Energy source and dissipation}					\label{s42}
Regarding the TW-field, its energy source (or loss) is given by the right hand side of (\ref{energy-s223}): 
\begin{equation}
	S[\bj] = - \bj \cdot\be = - \rho\bv \cdot\be - \bj_d \cdot\be.   \label{S-bj}
\end{equation}
The current flux is expressed as $\bj= \rho\bv +\bj_d $ by (\ref{TW-S-34}).  If $S[\bj]>0$ (or $<0$), 
it is a gain (or a loss).  However,  the first of (\ref{TW-bf}) gives another expression for $\bj$ by
\begin{equation}
 \bj= \mu^{-1} \nabla \times \bfb - \eps\,\pl_t \be.  \label{TW-S1}
\end{equation}
where $\bd = \eps \,\be$ and $\bh = \mu^{-1} \bfb$.  Bulk energy source of the TW-field due to 
the current $\bj$ is defined  by the integration of $S[\bj]$ of (\ref{S-bj}) over a volume $V$:
\begin{equation}
\overline{S}[\,\bj]  \equiv  - \int_V \bj \cdot \be\, \dd^3\bx \quad 
				= \frac{\dd}{\dd t} \int_V w_e \ \dd^3 \bx,	\label{TW-bS}
\end{equation}
where $w_e$ is the energy density defined in (\ref{en-mom}).    The last expression is obtained  as follows. 
Substituting the \underline{\it rhs} of (\ref{TW-S1}) to $\bj$ in the first integral, the integrand is 
\begin{equation}
  - (\mu^{-1} \nabla \times \bfb - \eps\,\pl_t \be) \cdot \be = - \mu^{-1} (\nabla \times \bfb )\cdot \be
	+ \half \eps \,\pl_t (\be\cdot\be) \,).			\label{rhs-49}
\end{equation}
Replacing $\be$ of the first term on \underline{\it rhs} with its definition $ -\pl_t\ba -\nabla \phi_a$, and 
integrating the resulting expression $\mu^{-1}(\nabla \times \bfb)\cdot (\pl_t\ba + \nabla \phi_a)$ by parts 
and neglecting terms of surface integration, we find the first term of (\ref{rhs-49})  reduces to 
\[  \int_V \mu^{-1} \bfb \cdot \pl_t\bfb \ \dd^3 \bx= \int_V \half \mu^{-1} \pl_t(\bfb \cdot \bfb )\ \dd^3 \bx. \]
Using $\eps \,\be =\bd$ and $\mu^{-1} \bfb = \bh$, we obtain finally the \underline{\it rhs} of (\ref{TW-bS}).

The second term of (\ref{S-bj}), \ie $S_d = - \bj_d \cdot \be$, describes dissipation if the {\it turbulence}-Darcy 
effect $\bj_d = \sg \be$ is taken into account: $ S_d = - \sg |\be|^2  <0$ (see (\ref{J1-loss})).	
This effect influences wave propagation  as  damping of wave amplitude.  This damping effect is considered 
in the sections \S\ref{s71} and \ref{s72}.  It is  remarkable to find that the expression (\ref{TW-bS}) 
leads to an expression analogous to that of eddy viscosity models. Namely, the {\it turbulence}-Darcy effect 
can describe the enhanced dissipation analogous to the eddy viscosity models.

\subsection{Momentum exchange}			\label{s43}

When the TW-field $\ba$ is excited, a small fluid particle of density $\rho$ in turbulence is acted on by the 
force $\bF_{\mbox{\tiny L}} = \rho \bff_{\mbox{\tiny L}}$ of (\ref{FF-m-s23}) and (\ref{TW-m-s23}), with its 
acceleration $\bff_{\mbox{\tiny L}} [\ba]$ defined by
\begin{equation}  
	\bff_{\mbox{\tiny L}} [\ba] = \be + \rho^{-1} \bj \times \bfb,	\hskip8mm 	
	\be = - \pl_t \ba - \nabla \phi_a, \qquad \bfb =  \nabla\times \ba. 	\label{Mom-42}
\end{equation}  
Conversely, the TW-field receives  back reaction  from the  FF-field by the  force $- \bF_{\mbox{\tiny L}}[\ba]$,
which gives the rate of change of the momentum density $\bg \equiv \bd \times \bfb$ of (\ref{TW-m-s23}).

Thus, the TW-field could be a dynamically active agent working in the turbulence field, and possibly long-living 
robustly within turbulent flow unless interaction modifies it.


\subsection{Energy supply by phase shift due to periodic perturbation}			\label{s44}

In parallel with the momentum exchange of \S\ref{s43}, energy is also supplied from the FF-field  to the TW-field 
by $S[\rho\bv] \equiv - \rho (\bU+\bu) \cdot \be$, neglecting the dissipation term $- \bj_d \cdot \be$, 
considered in \S\ref{s232}.  There are two source terms of $S[\rho\bU] = - \rho \bU \cdot \be$ 
and $S[\rho\bu] =- \rho \bu \cdot \be$. Here, we are interested in the second term only, because the energy supply
by the term $S[\rho\bU]$ is already considered in \S\ref{s34} and also because  $S[\rho\bU]$ does not give net 
effect if the flow $\bU$ is steady and the field $\be$ is periodic with respect to time and  in addition if 
time average is taken.

In the presence of the periodic perturbations of $\bu(t)$ and $\ba(t)$,  energy is supplied to the TW-field  
when $\bu\cdot \be <0$. In \S\ref{s53} below, we will see that the drift current causes a phase shift  between 
the two fields $\bu$ and $\ba$.  If there is a {\it phase shift}  such that $\bu(t) = - \overline{D}\, \ba(t-\del)$ 
for a small positive $\del$ and a positive constant $\overline{D}$, then  we have $ \bu(t)/\overline{D}  = 
- \ba(t) + \del\, \pl_t \ba + O(\vep^2)$ by expansion, and the source term $ \rho^{-1} S[\rho \bu]$ divided by
$\overline{D}$ is given by
\begin{eqnarray}  
 (\rho \overline{D})^{-1} S[\rho \bu] = - (\overline{D})^{-1} \bu \cdot \be 
 	& =  &  - \ba \cdot \pl_t \ba + \del\, |\be|^2 +  O(\vep^2) 			
        = - \half \pl_t |\ba|^2   + \del\, |\be|^2 +  O(\vep^2),  \hspace*{5mm}		\label{S-ru}
\end{eqnarray}
where $\be= -\pl_t\ba$ is used, assuming that $\phi_a$ is constant (see \S\ref{s5}, below (\ref{div-ea})). Time 
average of the first term $\overline{\half \pl_t |\ba|^2}$ over a period vanishes, whereas time average of the 
second term $\overline{\vep |\be|^2}$ gives a positive  energy gain of the wave field if $\del>0$. Thus, existence 
of the phase shift enables energy supply from the flow field to the TW-field.


\section{Traveling waves and wave dynamics (large scale motion)}  \label{s5}
In \S3, we considered possible dynamical mechanisms of excitation of {\it TW}-waves. According to the 
review of recent  observations  in \S1 ($b$) and ($c$),  transverse traveling waves are triggered either 
by a transiently amplified  disturbance  (but decaying later) or  by a packet of hairpins appearing 
spontaneously in wall shear layer.

In the first section \S\ref{s51}, we investigate propagation of {\it TW}-waves traveling through channel 
turbulence, and  in the subsection \S\ref{s512}, we describe two large scales LSM and VLSM, which are  characteristic 
features of the energy spectrum  at low wave numbers,  observed experimentally in wall turbulence.  In the 
section \S\ref{s52},  spatial and temporal damping during the propagation are investigated. 
The section 5.3 considers wave dynamics of growth and decay, and its subsection \S\ref{s532}  describes an 
interesting property that the resistive drift current $\bj_d= \sg \be$ causes a phase shift between the flow 
perturbation $u_x(t)$ and the wave field  $a_x(t)$, enabling energy transfer from the flow field to the wave field.

\vspace{-3mm} 
\subsection{Waves traveling through  turbulence and large scales}		 \label{s51}	
\subsubsection{Wave equation}  \label{s511}  \vspace{-1mm}
Let us consider wave propagation through  turbulent flows along a plane channel (its channel width $2h$). The waves 
traveling through  the  turbulence are governed by the equations (\ref{WE-be}) and (\ref{WE-bfb}), where the 
relations $\bd = \eps \,\be$ and $\bh= \mu^{-1}\bfb$ are assumed.  This problem may be reduced to that of the wave 
guide filled with a medium characterized by the parameters $\eps$ and $\mu$, assumed constant. The density 
is also assumed  constant (for simplicity): $\rho=$\,const. In \S\ref{s221}, we defined 
\begin{equation} 
\be= -\pl_t\ba -\nabla \phi_a,  \hskip10mm \bfb =\nabla\times\ba, \hskip10mm \bd= \eps \be, 
\hskip10mm \bh= \mu^{-1}\bfb. 				\label{e-h-b}
\end{equation}
By  (\ref{TW-a}) and  (\ref{TW-bf}), these are governed by 
\begin{equation} 
\nabla \times \be = -\mu\,\pl_t\bh, \hskip10mm 
\nabla \times \bh =  \eps\,\pl_t \be  + \bj_c + \bj_d, 			\label{TW-eh}  
\end{equation}	
 where  $\bj = \bj_c + \bj_d$.  Owing to the drift current $\bj_d$, there is  damping in the wave propagation.  
 Using (\ref{drift-C}) and taking {\it curl} of the two equations of (\ref{TW-eh}) and noting the identity 
 ($ii$) at the footnote below Eq. (\ref{NS-b-c}), the above two equations are transformed to
\begin{equation} \hspace*{-6mm} 
\nabla^2 \be - c^{\,-2} \pl_t^{\,2}\be  =  \mu\, \pl_t \bj_c +  \mu \sg \pl_t \be,  \hskip10mm 
\nabla^2 \bh  - c^{\,-2} \pl_t^{\,2}\bh = - \nabla \times \bj_c +  \mu \sg \pl_t \bh, \hspace{5mm} \label{TW-eh-2}  
\end{equation}  
since $\nabla (\nabla\cdot\be)= \eps^{-1}  \nabla\,\rho = 0$ (by the assumption) and $\nabla (\nabla\cdot\bh)=0$, where 
\begin{equation}   	c =1/\sqrt{\eps\mu} \,,		\label{phase speed}  \end{equation}
is the phase speed of the wave.\footnote[1]{The speed $c$ is used instead of $c_t$ in \S5 for simplicity} 
Corresponding wave equation of $\ba$ and $\phi_a$  are given by (\ref{WE-aphi}): 
$[ \nabla^2 - c^{-2} \pl_t^{\,2} ] \ba = - \mu\, \bj$ and $[\nabla^2 - c_t^{-2} \pl_t^{\,2} ]\,\phi_a =  0$.

Each of the  \underline{\it lhs} (left hand side)  expresses wave propagation of $\be$ (or $\bh$) with a phase speed 
$c$. The propagation proceeds under the two effects on the \underline{\it rhs}: ($i$) damping effect expressed by 
$ \mu \sg \pl_t \be$ (or  $\mu \sg \pl_t \bh$)  and ($ii$)  wave source expressed by the term $\mu\, \pl_t \bj_c$  
(or $ - \nabla \times \bj_c$).  Neglecting those two effects, we obtain the  wave equations for  $\be$ (or $\bh$):
\begin{equation} 
\nabla^2 \be - c^{\,-2} \pl_t^{\,2}\be  = 0,  \hskip20mm  \nabla^2 \bh  - c^{\,-2} \pl_t^{\,2}\bh = 0. \label{TW-eh-3}  
\end{equation}  
Suppose  that this unforced (\ie \,{\it natural}) wave is propagating one-dimensionally, along the streamwise 
direction denoted by the $x$-axis with a frequency $\om$, and that the wave amplitude is expressed by a factor 
proportional to $e^{i(k_0 x -\om t)}$, representing a travelling wave of the wave length $\lam_0=2\pi/k_0$. 
The channel cross-section (with $h$ its half-width) is described by the wall-normal 
$y$ and spanwise $z$ coordinates where  $0 < y < 2h$.  Let us  define the wave amplitude $\Psi$ 	
to denotes one of the components of $\be$ (or $\bh$) and is represented by a form of traveling wave,
\begin{equation} 
\Psi = A(y, z) \, e^{i(k_0 x -\om t)}, \hskip10mm (\mbox{wave speed,\ wave length}) = (\,c, \ \lam_0 \,).
  \label{TW-ps1} 
\end{equation}
Namely, this describes a wave of wavelength $\lam_0$ propagating naturally through a turbulent 
shear flow with the phase speed $c$.

\subsubsection{Large scale motions: experimental aspects}		   \label{s512}
\vspace{-1mm} 
\centerline{\it Two large scales: LSM and VLSM } \label{s521}    \vspace{1mm} 
Recent experiment studies of shear flow turbulence (\S1($c$))  recognize existence of two large scales
of the streaky  structures:  LSM (large-scale motions) and VLSM (very-large-scale motions), characterizing 
the streamwise streaks and long meandering structures. 

It is generally accepted  that the streamwise scale $l_p$ of the vortex packets consisting of hairpin-like 
structures in the wall shear layer characterizes the scale  LSM.  This would trigger generation of waves 
extending over the whole cross-section  in the surrounding space. It is proposed that the wave length of the 
traveling wave, $\lam=2\pi/k_0$ of (\ref{TW-ps1}), is of the order of the channel half-width $h$ which is 
supposed to be the LSM  observed in the experiments. Thus,
\begin{equation}  \mbox{LSM} \ : \quad   \lam_{\rm lsm} =  h \sim 3 h \quad \approx \lam_0=2\pi/k_0. 	\label{lsm}
\end{equation}   
This defines a natural frequency intrinsic to the LSM phenomenon: $\om_0= c\,k_0 \approx 2\pi\,c/ \lam_{\rm lsm}$.

Furthermore, the wave propagation is modulated by another waves of longer wavelengths. 
In this regard, the study of Del \'{A}lamo \& Jim\'{e}nez (2006)\footnote{They used  a turbulent  viscosity $\nu_{\rm T}$ instead 
of the molecular viscosity $\nu$ to solve a modified Orr-Sommerfeld equation. This  implies that current 
theory  should be reconsidered in order to be able to interpret observed large-scale structures.}
is worth being mentioned. They found that there exist two scales of disturbances in turbulent channel 
flow which are  transiently amplified  sufficiently according to linear perturbation equations: one corresponds
to sublayer scale and the other  to the larger-scale structure spanning the full channel. Their study hints 
that the sublayer scale disturbance grows self-similarly in the logarithmic layer (at the overlap region) up 
to the scale of vortex  packets, namely up to  LSM.  It is likely that the second larger-scale of the  waves 
amplified by the transient growth  corresponds to the VLSM scale:
\begin{equation}	\mbox{VLSM} \ : \quad  15h \sim 20h . 	\label{vlsm}
\end{equation}
This observation is considered in \S\ref{s6} again.
The whole length of the TW-wave train should be finite because of its damping effect owing to the {\it D}-effect 
to be considered in the next section.

\vspace{-5mm} 
\subsection{\bf Propagation under damping effects}    \label{s52}
We investigate wave propagation under damping effects according to (\ref{TW-eh-2}) with the wave form,
\begin{equation} 
\be, \ \bh  \propto  e^{i(k x -\om t)}.   \label{TW-eh3} 
\end{equation}
 Replacing $\pl_t$ with $-i\om$ (except the $\bj_c$-term), the  two equations of (\ref{TW-eh-2})
are written as
\begin{equation}  \hspace*{-10mm} 
\nabla^2 \be +(k_0^{\,2}+ i\om \mu \sg) \be =  \mu\, \pl_t \bj_c, \hskip10mm 
\nabla^2 \bh +(k_0^{\,2}+ i\om \mu \sg) \bh = - \nabla \times \bj_c, 			\label{TW-eh3}  
\end{equation}  
where $k_0=\om/c$. It is known in the corresponding wave guide problem of electromagnetism (Appendix \ref{dd}; and 
also Jackson (1999))  that, given the convection current $\bj_c$, the $y$ and $z$ components of $\be$ 
and $\bh$ in the cross-sectional plane are determined once the axial components $e_x$ and $h_x$ are 
known.  Therefore we consider only the $x$-components of  (\ref{TW-eh3}).  The derivative $\pl_x^2$ can be 
replaced by $(ik)^2=-k^2$. Thus, we obtain
\begin{eqnarray} 
(\nabla_\perp^{\ 2}  + K^2 ) \left\{ \begin{array}{c} e_x \\ h_x \end{array} \right\} & = & 
	\left\{\begin{array}{c} \mu\, \pl_t j_{c,x} \\ -(\nabla \times \bj_c)_x \end{array}\right\}, 
        								\label{TW-eh4} 		\\
 \nabla_\perp^{\ 2} =\pl_y^2+ \pl_z^2, \hspace*{13mm} &&   K^2=k_0^{\,2} - k^2 + i \om \mu \sg,   \nonumber
\end{eqnarray}  
Similarly,    the equation of $a_x$ is given by
\begin{equation}  (\nabla_\perp^{\ 2}  + K^2 ) \,a_x  = - \mu\, j_{c,x}.  \label{TW-ax}    \end{equation}

\subsubsection{Spatial damping} 	 \label{s521} 
Let us consider spatial damping of a traveling wave in the absence of wave source by neglecting the term on the 
\underline{\it rhs} of (\ref{TW-eh4}). For the purpose to account for the damping effect owing to $\bj_d$,  the 
wave-number $k$ is expressed with  a complex form $k= k_r + i k_i$. Substituting $k=k_r+i k_i$ in (\ref{TW-ps1}), 
we obtain the following  damped  wave, traveling to the positive $x$ direction for $k_i>0$ and $k_r>0$: 
\begin{equation} 
\Psi =  \psi(y) \,e^{i \beta z}\,e^{-k_i x}\, e^{i(k_r x -\om t)}, \hskip15mm  \lam=2\pi/k_r, 	\label{TW-ps2} 
\end{equation} 
where $\Psi= e_x,\ h_x$ or $a_x$, and $\lam$ is its wave length, and spanwise variation $\Psi \propto e^{i \beta z}$
is assumed. In regard to (\ref{TW-eh4}), we have  $ K^2 = k_0^{\,2} - (k_r + i k_i)^2 + i \om \mu \sg$.
Thus, the equation of $e_x$ reduces to
\begin{equation}  \hspace*{-5mm} 
 \Big[ \dd_y^{\,2}  -  \beta^2 + \kappa^{\,2} + i(\om \mu \sg - 2k_rk_i) 
	\Big] \, \psi(y)  = S_x,   \quad \kappa^2= k_0^{\,2} - k_r^2 + k_i^2, 	 \label{TW-eh-x1}  
\end{equation}
where  $\dd_y=\dd/\dd y$  and $S_x = \mu\, \pl_t j_{c,x}$ (for the $e_x$ case).  If $S_x = -\mu\, j_{c,x}$ is used, 
the equation (\ref{TW-eh-x1}) reduces to that of $a_x$. Suppose that  the terms such as $\kappa$, $\beta$ and 
$S_x $ are  real and that  the real part of $\psi$  is taken (without losing generality).  Then the imaginary part 
in the brackets $[\,\cdots\,]$ is required to vanish.  Hence, requirement of reality of (\ref{TW-eh-x1}) results in
\begin{equation} 
k_i = \om \mu \sg /2k_r = (1/2 \kappa)\,\mu \sg c, \hskip10mm \kappa = k_r/k_0, \quad  k_0 \equiv \om/c. \label{k-i}  
\end{equation}
where $k_0$ is assumed to be related to LSM ($ k_0 = 2\pi/\lam_{\rm lsm}$).  Thus for $e_x$, the equation 
(\ref{TW-eh-x1}) reduces to  $[ \dd_y^{\,2} -  \beta^2 + K^{\,2} ] \, e_x(r) = \mu\, \pl_t j_{c,x}$.	 Now, 
neglecting the term $\mu\, \pl_t j_{c,x}$ on the \underline{\it rhs},  we seek a wave solution $\Psi$ traveling  
purely under  damping.  The equation governing $\psi$ is 
\begin{equation} 
\Big[ \dd_y^{\,2}  -\beta^2 + K^{\,2} \Big] \,\psi(y)= 0, \hskip10mm  K = \sqrt{k_0^2-k_r^2+k_i^2}, 
 \label{hom-eq}  
\end{equation}
Thus, a general solution of {\it damped} traveling wave  satisfying this is given by
\begin{equation} 
\Psi(x, y, z, t) =  C(y, z) \, e^{- x/d}\, e^{i(k_r x -\om t)}, \qquad  C(y, z) = \psi(y) e^{i \beta z}.
\label{x+Trav-wave}  
\end{equation}		
This implies that the damping distance $d$ of the wave (wavelength $\lam=2\pi/k_r$) is given by 
\begin{equation} 
d = 1/k_i = \kappa \,\frac{2}{\mu\sg c},  \qquad \kappa  = \frac{k_r}{k_0} = \frac{\lam_{\rm lsm}}{\lam}, \label{d-ki} 
\end{equation}	
This means that larger waves ($\lam > \lam_{\rm lsm}$) suffer stronger damping, resulting in reduced $d$.

\subsubsection{Temporal damping} 	 \label{s522}

Next, we consider temporal damping by using a complex frequency $\om =\om_r + i\om_i$, assuming 
that $k$ is real, $\om_i < 0$ and $\om_r>0$: 
\[	e_x,\ h_x \propto e^{\om_i t} \,e^{i(k x -\om_r t)},		\]
where the oscillation frequency is $\om_r$. In this case, the time derivative $\pl_t$ is replaced by
$-i\om = -i \om_r + \om_i$. Then the term $k_0^{\,2}+ i\om \mu \sg$ of (\ref{TW-eh3}) is replaced by
\[  \hspace*{-10mm} (\om_r + i\om_i)^2/c^{\,2}  + i \mu \sg (\om_r+ i\om_i) 
     = c^{\,-2} (\om_r^{\,2} -\om_i^{\,2}) - \mu \sg \om_i + i \om_r( 2\om_i c^{\,-2}+\mu\sg). \]
Thus the requirement of reality of an equation corresponding to (\ref{TW-eh-x1}) is satisfied by
\[ \om_i = - \half c^{\,2} \mu \sg = - c/d_*  \hskip10mm (d_* = 2/c\mu\sg). 	\]  
One may interpret this as follows. An initial uniform wave of wave number $k$ decays exponentially 
with a time constant,
\begin{equation} 
\tau_d = 1/|\om_i| = \frac{d_*}{ c} = \frac{2}{\mu \sg c^{\,2}} . 			\label{tau-d} 
\end{equation}
In this temporal-decay case,  a general wave solution of the equation (\ref{TW-eh4}) without the \underline{\it rhs}
forcing is a traveling wave (to the positive $x$),  given by
\begin{equation} 
\Psi =  \psi(y) e^{i \beta z}\, e^{- t/\tau_d}\, e^{i(k x -\om_r t)},   \label{t-Trav-wave}  
\end{equation}
with a characteristic decay time $\tau_d = d_*/c$.

\subsubsection{Wave excitation as a boundary value problem} \label{s523}
Let us try to consider the wave excitation  as a boundary value problem.  Suppose that a packet of hairpin-like
structure in the local wall layer was formed at a section around  $x=0$,  and that this triggered to  excite
transverse waves spanning the whole cross section.  Subsequent development is modeled as 
a boundary value problem of the first equation of (\ref{TW-eh-2}) for the field $\be$:
\begin{equation} 
\nabla^2 \be - c_t^{\,-2} \pl_t^{\,2}\be =  \mu \sg \pl_t \be ,  \hskip6mm 
		\mbox{for} \quad x > 0 \quad (\mbox{or} \ x<0 ).  	\label{TW-ex}  
\end{equation}
where the source term $\pl_t \bj_c$ on the \underline{\it rhs} is omitted by the understanding that a wave was 
excited by the action at a small section  localized  around $x=0$, and that there is no forcing source for $x>0$.  
It is assumed that waves are excited at a single angular frequency $\om$ for simplicity.  

We try to find a solution of (\ref{TW-ex}) for $x>0$, satisfying the boundary conditions of $e_x$ and $\pl_x e_x$ 
at $x=0$: \ $e_x(t)|_{x=0} = C(y, z)\,e^{ -i \om t}$, and  $\pl_x e_x(t)|_{x=0}=ik\,C(y, z)\,e^{ -i \om t}$, 
where real parts of \underline{\it rhs} are understood for their physical expression. Its solution is 
immediately found as  
\begin{equation} 
e_x(x,t) = C(y, z) \,e^{- x/d}\, e^{i(k_r x -\om t)} 	 \quad (x>0), 	 	\label{sol-ex+}  
\end{equation}
which satisfies the equation  (\ref{TW-ex}) for $x>0$ and the boundary conditions at $x=0$.

 This solution is characterized with two scales: a wave length $\lambda=2\pi/k_r$ and a damping distance  
 $d =1/k_i = 2\kappa /(\mu\sg c)$ from (\ref{d-ki}) where $\kappa =k_r/k_0$.
 

\subsection{Dynamical process of TW-field: Equation of growth and decay}  \label{s53}

Let us investigates  the dynamical process of TW-field excited in a streaky flow $\bU$ with streamwise vorticity 
(\S\ref{s3} and \S\ref{s6}) by deriving a model equation taking account of both terms of energy supply  and 
dissipation.  If the wavelength $\lam$ is larger than a natural wavelength $\lam_0=2\pi/k_0= 2\pi c/\om$ 
(\ie if  $\lam > \lam_0$), the TW-wave gains energy from the flow field $\bU+\bu$ (see \S\ref{s532}).  
This could be a prolongation mechanism. Namely, if there is no supply from the FF-field, the traveling wave just 
decays with a time constant $\tau_d$. If there is energy supply for perturbations of long wavelength 
from the FF-field owing to the effect of phase shift, the decay time is prolonged as $\tau>\tau_d$.

\vspace{1mm}
\subsubsection{Wave equation with forcing terms}   \label{s531}  \vspace{-1mm}

Once the TW-field is  developed sufficiently, the second dissipative term of (\ref{TW-S-34}) would  become 
significant. In order to investigate this situation,  let us consider the first of equation (\ref{TW-bf}) (or the 
second  of (\ref{TW-eh})) with $\bd=\eps \be$ and $\bj= \rho (\bU + \bu) + \bj_d $:
\begin{equation} 
- \eps \,\pl_t \be +  \curl \,\bh = \rho \bU + \rho \bu + \bj_d,   \qquad  \bj_d= \sg \be,  \label{TW-bf1}  
\end{equation}
by (\ref{drift-C}), under the condition of $\rho=\,$const and $\divg \bv=0$  for the sake of simplicity.   This 
reduces to $ \curl \,\bh_0 = \rho\,\bU$, for unperturbed state 
$\bv=\bU(\bx)$ which is assumed to be {\it time-independent} in the absence of perturbations (\ie $\bu=0$ and 
$\ba=0$).  The field $\bh_0$ solving this is a steady vector potential induced by the steady current $\bU$.

In the presence of perturbations, by setting $\bh = \bh_0 + \mu^{-1} \bfb$, the above equation becomes
\begin{equation} 
	- \eps \,\pl_t \be  + (1/ \mu)\, \curl \,\bfb =  \rho \bu +  \sg \be.    \label{TW-bf2}  
\end{equation}
Taking time derivative of both sides and using $\pl_t \bfb = - \curl \,\be$ from (\ref{TW-a}) and 
multiplying $\mu$ on both sides, one can transform (\ref{TW-bf2}) to the following\footnote[3]{\ $\curl \,\pl_t \bfb
=- \curl(\curl\,\be ) = \nabla^2 \be$ since $ \nabla(\nabla\cdot \be) = \eps^{-1} \nabla(\nabla\cdot\bd) = \eps^{-1} 
\nabla \rho=0$ by the assumption $\rho=$const.}:
\begin{equation} 
\nabla^2 \be - c^{\,-2} \pl_t^{\,2}\be  = \mu \rho \, \pl_t \bu + \mu \sg \pl_t \be,   \label{TW-e2}  
\end{equation}  
where  $c=1/\sqrt{\eps\mu}$.  The first term  on \underline{\it rhs} can act as a  wave source, while the second 
term acts as  damping.  In view of vanishing density perturbation $\rho'=0$ (assumed), the perturbation part 
of the second equation of (\ref{TW-bf}) reduces to
\begin{equation} 
\divg \be' =  -\pl_t \divg \ba' - \nabla^2\phi' =0.      \label{div-ea}  
\end{equation}  
where $\be' = - \pl_t \ba' - \nabla \phi'$.  By the Lorenz condition $\divg \ba' + c^{-2} \,\pl_t \phi' = 0$
of (\ref{Lorenz-c}), the above becomes  \ $c^{-2} \,\pl_t^2 \phi'  - \nabla^2\phi' =0$.
One can assume $\phi'=\,$const  and we have $\be = -\pl_t\ba$. 

Energy is supplied to the TW-field when $\bu\cdot \be <0$\, (\S\ref{s44}). Suppose that the 
state is maintained for some period in the presence of periodic perturbation waves of $\bu(t)$ and 
$\ba(t)$.  Taking  the $x$-component of (\ref{TW-e2}):
\begin{equation} 
\nabla^2 e_x - c^{\,-2} \pl_t^{\,2} e_x  = \mu \rho\,\pl_t u_x  + \mu \sg \pl_t e_x, 	 \label{TW-eu1}  
\end{equation} 
where the variables $e_x$ and $u_x$ denote the $x$-components of $\be$ and $\bu$.  The drift current term 
$\sg  e_x$ on \underline{\it rhs} causes a {\it phase shift} between $u_x$ and $e_x$. This is studied  next.

\vspace{1mm}
\subsubsection{Phase shift caused by the drift current and energy supply}  \label{s532}  \vspace{-1.5mm}
The drift current $\bj_d= \sg \be$ causes a phase shift between the flow perturbation $\bu(t)$ and the TW-field 
$\ba(t)$. Let us consider  monochromatic perturbation waves represented by 
\begin{equation}  
	\bu(t,\bx) = \bu_0 \, e^{i \theta}, \quad  \be(t,\bx) = \be_0 \, e^{i \theta},  \quad 
	\ba(t,\bx) = \ba_0 \, e^{i \theta},   			\label{TW-uea}  
\end{equation} 
where $\theta = \om t - kx$.  The coefficients $\ba_0$ and $\bu_0$ are complex constants in general and may take 
different phase arguments. Choosing appropriate origins of $t$ and $x$, the coefficient $\ba_0$ can be
assumed to be real without losing generality, while $\bu_0$ may be a complex. Thus, phase difference 
between $\ba$ and $\bu$ are taken into account naturally. Since $\be = -\pl_t\ba$, we have 
$\be_0= -i \om \ba_0$. Hence, the field $e_x$ is linearly-related to $a_x$:\, $e_x = -i \om a_x$. 

In addition, because of the linear form of the differential equation (\ref{TW-eu1}), one can assume a linear 
relation as well between  $u_x$ and $a_x$ as $u_x = i\om C\,a_x$ with $C$ a complex constant to be determined. 
Thus, by eliminating $u_x$ and $e_x$ with using this and $e_x = -i \om a_x$,  the equation (\ref{TW-eu1}) 
reduces to (after dividing both sides by $i \om$), 
\begin{equation}  
- \nabla^2  a_x + c^{\,-2}\pl_t^{\,2} a_x = c_\rho \,C \,\pl_t a_x - c_\sg \,\pl_t a_x, 	\label{TW-eu2}  
\end{equation} 
where  $c_\rho = \mu \rho$ and $c_\sg=\mu \sg$.  A solution to (\ref{TW-eu2}) 
is sought with a wave form of $a_x \propto e^{i(\om t - kx)}$  traveling to the $x$-direction.
Substituting $a_x = (a_x)_0 \,e^{i(\om t - kx)}$  into (\ref{TW-eu2}), and  rearranging the resulting equation, 
we find an equation to determine the unknown constant $C$.  Thus,   
\begin{eqnarray}
i \om \,C  & = & - \frac{1}{c_\rho} \big(k_0^{\,2}- k^2 -i\,\om c_\sg \big) 
		\equiv  - D \,e^{- i \,\om \del},			\label{om-C}  \\
&&  \hspace*{-25mm}  k_0 = \om/c, \quad \del \equiv \frac{1}{\om} \tan^{-1} \big[ (\om c_\sg) /(k_0^{\,2} - k^2) \big], 
	\quad   D \equiv \frac{1}{c_\rho} \sqrt{ (k_0^{\,2} - k^2)^2 + (\om c_\sg)^2 }.    \label{ep-ga}
\end{eqnarray}
If $k <k_0$, we have $\del > 0$ (for sufficiently small value of $c_\sg$), and the perturbation wavelength 
$\lam=2\pi/k$ is larger than the natural wavelength $\lam_0=2\pi/k_0$.   Thus it is found that 
\begin{equation}
u_x(t,\bx) = i\om C\,a_x = - D \,e^{- i \,\om \del} (a_0)_x e^{i (\om t - kx)}
	= - D\, a_x(t-\del, \bx).	\label{bu-ba-53}    
\end{equation}
This is the relation investigated in \S\ref{s44}, and the existence of such a phase shift enables energy supply 
from the flow field to the TW-field.  Since $c_\sg=\mu \sg$, the phase shift $\del$ defined by (\ref{ep-ga}) 
is caused by non-zero value of the constant $\sg$ of the fluid-Ohm's law (\ref{drift-C}) (in other word, 
{\it D}-effect).  This implies that, if the perturbation wavelength $\lam$ is larger than the 
natural one $\lam_0$, the wave field $\be$ gains energy from the flow field $\bu$, according to \S\ref{s44}.

If $k > k_0$ on the other hand,  we have $\del < 0$, and the perturbation wavelength $\lam=2\pi/k$ is smaller than 
$\lam_0=2\pi/k_0$. The energy flow is reversed such that it is from  the TW-field to the flow 
field if the perturbation wavelength is sufficiently short.   

Anyway, the resistive drift current $\bj_d= \sg \be$ causes the phase shift between the flow perturbation 
$u_x(t)$ and the wave field potential $a_x(t)$.

\section{Streaky wall turbulence}  \label{s6}
One of the important  areas of application of the present formulation  would be 
the streaky shear-flow turbulence.  To begin with,  we first review some experimental facts which are well-known 
but viewed from the light of the present scenario, in which  significance of the travelling wave component 
is emphasized. In \S\ref{s35}, we considered the {\it transient growth} mechanism of small disturbance waves 
in {\it laminar} shear flows as being possible seeds for turbulent motions in channel flows.  When we consider 
the streaky channel turbulence, we cannot proceed without mentioning the scenario of transient growth mechanism  
(Gustavsson 1991;  Butler \& Farrell 1992;  Henningson, Lundbladh \& Johansson 1993;  Trefethen,  Trefethen,  
Reddy \& Driscoll 1993) for laminar wall flows.

The streaks in actual  turbulence are wavy and non-uniform, and surrounded by a sea of incoherent 
turbulent motions. For {\it turbulent} shear flows too, the transient amplification mechanism  was investigated 
for infinitesimal disturbances.  Here, two  studies by Schoppa \& Hussain (2002) and  Del \'{A}lamo \& Jim\'{e}nez 
(2006) are cited. Then we  consider how those are re-interpreted in terms of the new scenario in \S\ref{s62}, 
\S\ref{s63} and \S\ref{s64}.

The streak structure in the wall turbulence is considered to be a {\it dissipative structure}, analogous to the 
convection cells in the thermal convection where thermal energy is transferred from heated bottom  to cooled top 
surface.  In the present problem,  however, energy is transferred from the flow field to the TW-wave field and 
dissipated  partly (\S\ref{s63}).

\subsection{Experimental features of wall turbulence} 	\label{s61}  
\centerline{($a$) {\it Triple decomposition}}   \label{s61a}
\vskip1mm
Near-wall turbulence is characterized by streaky structures which are wavy  in a sea of turbulent 
fluctuations. In early times of experimental studies of  turbulent shear flows, Hussain \& Reynolds (1970, 72, 75)
investigated channel flows which were fully developed turbulence and in addition excited weakly by a periodically 
vibrating ribbon.  Thus by introducing a weak sinusoidal wave at an upstream position and then extracting signals of 
weak periodic motions at downstream stations from the background  turbulent flow, they expressed the detected 
time-dependent signal $f(t)$  with a triple decomposition, consisting of ($i$) time-averaged  mean component  
$\overline{f}$, ($ii$) time-periodic component $\tilde{f}_p$,  and ($iii$)  incoherent turbulent fluctuation $f'$:
\begin{equation}
f(t) = \overline{f} + \tilde{f}_p(t) + f'(t).  		\label{triple-s4}	 
\end{equation}
Magnitude of each component was as follows (Hussain \& Reynolds, 1972).  Regarding the periodic component, the 
amplitude $|\tilde{f}_p|$ was typically about $10^{-3}$ of the mean center-line velocity $\overline{U}_0$,  or a few 
hundredths of the {\it rms} velocity of turbulent component $|f'|_{\rm rms}$.  Hence the organized periodic 
component was very weak in the background turbulent fluctuations.  

More recently,  Schoppa \& Hussain (2002) proposed a triple decomposition of the velocity field of near-wall 
turbulence into ($i$) a {\it time-independent} mean streaky flow $U_s(y,z)$, ($ii$) a  wavy  perturbation superposed 
on the streak which is evolving  from the evolution (growth, vortex-dynamics, and decay) to regeneration cyclically, 
and ($iii$) incoherent turbulence, 

It is noteworthy that the periodic  wave detected in the former case was long-living and {\it robust} in 
the background irregularly fluctuating  flow.  In fact, two features are particularly  noted for this study of  
turbulent channel flow. First one is the robustness of the periodic wave component just mentioned above. 
The present approach may support these observations.  It is one of its essential features that a {\it TW}-wave 
has its own momentum and energy (\S\ref{s2}) like the electromagnetic waves. Hence the wave keeps existing 
unchanged unless it loses (changes) its momentum and energy by interaction with other components. By such 
interaction, total momentum and energy of all the interacting components must be conserved. This  explains the 
robustness of the periodic wave component $\tilde{f}_p$ in the turbulent channel flow.  Second one 
is considered in the next ($b$).

\vskip3mm 
\centerline{($b$) {\it Enhanced diffusivity and dissipation}}    \label{s61b}
\vskip1mm
Reynolds \& Hussain (1972) found that their {\it eddy}-viscosity representation served very well.  To calculate 
eigenfunctions  by solving their linear perturbation equation for small disturbances in the turbulent shear flow, 
they took into account the wave-induced oscillations in the Reynolds stresses. It is essential to model the interaction 
of the wave component with background turbulence by a turbulent {\it eddy}-viscosity representation,  for which 
an empirical eddy viscosity model was used to obtain reasonable agreement with experimental observations.
This was also confirmed by the study of  Del \'{A}lamo \& Jim\'{e}nez (2006).

The present theory is equipped with an additional  mechanism of enhanced dissipation, which is described 
compactly in \S1 (Introduction), \S\ref{s232}, and in  \S\ref{s7}.  The dissipation is caused by a drift current 
$\bj_d$ driven by the {\it TW}-field acting on the turbulence field (\eg see (\ref{drift-C}), 
(\ref{J1-loss})). This is called the {\it turbulence} Darcy effect, in which the fluid Lorentz force acting on 
the turbulent medium (consisting of a number of turbulent eddies) plays a role, analogous to  the pressure 
gradient acting on a viscous medium causing the Darcy current through a porous medium.  This effect is called 
shortly as  \ {\it D}-effect, and  resembles the Ohm's law  in the electromagnetism.

As shown in \S\ref{s71}, energy dissipation by the {\it D}-effect can be expressed in a form analogous 
to the eddy-viscosity, and its magnitude  is much larger than the dissipation of molecular 
viscosity and comparable to  that of  eddy-viscosity model. Even within the framework of current theory, 
there are studies (Reynolds \& Hussain 1972;   Del \'{A}lamo \& Jim\'{e}nez 2006),
in which the turbulent eddy diffusivities were taken into account in  linear 
analysis for disturbances in  turbulent shear flow to obtain results consistent with experimental observations.  

\subsection{Streaky channel turbulence and large scales}   \label{s62}
Two studies are particularly noted here. First one is that by Schoppa \& Hussain (2002),  who proposed 
streak transient growth mechanism for generation of streamwise vortices in a streaky   near-wall turbulence. 
Second one is that by  Del \'{A}lamo \& Jim\'{e}nez (2006), who investigated the stability of the 
mean velocity profile of turbulent channel flow by using an eddy viscosity, suggesting that the modes selected with 
the largest transient growth could be {\it seeds} for structures of the streamwise velocity in the 
turbulent flow.  The present study provides a mechanism  supporting these but with a new scenario. Each subsection 
here gives some of supporting evidence.

A conceivable approach is a triple decomposition of the total velocity  $\bv$ of the fully 
turbulent flow into ($i$) a mean  flow $\bU_m(y)$, ($ii$) a  wavy  perturbation $\bu_w=(u_j)$ superposed on 
the mean flow, and ($iii$) incoherent turbulence $\bu'$.    Thus, the total velocity is $\bv= \bU_m + \bu_w + \bu'
= \bU_m +\bw$, where  $\bw= \bu_w + \bu'$ is the {\it time-dependent} part.  Let us consider turbulent channel 
flow of an incompressible fluid of density ($\rho=$ const), where the mean shear flow is directed to the 
$x$-direction with the wall-normal $y$ and spanwise $z$  ($0 < y < 2h$).   Experimental 
studies of shear flow turbulence  recognize existence of two large scales of the streaky  structures:  LSM  and VLSM, 
characterizing the streamwise streaks and long meandering structures. 

\subsubsection{Transient growth mechanism of laminar channel flow} \label{s621}
\vspace*{-1.5mm} 
In the transient growth mechanism studied by Gustavsson (1991),   Butler \& Farrell (1992) and  Henningson, 
Lundbladh \& Johansson (1993),  small disturbances  
grow rapidly and robustly in {\it laminar} channel flow (\ie the plane Poiseuille flow).  An initial state is usually 
made up of many non-orthogonal modes,  the combination of which can result in dramatic growth of $O(10^3)$ times 
of initial energy for three-dimensional disturbances of spanwise wavelength comparable with the channel width. 
Such disturbances that can grow sufficiently are  characterized by  elongated structure in the streamwise 
direction (expressed by very small or zero  wavenumber in the streamwise direction). In the cross-stream section 
of this disturbance, there exists spanwise variation and wall-normal variation of cross-stream velocities which 
define streamwise vorticity. Once generated, perturbations with streamwise vorticity have a significant effect 
on the flow through formation of {\it streaky} structures. 

In fact, Butler \& Farrell (1992) gave the value $\beta =2.04$ for the spanwise wavenumber of the optimal 
perturbation (streak width,  $\lam_s/h = 2\pi/\beta $). The computation of  Del \'{A}lamo \& Jim\'{e}nez (2006) 
described in the next \S\ref{s622} predicts the streak 
width of about $3h$ corresponding to their $\beta$ of about 2.09. In order that the optimal perturbation wave predicts 
the scales of LMS, or VLSM, its streamwise wavenumber $\al$ must take a value close to $2 \sim 6$ from Eq.(\ref{lsm}), 
or $0.3 \sim 0.4$ from Eq.(\ref{vlsm}).  However, the {\it laminar} flow (Poiseuille) profile studied by 
Butler \& Farrell gave the optimal perturbation the value $\al \approx 0$. 

It is noteworthy however that Butler \& Farrell (1992)  gave a diagram  (their Fig.14) of the cell pattern of developed 
streamwise velocity $u$.  By the present linear analysis of \S\ref{s34}, there exists an excitation mechanism of the  
$\ba$-field from the transient growth scenario.  In fact, the analysis there implies that the streamwise velocity 
$u$ predicts a streamwise $\ba$-field by the first  of (\ref{sol-ua}).  Seeing the energy equation (\ref{TW-e-s23}), 
its \underline{\it rhs}  can be positive locally  where the $\ba$-field is growing to the direction of $\bU$ 
(see (\ref{Sw})).  Thus, it is likely that the {\it TW} field can be excited in shear flows by the perturbations 
of the transient growth scenario.

\subsubsection{Large scales of turbulent channel flow from Del \'{A}lamo \& Jim\'{e}nez} \label{s622}
\vspace*{-1.5mm} 
In regard to {\it turbulent} shear flows too, the transient growth scenario was examined by Del \'{A}lamo \& 
Jim\'{e}nez (2006), who first computed  the mean velocity $\bU_{\rm m} =(U_t(y), 0, 0)$ of turbulent channel 
flows satisfying the following mean-flow equation ${\rm M}[\bU_{\rm m}] =0$:
\begin{equation}
{\rm M}[\bU_{\rm m}]\,_x = \bU_{\rm m}\cdot\nabla U_t + \rho^{-1} \pl_x \overline{p} + \pl_j (\overline{ w_x w_j} ) 
		\quad ( \, \mbox{\small $ - \nu \,\nabla^2 U_t$ } \,) = 0.			\label{MeanEq}
\end{equation}
(the overline denoting  {\it time-mean}). The first term $\bU_{\rm m}\cdot\nabla U_t$ 
vanishes by the assumed form of $\bU_{\rm m}$. The channel turbulence is driven by a constant {\it negative} 
mean gradient of pressure $\pl_x \overline{p}$\,: 
\[ 		\pl_x \overline{p} = - \tau_w /h = - \rho \frac{u_\tau^{\,2}}{h} \quad (\mbox{constant}), 	\]
where $u_\tau$ and $\tau_w = \rho u_\tau^2$ are the friction velocity and wall shear stress 
respectively (Pope 2000), and the viscous friction length is defined by $\del_\nu \equiv \nu/u_\tau$. 
Normalizing  $w_j$ and $\overline{p}$  by $u_\tau$ and $\rho u_\tau^2$ respectively, the above (\ref{MeanEq}) 
is reduced to the following (with dividing by $\tau^2/h$),
\begin{equation}  
 -1  +  \frac{\dd}{\dd \eta} \Big( \nu_t(\eta)\,  \frac{\dd U_t}{\dd \eta} \Big) = 0,	 \qquad
 	\eta = \frac{y}{h}. 		\label{MeanEqN}
\end{equation}
The gradient of Reynolds stress term $\pl_j (\overline{ w_x w_j} )$  of (\ref{MeanEq})  was  replaced by a 
model of  eddy-diffusivity form $ - \pl_y(\nu_t \,\pl_y U_t)$. Its magnitude is about $10^2$-times larger than that of 
 molecular-viscous-stress term $\nu \,\nabla^2 U_t$ in the parentheses of (\ref{MeanEq}). Hence the latter was omitted.

They studied the stability of the turbulent channel flow $\bU_{\rm m}$, using a variable eddy-viscosity 
$\nu_t(y)$, and proposed that  the modes with the largest transient growth are related to the large scale 
structures.  They assumed wave-like perturbations, 
\begin{equation}
\bq= [u_y, \,\om_y] = [\hat{u}_y(y), \,\hat{\om}_y(y)] \exp[ i (\al x + \beta z -\gam t)],  \label{wave-622}
\end{equation}
for the $y$-components of both the velocity $\bu_w$ and the vorticity $\bom =\nabla \times \bu_w$.  Their 
analysis is based on the perturbation  solutions satisfying the modified Orr-Sommerfeld equation for $u_y$  
and the Squire equation for $\om_y$.

It was found that there exist two scales of disturbances in turbulent channel flow which are  transiently 
amplified  sufficiently according to linear perturbation equations. One corresponds to a sublayer scale and 
the other  to the larger-scale structure spanning the full channel.   The disturbance of a sublayer scale grows 
self-similarly in the logarithmic (or overlap) region up to the scale of vortex packets.  It is generally 
accepted  that the vortex packets in the wall  shear layer are composed  of hairpin-like structures and their typical 
streamwise scale $l_p$ characterizes the scale of LSM of (\ref{lsm}). It is proposed that  the wave in the 
logarithmic region triggers this wave of the scale $\lam_{\rm lsm}$, which extends over the whole cross-section  
and travels to the surrounding space. This is considered to be the {\it TW}-wave described next in \S\ref{s623}.

In addition, they found another larger-scale  waves of streamwise wavelength of 
$20h \sim 60h$, corresponding to the VLSM scale. Its spanwise wavelength of the streak is $\lam_s/h \approx 3$
(where $\lam_s/h =2\pi/\beta$).  These are  amplified by the transient growth mechanism. 
From these observations, their study hints that these waves act as seeds for the longer-lived and stronger 
structures of streamwise velocity, surrounded by a sea of turbulent fluctuations.

Regarding  this interpretation, there remains one question why those structures exist robustly in the sea of 
turbulent fluctuations. This is resolved by the fact that the waves of  scales  LSM and VLSM  are 
connected with the {\it TW}-waves which have their own characteristic energy and momentum. 

\vskip1mm  
\subsubsection{Our system with new TW-field} \label{s623}
\vspace{-1.5mm}
In our system,  the mean flow equation is supposed to take the same form as (\ref{MeanEqN}). The eddy-viscosity 
may be somewhat different from the $\nu_t(y)$ of  Del \'{A}lamo \& Jim\'{e}nez (2006).  However, for the sake of 
interpretation of the present scenario, we consider the case of the same viscosity as theirs.

Suppose that the sublayer scale disturbance found by them has grown  up to the scale of vortex 
packets of the wall-shear layer.  Then  the convection current $\bj_c = \rho \bv$ would have become sufficiently 
large to excite TW-field.  According to the equations of TW-waves (\ref{TW-eh-2}):
\begin{equation} 
\nabla^2 \be - c^{\,-2} \pl_t^{\,2}\be  -  \mu \sg \pl_t \be =  \mu\, \pl_t \bj_c ,  \hskip8mm 
\nabla^2 \bfb  - c^{\,-2} \pl_t^{\,2}\bfb  - \mu \sg \pl_t \bfb= - \mu\, \nabla \times \bj_c ,   \label{TW-eh-623}
\end{equation}
both of the $\be$- and $\bfb$-fields are excited simultaneously by the time derivative $ \pl_t \bj_c $ and its 
rotational property $\nabla \times \bj_c$, respectively. The excited wave fields $\be$ and $\bfb$  propagate 
through the turbulent field with the phase velocity $c=1/\sqrt{\eps\mu}$ and the wavelength $\lam_{\rm lsm}$, 
where the turbulent field is characterized with the parameters $\eps$, $\mu$ and $\sg$.  The waves exchange 
energy with the flow field $\bv$, and also lose it during propagation owing to the {\it D}-effect. 

\vskip1mm \noindent 
($a$) {\it Governing equations}: \ Once the TW-field is excited sufficiently, the  flow field is governed  by  the  
equation (\ref{NS-fLa}),  where the total velocity  $\bv$ is composed of  the mean  flow $\bU_m$ and the 
time-dependent part $\bw= \bu_w + \bu'$ consisting of the wavy part $\bu_w$ and incoherent turbulent part $\bu'$.
The time-mean part $\bU_m$ is governed by (\ref{MeanEq}). Subtracting this mean equation from the total equation 
(\ref{NS-fLa}) for $\bv$, the equation of the time-dependent component $\bw$ is given by 
\begin{equation}  
\pl_t\bw + (\bU_m \cdot\nabla) \bw + (\bw \cdot\nabla) \bU_m  + (\bw \cdot\nabla) \bw =
 		- \nabla p_w  + \bff_{\mbox{\tiny L}}[\ba], 				\label{Eq-w-623} 
\end{equation}
where $\bff_{\mbox{\tiny L}}[\ba] = \be + \rho^{-1} \bj\times \bfb$, with $\be$ and $\bfb$ are governed by 
(\ref{TW-eh-623}), and $p_w$ is the pressure associated with the $\bw$ motion,  and  $\nabla\cdot \bw=0$ is assumed.
The total current is given by $\bj = \bj_c + \sg \be$ where $\bj_c = \rho (\bU_m +\bw)$.  The wave component $\bu_w$ 
is supported by the $\ba$-field through the interaction term $ \bff_{\rm L}[\ba]$.  The  TW-wave has its own 
momentum and energy and  keeps them unchanged unless the energy and  momentum are changed according to 
\begin{equation}
\pl_t w_e  + \divg \bq_{\mbox{\tiny fP}} = - \bj_c \cdot\be -  \sg |\be|^2,  \hskip10mm 
\pl_t \bg  + \nabla\cdot M = - \rho\,\bff_{\mbox{\tiny L}}[\ba],     \label{TW-623}   
\end{equation}
(see (\ref{TW-e-s23}) and (\ref{TW-m-s23})).   This is an advantage of the present scenario because it can 
explain why the wave component exists robustly  in the turbulent environment.

\vskip1mm \noindent 
($b$) {\it Important notes}: \ In the equation (\ref{Eq-w-623}) for the time-dependent  $\bw$, the viscosity term 
$\nu \nabla^2 \bw$ is omitted. Instead, much larger dissipation term $\sg |\be|^2$ is included in the first equation  
of (\ref{TW-623}) for the energy density  $w_e$.    One reason is that the role of molecular viscosity term is 
unimportant in the  streaky turbulence as interpreted in the item (i) of the next section \S\ref{s63} and in 
addition that the magnitude of dissipation due to the term  $\nu \nabla^2 \bw$  is much smaller than that of 
$\sg |\be|^2$ explained in \S\ref{s7}.  Second is a fundamental aspect of the theoretical physics, which requires 
{\it causality}, namely signals should propagate with a finite speed. The diffusion-type equation of the form 
$\pl_t u = \nu \nabla^2u$ predicts that a signal of $u$ propagates at infinite speed, and the Navier-Stokes equation 
has such a property.  This is remarked by Scofield \& Huq (2014).

\subsubsection{Coherent structures in wall turbulence by Schoppa and Hussain} \label{s624}
\vspace{-1.5mm}
 By the transient growth mechanism  acting for initial small amplitude perturbations in the streak-less flow 
$U_*(y)$, the most amplified perturbation grows into an $x$-independent  finite-amplitude streak. The
spanwise wavelength thus obtained  for a laminar base flow of  plane Poiseuille profile  was found as 
$\lam_s/h = 2\pi/\beta \approx 3.1$ with $\al \approx 0$ \,  (Butler \& Farrell 1992;  Reddy \& Henningson 1993). 
For such $x$-{\it independent} flow, however, streamwise vorticity decays monotonically. 

In contrast,  for a streaky  flow $U_s(y,z)$, Schoppa and Hussain (2002) proposed a scenario of {\it streak transient 
growth} concerning growth-decay evolution. They considered an $x$-{\it dependent} perturbation $\bu_w=(u_x, u_y, u_z)$
in the streaky flow, $u_z \propto   g(y)\,\sin (\al x)$ (with $g(y)=y \exp(- c y^2)$),  \ie {\it wavy} 
perturbations with spanwise motion superposed on $U_s(y,z)$. They deduce that this generates {\it streamwise 
vortices} on the basis of the  evolution equations for vorticity perturbations and those for perturbation 
kinetic energy.  The wave component $\bu_w$ is supported by the $\ba$-field through the interaction term 
$ \bff_{\rm L}[\ba]$ in the equation (\ref{Eq-w-623}), and exists robustly since the TW-wave has its own momentum 
and energy, while the vorticity dynamics is governed by the flow equation (\ref{NS-fLa}): \, 
${\rm NS}[\bv] = \bff_{\rm L}[\ba]$.

Their scenario is summarized as follows: ($i$) transient growth of $x$-independent perturbations to the base flow
$U(y)$ which  generate a finite-amplitude $z$-varying streak $U_s(y, z)$, and followed by ($ii$) transient growth of 
$x$-dependent perturbations to the streaky flow $U_s(y,z)$ which generate new streamwise vortices repeatedly.

It is proposed that  this wavy streak with streamwise vortices would be the wave of the larger scale $\lam_{\rm vlsm}$ 
of (\ref{vlsm}). This could be associated with the larger-scale  waves of streamwise wavelength, $20h \sim 60h$, 
found by Del \'{A}lamo \& Jim\'{e}nez (2006) which were amplified transiently within turbulent flows. 
Again, the transiently amplified waves are captured by  the system (\ref{TW-eh-623}) of TW-field. 
Finally this  results in sustained streaky turbulence. 

According to Schoppa and Hussain (2002), the streamwise vortices thus generated are similar to the coherent structures 
educed from the numerical experiment (Jeong, Hussain, Schoppa \& Kim, 1997),

\subsection{Dissipative structure}   \label{s63}
The previous section \S\ref{s62} has summarized  how the streaky  structure of near-wall turbulence is 
understood by the current theory, and then how the streaky  turbulence is reinterpreted by our new scenario.
We recognize three characteristic features of the new scenario as follows.
\begin{list}{}{\leftmargin=3mm}
\item[($i$)] \  Firstly, no major role is played by the molecular viscous term in the streaky turbulence.\footnote{There 
is an exceptional layer, \ie the viscous sublayer adjacent to the wall $y^+= y/\del_\nu = y u_\tau/\nu <5$, where the 
viscous boundary layer is formed by the no-slip condition for $Re_y \equiv u_\tau y/\nu <5$.}   In  the main shear 
layer above $y^+ \approx 5$, the dynamics of disturbances is controlled by the Reynolds stress term, and the turbulent 
eddy-viscosity $\nu_t$ was used to describe the {\it turbulence-induced} diffusivity of the perturbations.
The perturbation waves grow transiently (with an inviscid mechanism) within the turbulent environment sufficiently 
even for conditions under which  the flow Reynolds number is below a critical value for stability from the  
normal linear  theory.  The transiently amplified wave is captured by the {\it TW}-field. The TW-field 
would be maintained if energy is gained from the flow field sufficiently, but decays by the dissipation 
mechanism mentioned in \S\ref{s623} ($b$).
\item[($ii$)] \  Each TW-wave has its own energy and momentum.  Hence, it keeps its own state unless changed by 
interactions with other components satisfying  conservation laws (\ref{TW-623}). This explains the second feature, 
namely  the streaky structure exists robustly in turbulent environment,  maintaining itself.   
\item[($iii$)] \  The resistive drift current $\bj_d= \sg \be$ causes the phase shift between the flow 
perturbation $\bu(t)$ and the wave field potential $\ba(t)$ (\S\ref{s532}). Existence of the phase shift 
enables energy supply from the flow field to the TW-field (\S\ref{s44}). If the perturbation wavelength $\lam$ 
is larger than the natural one $\lam_0 \approx \lam_{\rm lsm}$, which is proposed to be $h \sim 3h$ by
(\ref{lsm}), the wave field $\be$ gains energy from the flow field $\bu$. It is natural to consider that the waves 
of $\lam$ in the range $\lam_{\rm lsm} \lesssim \lam  \lesssim\lam_{\rm vlsm}$ are able to gain energy from 
the flow field, because those waves are observed in the energy spectrum obtained experimentally. 
Initially, these waves are amplified transiently by the {\it transient growth mechanism} investigated by 
Del \'{A}lamo \& Jim\'{e}nez (2006).
\\[-5mm]
\end{list}
In addition,      \\[-5mm]
\begin{list}{}{\leftmargin=3mm}
\item[$\bullet$] \  The above mechanism and dynamics  imply that the streaky structure of wall-bounded 
turbulence is  a {\it dissipative structure}. Energy is supplied from the main flow $\bU$ to the TW-wave field, 
and some part of the energy is dissipated into heat. If there is energy balance between 
supply and dissipation, the structure is maintained. 
\item[$\bullet$] \ The TW-field is equipped with  a  mechanism of energy dissipation,  called  {\it turbulence}-Darcy 
effect (\S\ref{s232}).  In fact, the rate of energy dissipation takes a form resembling the models of eddy-viscosity, 
and its magnitude is comparable  with that of eddy viscosity (see \S\ref{s7}).
\item[$\bullet$] \ The streaky wall flow $\bU$ with streamwise vorticity is basically unstable (\eg Schoppa \& Hussain 
(2002); Wedin \& Kerswell (2004)) with respect to cross-stream perturbations. Transient growth of those $x$-dependent 
perturbations  generate new streamwise vortices. This process continues cyclically, and thus the streaky structures
 are maintained in turbulent environment of shear flows.
\end{list}
In the present scenario, the energy of excited TW-field is described by the energy equation (\ref{TW-e-s23}) for 
its energy density $w_e$ (\S\ref{s231}). The excited $\be$-field is a traveling wave,  described by the wave 
equation (\ref{TW-e2}),
\begin{equation} 
\nabla^2 \be - c^{\,-2} \pl_t^{\,2}\be  = \mu \rho \, \pl_t \bu + \mu \sg \pl_t \be,  	   \label{TW-e-63}  
\end{equation} 
where the first term on \underline{\it rhs} is a source, while the second term acts as wave damping.  If the 
both terms on \underline{\it rhs} balance exactly and cancel out, one obtains just simple wave propagation, 
$\nabla^2 \be - c^{\,-2} \pl_t^{\,2}\be=0$.  Thus the structure is maintained as far as the basic main flow 
$\bU$ is kept unchanged by external means.

The energy flows down from the main flow field, passing through the structure,  to the transverse wave field, 
and finally dissipates into heat, whereas the structure itself is maintained.  This is the reasoning why 
the streaky structure is called a {\it dissipative structure}. 
This dynamical phenomenon is illustrated conceptually as a diagram in Fig.1.

\newpage
\begin{figure}[h]
\begin{center}
\includegraphics[width=0.8\textwidth]{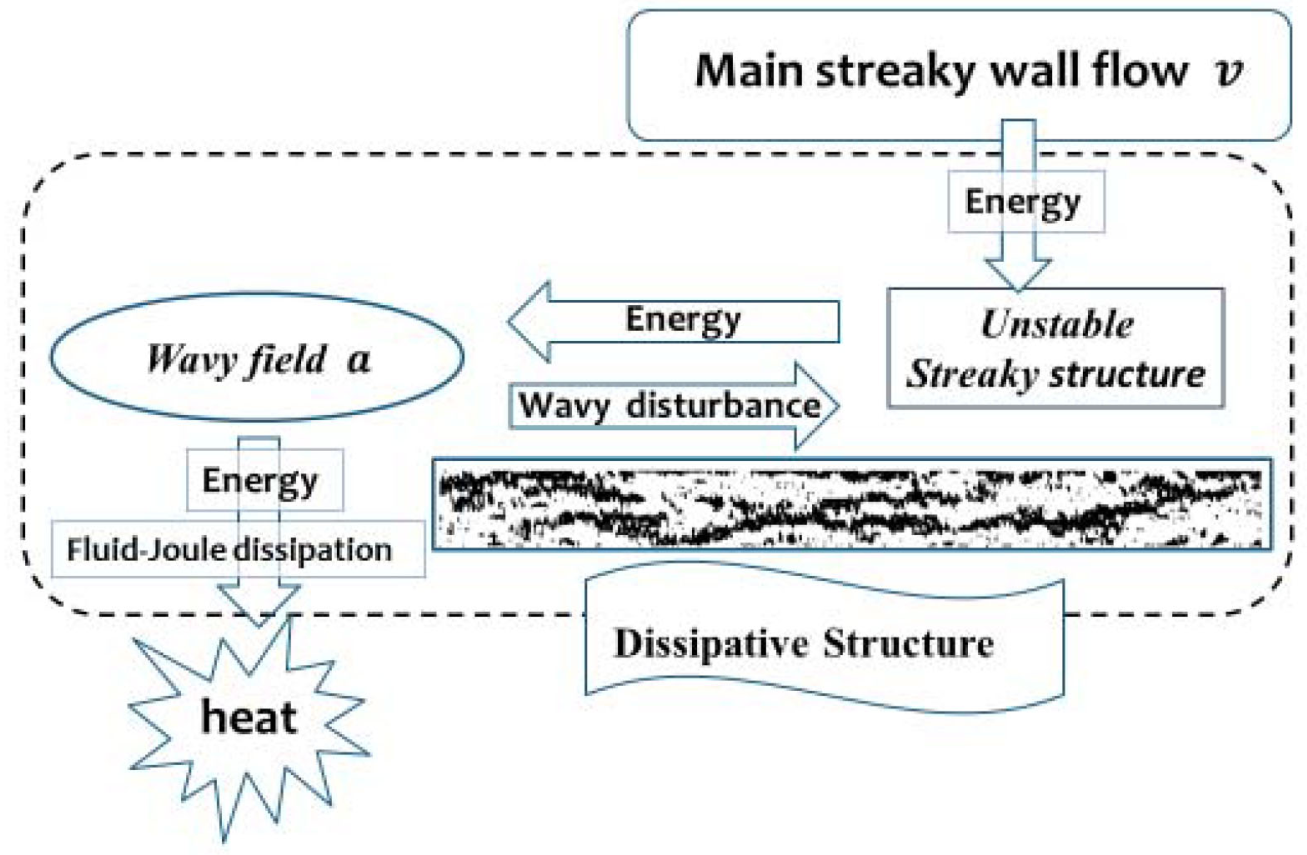} \par		
\vspace{-10mm}  
\baselineskip=3.3mm    \vspace*{6mm} 
\begin{minipage}{160mm}  
\caption{	
Conceptual diagram of sustaining mechanism of streaky structure, showing that
the streaky structure of wall-bounded turbulence is a dissipative structure. \\ 
\hspace*{5mm} The inset in the lower right corner shows a meandering (low-speed) streaky 
structure observed experimentally in pipe turbulence (from Monty et al. (2007)). } 
\end{minipage}  
\end{center}
\end{figure}

\baselineskip=4.5mm

\vspace{-9mm}
\subsection{Energy spectrum  $k_x^{-1}$ between LSM and VLSM }   \label{s64}
\vspace{-2mm}
Experimental studies of pipe turbulence were carried out concerning the streamwise energy spectrum 
$E_\parallel(k)$, which is reviewed in \S\ref{s1} ($c$).  It is found that the pre-multiplied energy spectrum 
$k_xE_{\footnotesize \parallel}(k_x)$ has two characteristic scales at $k_{\rm lsm}= 2\pi/\lambda_{\rm lsm}$ 
corresponding to LSM of  $\lambda_{\rm lsm}/R \approx 1\sim 3$ and at $k_{\rm vlsm}= 2\pi/\lambda_{\rm vlsm}$ 
corresponding to  VLSM  (where $R$ is the pipe radius and $k_x$ the streamwise wave number), and decays 
beyond the VLSM. The energy spectrum $E_{\footnotesize \parallel}(k_x)$ takes a scaling form as
\begin{eqnarray} 			 
E_{\footnotesize \parallel} \ & \propto  \ k_x^{\,-1} \qquad \qquad  & {\rm A}: \ ( k_{\rm lsm} \, \gtrsim \ k_x \ 
		\gtrsim  \ k_{\rm vlsm} ),			\label{TW-range}		\\
E_{\footnotesize \parallel} \  & \ \ \propto  \ k_x^{\,-5/3} \qquad \qquad & 
		 {\rm B} : \  (k_x \, \gtrsim \ k_{\rm lsm}). \label{inert-range}	 
\end{eqnarray}
The second spectrum $k_x^{-5/3}$ for higher wavenumbers  B:\, ($k_x \, \gtrsim \ k_{\rm lsm}$) is the well-known 
power law of developed turbulence, and further remark would be unnecessary. However, the first $k_x^{-1}$-spectrum 
for lower wavenumbers A needs additional interpretation.

In the standard theory, turbulence is regarded as composed of a number of eddies of continuously different 
scales.  Scaling estimates are carried out {\it phenomenologically} to obtain scaling laws of turbulence dynamics 
by denoting  the eddy scale as $\ell$   and its wavenumber $k$ expressed as $\ell^{-1}$ in the 
order-of-magnitude arguments.  A representative velocity and kinetic energy (per unit mass) of the eddy $\ell$ 
is expressed by $v_\ell$ and $e_\ell$ respectively. Then we have the scaling relation $e_\ell \sim v_\ell^{\,2}$, 
and the time scale $\tau_\ell \sim \ell/v_\ell$ with respect to  the eddy $\ell$. 

Rate of energy transfer through different wavenumbers is defined by a scaling law, $\vep_\ell \sim
e_\ell/\tau_\ell \sim v_\ell^{\,3}/\ell$. Hence, we have $v_\ell \sim (\vep_\ell\,\ell)^{1/3}$. The fully developed 
turbulence is characterized by the law of constant energy transfer defined by $\vep_\ell= \vep = \const$. Then the 
kinetic energy per unit mass is given by $e_\ell \sim v_\ell^{\,2} \sim \vep^{2/3} \ell^{2/3} \sim \vep^{2/3} \,
k^{-2/3} $.  Equating this to $k\,E(k),$\footnote{Expressing the scale-$\ell$ energy by $e_\ell \sim E(k) \Del k$ 
by the energy spectrum $E(k)$, we have $E(k) \Del k \sim   E(k)\,k \,\big(\Del k)/k \big)$.   We 
obtain the relation $e_\ell \sim E(k) k $ if  $ \Del k /k = \Del \log k \sim 1$  (with the logarithmic scale).}\  
 we obtain the well-known Kolmogorov's energy spectrum $E(k) \sim \vep^{2/3} \,k^{-5/3}$, where the $\vep$ is 
 also equal to the 
 
\newpage 
\noindent
rate of energy dissipation in a statistically stationary state.

On the other hand, for the $k_x^{-1}$-spectrum at lower wave numbers A, let us suppose that $\vep_\ell \sim k^\al$ 
for a  parameter $\al$, instead of constant $\vep_\ell$.  Then we have $E(k) \sim k^{(2/3)\al} \,k^{-5/3}$,
which is given by $k^{-1}$ for the wavenumbers A. This requires that $\al=1$, hence we obtain
$\vep_\ell \sim \ell^{-1}$ and  $v_\ell \sim \const$. The last means that $v_\ell$ is independent of $\ell$.

In the scale range A of large-scale waves, the magnitude of eddy of scale $\ell$ would be reinterpreted as the 
magnitude of waves of wavelength $\lam$, and the wave magnitude is independent of $\lam$ according to the last result.
This property does not contradict with the finding of Del \'{A}lamo \& Jim\'{e}nez (2006). By the transient growth
mechanism, large scale waves are amplified from the  channel turbulence acting as a source of random disturbances of
all sizes.  Most amplified waves of wavelength $\lam \gtrsim 10h$ become constant in magnitude for very large $\lam$, 
namely independent of $\lam$, where the spanwise widths of those waves are of the order of $3h$ for all modes.
Thus, the spectrum law  $k_x^{\,-1}$  does not imply the energy cascade, but that the energy of each spectrum 
component is independent of its wavelength and is likely amplified from the background turbulence 
by the transient growth mechanism.

\section{Turbulence-Darcy effect}  \label{s7}
Here we try to make {\it scaling estimate} of  magnitude of the rate of dissipation due to the {\it turbulence}-Darcy 
effect (called {\it D}-effect, which is analogous to the Joule effect). This can be done by estimating  the 
coefficient $\sg$ of the {\it fluid} Ohm's law $\bj_d=\sg \be$. It is found in this section rather unexpectedly 
that the bulk rate of {\it D}-effect dissipation takes a form resembling the  eddy-viscosity models and its 
coefficient is given as $\nu_{\mbox{\scriptsize D}}  \sim c_t\,d$ where  $c_t=1/\sqrt{\eps\mu}$.

\subsection{Scaling estimate of $\sg$ and rate of dissipation} \label{s71}
The {\it D}-effect was considered in  \S\ref{s1} ($e$),  \S\ref{s232} and \S\ref{s61} ($b$) as 
a new mechanism of energy dissipation. This {\it D}-effect is caused by the drift current $\bj_d$ existing 
in the traveling wave. The rate of dissipation due to the  effect is given by $Q_{\rm D} (=|S_d|) = \bj_d \cdot \be$ 
defined by (\ref{J1-loss}). 

\subsubsection{Two expressions of energy dissipation rate} \label{s711}
\vspace{-1.5mm}
To consider the bulk energy loss, let us choose a spatial volume $V$ chosen arbitrarily 
in the flow space. Substituting $\bj_d = \sg \be$ (suffix $a$ is dropped off) to $\bj$ of (\ref{TW-bS}) 
for the bulk energy loss, we obtain 
\begin{equation}
\overline{S}[\,\bj_d] = -\int_V \bj_d \cdot \be\ \dd^3\bx = -\int_V \sg \,|\be|^2 \ \dd^3\bx <0. \label{J0-loss}
\end{equation}
The right-most side of (\ref{TW-bS}) gives the same  dissipation rate with a  different form by the time 
derivative of energy density  $w_e= \half (  \mu |\bh|^2 +  \eps\, |\be|^2 )$,  and a second expression is given by
\begin{equation}
\overline{S}[\,\bj_d] = \frac{\dd}{\dd t} \int_V w \ \dd^3 \bx 
	= \half \mu \frac{\dd}{\dd t} \int_V |\bh|^2 \,\dd^3\bx 
        	+ \half \eps \frac{\dd}{\dd t} \int_V | \be |^2 \,\dd^3\bx,	 \label{J2-loss}
\end{equation}
where $\bh \,(=\mu^{-1} \bfb)$ and $\be \,(=\eps^{-1}\bd) $ are given by the freely  traveling and decaying waves 
studied in \S\ref{s5}, in particular by $\Psi$ of (\ref{t-Trav-wave}) for $e_x$ and a similar
expression for  $h_x$. Since other $y$ and $z$ components  have the same factor of temporal decay (assuming 
$|\pl_x \bj_c| \ll |\pl_y e_x|,\, |\pl_z e_x|,\, |\pl_y h_x|,\,|\pl_zh_x|$; see Appendix \ref{dd}),  the decay 
time scale  of $|\be|$ (or $|\bh|$) is estimated as $\tau_d$. Thus, it is found that 
\[  |\bh|^2,\ |\be|^2 \propto  e^{- (2/\tau_d)\,t}\, |(\mbox{function of}\ (x, r, \theta)|^2.  \]
Therefore we have 
\begin{equation}
\overline{S}[\,\bj_d] = - \frac{1}{\tau_d} \int_V \mu\, |\bh|^2 \,\dd^3\bx
		- \frac{1}{\tau_d} \int_V  \eps \,|\be |^2 \,\dd^3\bx, 		\label{J3-loss}
\end{equation}
\subsubsection{{\it D}-effect viscosity $\nu_{\mbox{\scriptsize D}}$} \label{s712}
\vspace{-1.5mm}
The first expression (\ref{J0-loss}) gives the dissipation rate per unit volume due to the  {\it D}-effect  by 
$Q_{\rm D} = \bj_d \cdot \be$.  In the waves where density variation is negligible, we have $\be \approx - \pl_t\ba$
(\S\ref{s531}). Hence we obtain a scaling estimate as $|\be| \sim |\ba|/\tau_d$.
 With respect to the drift current $\bj_d= \rho \bv_d =\sg \be$, we have its scaling estimate as 
\begin{equation}
	|\bj_d| = \rho\,|\bv_d|  = \sg \,|\be| \sim \sg \,|\ba|_d/\tau_d. 	\label{jd-sge}
\end{equation}
In view of  $\be = -\pl_t\ba -\nabla \phi$, the equation (\ref{U-u-a}) of \S\ref{s31} implies that the field 
variable $\ba$ should have the same physical dimension as the (fluctuating) velocity $\bu$. In addition, the 
consideration of \S\ref{s32} and \ref{s33} suggests that the $\bu$-field and $\ba$-field are convertible to 
each other (under certain restricted conditions).  Thus, it is proposed that the drift velocity $|\bv_d|$ and 
the magnitude $|\ba|$ are of the same order in (\ref{jd-sge}): $|\bv_d| \sim |\ba|_d$. This is because the term 
$|\pl_t \ba|$ was replaced by $|\ba|_d/\tau_d$. Hence the term $|\ba|_d$ in (\ref{jd-sge}) is the part related 
to damping,  therefore might be comparable to $|\bv_d|$.  Thus we have  the following scaling estimate:
\begin{equation}
  \sg \sim \rho\,\tau_d		 	\label{sg-rt}
\end{equation}
from (\ref{jd-sge}).  Using this  estimate,  the {\it D}-effect loss  $Q_{\rm D} = \bj_d \cdot \be$ 
can be written in a form analogous to the viscous  dissipation. In fact, substituting 
$\be= \bj_d/\sg$ and $\bj_d=\rho \bv_d$, we have
\[	Q_{\rm D} = \bj_d \cdot \be = \sg^{-1} |\bj_d|^2 = \sg^{-1} \rho^2  v_d^{\,2}
	\ \sim \ \frac{\rho}{\tau_d} \,v_d^{\,2}.	\]
where $v_d=|\bv_d|$, and (\ref{sg-rt}) is used to obtain the last expression. Substituting $\tau_d = 
d/c_t$, the last expression (divided by $\rho$) reduces to 
\begin{equation}
Q_{\mbox{\scriptsize D}}/\rho \sim\ \nu_{\mbox{\scriptsize D}}\,\left(\frac{v_d}{d}\right)^2, 
	\hskip10mm   \nu_{\mbox{\scriptsize D}} =c_t\,d, \quad   d = c_t\tau_d.    \label{Qj-turb}
\end{equation} 
The coefficient $\nu_{\mbox{\scriptsize D}}= c_t\,\!d$ is analogous to the eddy-viscosity, usually 
composed of product of (velocity scale) and (length scale). Here, the velocity is the speed $c_t$ of 
transverse wave in turbulence and the length is the damping distance $d$. The coefficient $c_t\,\!d$ 
may be called {\it D}-effect viscosity.  Note that the molecular kinematic viscosity $\nu_m$ is expressed as 
$\nu_m \sim l_m c_s$, where $c_s$ is the sound speed and $l_m$ the mean free path of molecular motion.

In the air at room temperature under  1 atm, we have $c_s \sim 3\times 10^4$cm and $l_m \sim 7\times10^{-6}$cm.
Hence $\nu_m \sim 10^{-1}$cm$^2$/s.  On the other hand, from the experimental study of pipe turbulence 
at $Re =2RU/\nu_m \approx 10^5$ (Kim \& Adrian, 1999), we have an estimate of $c_t \sim 50$cm/s, and 
$d \sim 10R \sim  60$cm. Hence $\nu_{\mbox{\scriptsize D}} \sim 10^3$cm$^2$/s. Thus, we have 
\[	\nu_m \sim 10^{-1}\,\mbox{cm}^2/s, \qquad \quad \nu_{\mbox{\scriptsize D}} \sim 10^3\,\mbox{cm}^2/s. 	\]
It is found that the {\it D}-effect viscosity $\nu_{\mbox{\scriptsize D}}$ of pipe turbulence at $Re \approx 10^5$  
is much larger  than $\nu_m$ of the air (under 1 atm at room temperature) by some orders of magnitude.

\subsection{New aspects of dissipation and {\it D}-effect viscosity}  \label{s72}
Let us consider the second expression (\ref{J3-loss}) and examine the two terms on the \underline{\it rhs} 
separately. Consider the first term by noting 
$\bh= \mu^{-1}\bfb =\mu^{-1} \nabla\times \ba$. We write it as
\begin{equation} 
	\overline{S}_1 = - \eta_{\,b}\, \int_V \, |\nabla\times \ba|^2 \,\dd^3\bx.	\label{Sv-1}
\end{equation}
where $	\eta_{\,b} = 1/(\tau_d \,\mu)$.  Using (\ref{tau-d}) and (\ref{sg-rt}), \ie
definitions of $\tau_d$ and $\sg$,   its coefficient $\eta_{\,\!b}$ is 
\begin{equation}
 \eta_{\,b} = 1/(\tau_d \,\mu) = \half \sg\,c_t^{\,2} \sim  \rho\, \nu_{\mbox{\scriptsize D}},
 \hskip10mm  \nu_{\mbox{\scriptsize D}} = c_t\,d .  			\label{C1}
\end{equation}
since $\tau_d c_t =d = 2/(\mu\sg c_t)$. Thus again, the coefficient $\eta_{\,b}$ divided by $\rho$ 
is found to be the order of {\it D}-effect viscosity $\nu_{\mbox{\scriptsize D}} = c_t\,d$, and the magnitude 
of $S_{b,1}$ is expressed as
\begin{equation} 
	| \,\overline{S}_1 |\, \sim \nu_{\mbox{\scriptsize D}} \left(\frac{u}{d}\right)^2\,V, 	\label{Sv-1-s}
\end{equation}
where  $u$ is a representative scale of velocity fluctuation (assuming $|\ba| \sim u$),  and $V$ is the integration 
volume. Needless to say, the rate of viscous energy dissipation of an incompressible flow including turbulence is
expressed in an analogous form to (\ref{Sv-1}), like  $\rho \nu_m\int_V |\bom_v|^2\,\dd^3\bx$ where 
$\eta_{\,b}$ is replaced by $\rho \nu_m$ and $\bom_v=\nabla\times\bv$ is the vorticity. 
(See, \eg  Kambe (2007 \S4.3), Eq.(4.24); for turbulence, see \eg Frisch (1995 \S2.3), Eq.(2.23).)
 
Next, we consider the second term:
\begin{equation} 
	\overline{S}_2 = -  \frac{\eps}{\tau_d} \int_V  \,|\be |^2 \,\dd^3\bx,   \hskip10mm
        \be= -\pl_t\ba -\nabla \phi.	\label{Sv-2}
\end{equation}
This  $\overline{S}_2$ has a notable new property because the $\be$-field includes the time derivative term
 $-\pl_t \ba$ besides the potential term $-\nabla\phi$, while the previous  $\overline{S}_1$ depends on 
spatial derivatives only, common to the conventional viscous terms. 

However, it can be shown that magnitude of $\overline{S}_2$ is expressed also as $\nu_{\mbox{\scriptsize D}}\,
( u/d )^2 V$, so that it is comparable to $|\overline{S}_1|$, which is obtained as follows.
From (\ref{tau-d}), we have 
\[	(\eps/\tau_d) \sim \eps\, \mu \sg c_t^{\,2} \sim \sg \sim \rho \tau_d , 
	\hskip10mm  |\be |^2 \sim  (u/\tau_d)^2 \sim c_t^{\,2} (u/d)^2,		\]
where $\eps \mu = c_t^{\,-2}$, and Eq.(\ref{sg-rt}) is used for the first part, and $\tau_d =d/ c_t$
for the second part. In this way, we obtain 
\[	| \,\overline{S}_2 |\, \sim \nu_{\mbox{\scriptsize D}} \left(\frac{u}{d}\right)^2\,V. 	 \]
Thus, it  is found that the present formulation includes an essentially new effect, represented by 
$\overline{S}_2$. However, the first  $\overline{S}_1$  is also new in the sense that the magnitude 
of its coefficient $\nu_{\mbox{\scriptsize D}}$ is much larger than $\nu_m$, some orders of magnitude 
larger than the conventional viscosity coefficient $\nu_m$.

\section{Summary}  \label{s8}

 A new scenario of turbulence theory is proposed by introducing a new transverse wave field  to the
turbulence field. Any self-contradiction is not incurred by this formulation. As far as we have a law of 
{\it current conservation}, mathematics allows  transverse wave fields, called the TW-field, governed by 
a system of fluid-Maxwell equations. Summary is given here with two parts. First concerns the formulation of 
the present theory composed of fluid flow field  and  transverse wave field; second is its application to streaky 
turbulence characterized with large scales LSM and VLSM  which await theoretical interpretation.

\begin{list}{}{\leftmargin=3mm}
\item[]  \hspace*{-5mm}  (I) {\it Formulation of the present theory}:  \\[-3mm]
\item[$\bullet$] \  Whole field is composed of {\it fluid flow} field (FF-field) and {\it transverse wave} 
field (TW-field).   The TW-waves are excited by extracting energy and momentum from the FF-field,  
As a reaction,  the TW-field acts on the FF-field with a fluid-Lorentz-force $\bff_{\mbox{\tiny L}}$. 
Energy is exchanged between the two fields by the interaction term $\bj \cdot \be$.
\item[$\bullet$] \ This formulation 
is equipped with  a new mechanism of energy dissipation by a drift current $\bj_d$, called  a {\it D}-effect. The 
drift current arises as a response of the FF-field, acted by  the $\be$-field . The new dissipation 
is estimated by the {\it D}-effect viscosity $\nu_{\mbox{\scriptsize D}}$.  The {\it D}-effect is an abbreviation 
of {\it turbulence}-Darcy effect.\footnote{The well-known 
Darcy's law is a law to describe the current flux of a viscous fluid through a porous medium under an imposed 
pressure gradient.  In the present case, the fluid  in turbulent state is acted on by an additional fluid Lorentz 
force from the TW-wave field, The turbulent flow coexists with the TW-field, and this force  gives rise to an 
internal drift current $\bj_d$ through turbulent eddies, since any turbulent state is composed of a number  of eddies,
The turbulence-Darcy law is proposed to describe a relationship  $\bj_d=\sg \be$ between the drift-current flux 
$\bj_d$ and an acting TW-force $\rho\be$,  This resembles the Ohm's law in electromagnetism.}
\item[$\bullet$] \ 
Total velocity  $\bv$ of the turbulent channel flow is proposed to be expressed  by a triple decomposition : 
($i$) a time-mean velocity $\bU_m$, ($ii$) a  wavy  perturbation $\bu_w=(u_j)$,  and 
($iii$) incoherent turbulent velocity $\bu'$.  Thus, the total velocity is $\bv= \bU_m + \bu_w + \bu' = \bU_m +\bw$, 
where  $\bw= \bu_w + \bu'$ is  the {\it time-dependent} part. Together with the  time-dependent part $\bw$, 
a vector potential $\ba$ of the TW-field is excited simultaneously.  
 
From the experimental studies of  turbulent shear flows (Hussain \& Reynolds, 1970, 72, 75),  their periodic  wave was  
{\it robust} in the background irregularly fluctuating  flow field which acts on the wave with enhanced diffusivity
of fluctuating  Reynolds stress.  The present approach supports this observation.  
It is one of its essential features that a {\it TW}-wave has its own energy and momentum  (\S\ref{s2}) like the 
electromagnetic waves and it is governed by conservation equations of energy and momentum  with interaction terms with 
the flow field.  This {\it TW}-wave acts on the flow field by the fluid-Lorentz-force $\bff_{\mbox{\tiny L}}$. 
Thus, the wave keeps  unchanged unless those are changed  by the interactions.  This  explains the 
robustness of the periodic wave component  in the turbulent channel flow.
 
\item[$\bullet$] \ The present scenario can predict traveling waves (TW-field) in wall-bounded turbulence. 
Its dynamics   is studied by the equations  of growth and decay. One of them is 
\begin{equation} 
\nabla^2 \be - c_t^{\,-2} \pl_t^{\,2}\be  = \mu \rho \, \pl_t \bu + \mu \sg \pl_t \be.   	 	\label{TW-s8}
\end{equation}  
(see (\ref{TW-e2}), \S\ref{s5} and \S\ref{s6}).  The left hand side (\underline{\it lhs}) expresses propagation of 
a transverse traveling wave of the vector field $\be$ with a phase speed $c_t$, while the first term  on 
\underline{\it rhs} can act as a  wave source, and the second term acts as wave damping characterized with
a damping distance $d = \kappa /\mu \sg c_t$, where $\kappa= \lam_{\rm lsm}/\lam$ with $\lam$ its  wavelength.
\item[$\bullet$] \  Source term $S_w$ of the TW-energy  in the energy equation (\ref{TW-e-s23}) is given by 
\begin{equation} 
S_w = - \bj \cdot \be = - \rho\, \bU_m \cdot \be - \rho\, \bw \cdot \be - \bj_d \cdot \be
	\hskip8mm  \bj= \rho\, \bU + \rho\, \bw + \bj_d , 	 	\label{Sw-s8}
\end{equation}
where $\bj_d= \sg \be$, and  $\be=-\pl_t\ba -\nabla \phi_a$.  For infinitesimal perturbations of $\bw$ and $\ba$, 
the source $S_w$ is dominated by the first term, \ $ - \rho\, \bU_m \cdot \be \approx  \rho\, \bU_m \cdot \pl_t\ba$ 
(assuming  $\phi_a=$\,const),  stating that $S_w>0$ \ if\, $\bU_m \cdot \pl_t\ba >0$, \ie if $\ba$ is excited to the 
direction of $\bU_m$. In nonlinear state of periodic waves of $\bw$ and $\ba$, however, the first term gives net 
negligible contribution (after taking time average), but the second term $\rho \bw \cdot \be$ gives a net contribution 
on the time average (see \S\ref{s44}). Existence of phase difference between $\ba$ and $\bu_w$ (wave part of $\bw$) 
enables energy supply from the flow field to the TW-field.  The last term is the {\it D}-effect dissipation:  
$- \bj_d \cdot \be= - \sg |\be|^2 (<0)$, which is responsible for enhanced rate of energy dissipation.
\end{list}  			

\vspace{-4mm}
\begin{list}{}{\leftmargin=3mm}
\item[]  \hspace*{-5mm}  (II) {\it Present scenario is applied to streaky turbulence} \\[-4mm]
\item[$\circ$] \  One of the  application areas of the present formulation is the streaky shear-flow 
turbulence.  The streak structure in the wall turbulence is  a {\it dissipative structure}, as described below.
\item[$\circ$] \  The streaks in actual turbulence are wavy and non-uniform, and surrounded by a sea of incoherent 
turbulent motions.  In shear flow turbulence,    two large scales are recognized to exist 
in the streaky  structures:  LSM (large-scale motions) and VLSM (very-large-scale motions), characterizing 
the streamwise streaks and long meandering structures.  The vortex packets consisting of hairpin-like 
structures in the wall shear layer of channel turbulence are considered to characterize the scale  LSM:
$\lam_{\rm lsm} =  h \sim 3 h$, where  $h$ is the channel  half-width. It is likely that the waves are 
generated by the transient growth mechanism, and also those waves are captured and maintained by the TW-field.
\item[$\circ$] \  When we consider streaky channel turbulence, the scenario of transient growth mechanism  
(Gustavsson 1991;   Butler \& Farrell 1992;   Henningson, Lundbladh \& Johansson 1993) is helpful for understanding 
the phenomena. Originally, this was studied for laminar channel flows, but later it was investigated for 
turbulent shear flows too  (Del \'{A}lamo \& Jim\'{e}nez 2006); Schoppa and Hussain 2002).  It is proposed 
that  the wavy meandering streak with streamwise vortices would be the wave of the larger scale VLSM.  This could be 
associated with the larger-scale  waves of streamwise wavelength, $\lam_{\rm vlsm} \approx 10h \sim 60h$, found 
by Del \'{A}lamo \& Jim\'{e}nez (2006) which were amplified transiently within turbulent flows.  
\item[$\circ$] \ The waves in the range of the $k_x^{-1}$  spectrum of (\ref{TW-range})  are considered  to be 
the larger-scale  waves $\lam_{\rm vlsm} \approx 10h \sim 60h$  amplified transiently  which were found by 
Del \'{A}lamo \& Jim\'{e}nez. The scaling estimate of the $k_x^{-1}$  spectrum in \S\ref{s64} predicts that the 
magnitude of the waves is independent of the wavelengths, which is consistent with the computed result of Del \'{A}lamo 
\& Jim\'{e}nez (2006). Those large-scale waves  are captured by the TW-field by the equation (\ref{TW-eh-623}), which 
acts backward to the flow field by (\ref{Eq-w-623}).  Finally  dynamical interaction of streamwise vortices  
sustains streaky turbulence (Schoppa \& Hussain, 2002).
\item[$\circ$] \  
Perhaps, most remarkable outcome of the present scenario is the dissipation caused by the drift current $\bj_d$. 
The rate of energy dissipation takes a form resembling the models of eddy-viscosity, and its coefficient 
$\nu_{\mbox{\scriptsize D}}$ is estimated to be of the order of $c_t d$ from scaling estimate.  Its magnitude 
is comparable  with the models of eddy viscosity. In fact, the {\it D}-effect viscosity $\nu_{\mbox{\scriptsize D}}$
 in pipe turbulence at $Re  \approx 10^5$ is estimated as \ $\nu_{\mbox{\scriptsize D}} \sim 10^3\,\mbox{cm}^2/s$ 
(\S\ref{s71}), which is much larger than the molecular viscosity of the air under 1 atm at room temperature 
estimated as $\nu_m \sim 10^{-1}\,\mbox{cm}^2/s$.   The  {\it D}-effect  is derived analytically from the 
basic governing  equations. To the author's knowledge, no other theory is able to derive a law of energy dissipation 
 comparable with empirical models of eddy viscosity from fundamental governing equations.

\end{list}  			

\vskip1mm \noindent
{\it Acknowledgements}: 

The author  benefitted greatly from discussions with Andrew Gilbert.  The author appreciates it very
much for the valuable comments from  reviewers, which have encouraged to improve this manuscript significantly.

\vspace{1mm}
\centerline{\underline{\hspace*{70mm}}}	          
\vskip3mm    
\appendix
\centerline{APPENDIX}

\vspace*{-7mm}
\section{Reformulation of Maxwell equations by exterior algebra} \label{aa}
\vspace{-4mm}
Exterior algebra and  differential forms are now recognized as a powerful tool in mathematical physics.
In fact, reformulation of the system of Maxwell equations in terms of the exterior algebra has been 
studied by  mathematical physicists for past several decades of years.
The reformulation presented here is based on two fundamental assumptions: ($a$) There exists 
a field of 4-vector potential in the 4-dimensional space-time; ($b$) There exists a matter field 
satisfying current conservation law.  Resulting formulation states clearly that  transverse wave 
fields are excited as a result of dynamical evolutions of the current flux, mass density and 
vorticity field, governed by a system of dynamical equations,

Unified  presentation of  reformulation of the Maxwell equations of electromagnetism was given by 
Hehl \& Obukhov (2003) in their book, {\it Foundations of Classical Electrodynamics},  with collecting original 
works by a number of mathematical physicists. Those works were accomplished mainly in the second half 
of the 20th century although pioneering works on fundamental mathematical ingredients (such as 
{\it Poincar\'{e}} Lemma, {\it de Rahm} Theorem, {\it Hodge} operation, \etc) had been prepared in the 
first half of the century or earlier. The {\it Introduction} section of the book summarizes the history  
compactly together with a list of references. Another epoch, \ie\, its application to fluid dynamics, 
was marked very recently by Scofield \& Huq (2010, 2014).

This Appendix A is prepared  by the style of the author's own. The system of four Maxwell equations 
is divided into  two pairs: a force-free pair and a pair with external forcing by conserved current.
Derivation of each pair is presented in each of the following two 
subsections on the basis of the above premises.

\vspace*{-3mm}
\subsection{Field strength 2-form and a force-free pair of  equations} \label{a1}

Let us define a differential  one-form $\A^1$ by
\[  \A^1 = a_\mu \dd x^\mu =  \phi_a\, \dd t - a_1 \dd x^1 - a_2 \dd x^2 - a_3 \dd x^3.	\]
Taking external differential of $\A^1$, we obtain a two-form $\F^2 = \dd \A^1$:
\begin{eqnarray}   
&&  \hspace*{-15mm}  \ \F^2  = {\small - (e_1 \dd x^1 + e_2 \dd x^2 + e_3 \dd x^3)  \wedge \dd t  
   - (b_1\, \dd x^2\wedge \dd x^3 + b_2\, \dd x^3\wedge \dd x^1 + b_3 \,\dd x^1\wedge \dd x^2 )}     \label{A-F2-eb} \\
&&  \hspace*{-8mm} = \quad  - E^1 \wedge \dd t \hskip33mm - \hskip5mm \Bb^2,		\label{F2-xeb} \\[-2mm]
&&  \hspace*{-17mm} \mbox{where} 						\nonumber	\\[-1mm]
&&  \hspace*{-3mm} \be= (e_i) = - \pl_t \ba - \nabla \phi_a, \hskip20mm \bfb= (b_i)= \nabla\times \ba,  
 				\hskip10mm (i=1,2,3),				\label{A-eb}    \\
&&  \hspace*{-11mm} E^1= e_1 \dd x^1 + e_2 \dd x^2 + e_3 \dd x^3, \hskip4mm  \Bb^2 = b_1\, 
    \dd x^2\wedge \dd x^3 + b_2\, \dd x^3\wedge \dd x^1 + b_3 \,\dd x^1\wedge \dd x^2, 	\label{A-EB} 
\end{eqnarray}
with $\ba= (a_1, a_2, a_3)$. Equivalently, using $ F_{\mu\nu} = \pl_\mu a_\nu - \pl_\nu a_\mu$, 
$\F^2$ can be rewritten as   {\small 
\begin{equation}
  \F^2 = \sum_{\mu<\nu} F_{\mu\nu} \,\dd x^\mu \wedge \dd x^\nu,  \hskip5mm		
  (\, F_{\mu\nu}\,) = \left( \begin{array}{cccc}  
	0   &  e_1  &  e_2 &  e_3	\\	 -e_1 & 0   & -b_3  & b_2  \\	
	-e_2 &  b_3 & 0    & \,-b_1 	\\	 -e_3 & \,-b_2 & b_1 &  0  	
	\end{array}  \right), 							     \label{F_munu}  
\end{equation}   }
Once again, taking external differential  of $\F^2 = \dd \A^1$, we obtain the following identity,
\begin{equation}
	\dd \F^2 = \dd^2 \A^1 \equiv 0.  				\label{dF2-d2A}
\end{equation}
The equation $ \dd \F^2 =0$ can be written explicitly as 
{\small
\begin{eqnarray}
&&  \hspace*{-8mm}    0 = \dd  \F^2  = 
        (\pl_1 F_{23} + \pl_2 F_{31} + \pl_3 F_{12} )\,\dd x^1 \wedge \dd x^2 \wedge \dd x^3 	
  + (\pl_t F_{23} + \pl_2 F_{30} + \pl_3 F_{02} )\,\dd t \wedge \dd x^2 \wedge \dd x^3 \nonumber \\
&& \hspace*{0mm} + \,(\pl_t F_{31}+\pl_3 F_{10} + \pl_1 F_{03} )\,\dd t \wedge \dd x^3 \wedge \dd x^1 
  + (\pl_t F_{12} + \pl_1 F_{20} + \pl_2 F_{01} )\,\dd t \wedge \dd x^1 \wedge \dd x^2 . \label{dF2} 
\end{eqnarray} }
This results in one scalar equation (from vanishing of the coefficient of  $\dd x^1 \wedge \dd x^2 \wedge \dd x^3$) 
and one vectorial equation for two 3-vectors $\be$ and $\bfb$ (from the remaining three terms):
\begin{eqnarray}
 \nabla \cdot \bfb & = & 0,   				\label{A-db}  \\  
\pl_t \bfb + \nabla \times \be & = & 0 .		\label{A-dtb-de}
\end{eqnarray}
Thus, a force-free pair of Maxwell equations has been derived by the exterior  calculus.

\subsection{Excitation field  and another pair of Maxwell equations} \label{a2}

Another pair of  Maxwell equations can be derived as follows.  Suppose  we have a current 
4-vector $j^\mu$. Current conservation is expressed by
\begin{equation}
\pl_\mu j^{\mu} = \pl_t \rho + \pl_x j_x + \pl_y j_y + \pl_z j_z =0 , 		\label{A-C-cons1}
\end{equation}
where
\begin{eqnarray}
 j^\mu &  = & (\rho, j_x, j_y, j_z), \hskip15mm  \pl_{\mu} \equiv (\pl_t, \nabla).    	\label{current-j}   
\end{eqnarray}
From this conservation law, one can deduce  {\it excitation fields} (Hehl \& Obukhov, 2003). This subsection
aims to derive a set of  equations governing such excitation fields. 

Consider a 3-dimensional simply-connected space-region $V_3$ enclosed by 2-dimensional boundary surface 
$\pl V_3$.  An integration form corresponding to the above differential equation is expressed as 
\begin{equation}
\frac{\dd}{\dd t} \int_{V_3} \rho \ \dd x \,\dd y \,\dd z + \int_{\pl V_3} (j_x n_x +j_y n_y +j_z n_z) 
	\dd S = 0 ,	\label{A-intC0} 
\end{equation}
by the conventional vector analysis, where $(n_x, n_y, n_z)$ is a unit outward normal to the boundary 
surface $\pl V_3$.   According to the differential algebra, one can define a density 3-form $\rho_{(3)}$ and 
a current 2-form $j_{(2)}$ by
\[  \rho_{(3)} = \rho\, \dd x \wedge \dd y \wedge \dd z,  \hskip10mm 
	j_{(2)} = j_x \, \dd y \wedge \dd z + j_y \, \dd z \wedge \dd x + j_z \, \dd x \wedge \dd y. \]
Then, the above integration form (\ref{A-intC0}) can be represented as follows:
\begin{equation}	
\frac{\dd}{\dd t} \int_{V_3} \rho_{(3)} + \int_{\pl V_3} j_{(2)} = 0 ,	\label{A-intC} 
\end{equation}
where  three area 2-forms $\dd y \wedge \dd z, \,\dd z \wedge \dd x$ and  $\dd x \wedge \dd y$ are 
directed in such a way that the outflow is counted positively. Furthermore,  integrating (\ref{A-intC}) 
over a certain time interval $[t_1,t_2]$ ($t_1<t_2$), we obtain
\begin{equation}	
\int_{t_2, V_3} \rho_{(3)} - \int_{t_1, V_3} \rho_{(3)} \,+ \int_{[t_1,t_2] \times \pl V_3} 
		\dd t \wedge j_{(2)} =0.					\label{A-int2C}
\end{equation}
On the other hand, on a 4-dimensional manifold $x^\mu$,  one can define   a 4-volume form by 
\[	\V^{(4)} =  \dd t \wedge \dd x \wedge \dd y  \wedge \dd z.		\]
Furthermore, a current 3-form $J^{(3)}$ is defined by the interior product $- i_J \,\V^{(4)}$ where 
$J =  j^\mu = (\rho, j_x, j_y, j_z)$:
\begin{eqnarray}  \hspace*{-13mm} 
J^{(3)} = - i_J \,\V^{(4)} & = &  - \rho \,\dd x \wedge \dd y \wedge \dd z + ( j_x \, \dd y \wedge  \dd z 
	+ j_y \, \dd z \wedge \dd x + j_z \, \dd x \wedge \dd y)  \wedge \dd t,   \label{A-J3} \\
   & = &  - \rho_{(3)} + j_{(2)} \wedge	\dd t 		\nonumber 	
\end{eqnarray}
(see Frankel (1997, \S7.2) for the symbol $i_J$).\footnote{Alternatively, $J^{(3)} = i_J \,\V^{(4)}$ 
is the Hodge dual $^*J^{(1)}$ of $J^{(1)} = \rho \,\dd t + j_x \dd x + j_y \dd y +  j_z \dd z$.}   
Taking exterior differential of $J^{(3)}$, we obtain 
\begin{equation}
 \dd J^{(3)} = -(\pl_t \rho + \pl_x j_x + \pl_y j_y + \pl_z j_z) \,\dd t \wedge \dd x 
 		\wedge \dd y \wedge \dd z  = 0 ,					\label{A-dJ3}
\end{equation}
This vanishing is due to (\ref{A-C-cons1}).  This states that the 3-form $J^{(3)}$ is {\it closed}.

Let us consider a 4-dimensional simply connected region $\Om_4= [t_1,t_2] \times V_3$, enclosed by 
3-dimensional boundary  $\pl \Om_4$. Using the current 3-form $J^{(3)} = -(\rho_{(3)} + \dd t \wedge j_{(2)})$, 
the equation (\ref{A-int2C}) can be transformed to 
\begin{equation}	
\oint_{\pl \Om_4} J^{(3)} =0. 				\label{A-intJ3}
\end{equation}
[See Hehl \& Obukhov (2003, Part B) for the electrodynamics case].\footnote{For the validity of (\ref{A-intJ3}),
the condition of "simply connected region" is important.}   By the generalized Stokes theorem in the 
differential geometry, the expression (\ref{A-intJ3}) is transformed to $\int_{\Om_4} \dd J^{(3)} =0$ 
for an arbitrarily chosen $\Om_4$. This is  equivalent to (\ref{A-dJ3}).	

Having shown the properties (\ref{A-dJ3}) and (\ref{A-intJ3}) of the current 3-form $J^{(3)}$, we 
recognize  the equation (\ref{A-intJ3}) as the statement that the current $J^{(3)}$ is {\it exact} 
(de Rahm's Theorem):
\begin{equation}	
 J^{(3)} = \dd \Hm^2 ,					\label{A-J3dH}
\end{equation}
namely, there exists a certain 2-form $\Hm^2$ representing $ J^{(3)}$ by (\ref{A-J3dH}). This assures  
$\dd J^{(3)} (= \dd^2 \Hm^2) =0$ as an identity. Analogously to the electrodynamics (Hehl \& Obukhov, 2003), 
the 2-form $\Hm^2$ is termed here an {\it excitation}.

Reminding the derivation of the field strength 2-form $\F^2$ of (\ref{A-F2-eb}) in the previous section,
the excitation 2-form $\Hm^2$ may be defined in a like manner as  {\small 
\begin{eqnarray}  
\hspace*{-15mm} 
\Hm^2   &  = &   h_1 \dd x^1 + h_2 \dd x^2 + h_3 \dd x^3 \wedge \dd t  
         - ( d_1 \dd x^2\wedge \dd x^3 + d_2 \dd x^3\wedge \dd x^1 + d_3 \dd x^1\wedge \dd x^2  )   \label{A-H2-hd}  \\
\hspace*{-20mm}   &   = &  H^1 \wedge \dd t - \Dd^2,				\label{A-H2-xhd} \\
\hspace*{-10mm} 
&&  \hspace*{-10mm}  {\footnotesize  H^1 = h_1 \dd x^1 + h_2 \dd x^2 + h_3 \dd x^3, \hskip4mm  \Dd^2 = d_1\, 
    \dd x^2\wedge \dd x^3 + d_2\, \dd x^3\wedge \dd x^1 + d_3 \,\dd x^1\wedge \dd x^2, }	\label{A-hD} 
\end{eqnarray}  }
Writing it in a form of anti-symmetric matrix $(H_{\mu\nu})$, we have   {\small
\begin{eqnarray}
\Hm^2= \sum_{\mu<\nu} H_{\mu\nu}\, \dd x^\mu \wedge \dd x^\nu,  \hskip5mm
(\, H_{\mu\nu}\,) & = & \left( \begin{array}{cccc}  
	0 & -h_1 & -h_2 & -h_3  \\	 
         h_1 & 0 & -d_3 & d_2  \\	
	 h_2 &  \, d_3 & 0  & - d_1 	\\	
         h_3 &  \, -d_2 & d_1 & 0  	
	\end{array}  \right).					\label{G_munu}  
\end{eqnarray}  }
As carried out for $\F^2$ in \ref{a1}, we take external differential of $\Hm^{(2)}$, resulting in
$\dd \Hm^{(2)}= J^{(3)}$, where $\dd \Hm^{(2)}$ can be expressed like (\ref{dF2}) with $F_{\mu\nu}$ 
replaced by $H_{\mu\nu}$, and "$0=$" on its LHS  by "$J^{(3)}=$".  From this, we find one scalar equation and 
one vectorial equation for two 3-vectors $\bd$ and $\bh$:
\begin{eqnarray}
 \nabla \cdot \bd & = & \rho ,   			\label{A-dd}  \\
- \pl_t \bd + \nabla \times \bh & = & \bj , \hskip10mm \bj = (j_1, j_2, j_3).	\label{A-dtd-dh}
\end{eqnarray}
Thus, we have obtained a remaining pair of two equations.

\subsection{Hodge operator and  (positive-definite) scalar product}
We proceed a step further by taking  physical analogy with the electromagnetism. Introducing new physical fields  
$\F^2$ and $\Hm^2$ should be justified if one can construct scalar fields such as a Lagrangian functional which 
is invariant under coordinate transformation (or Lorentz transformation), or an energy which is positive definite.

Let us consider the exterior (wedge) product between $\F^2$ of (\ref{A-F2-eb}) and $\Hm^2$ of 
(\ref{A-H2-hd}):
\begin{eqnarray} 	\hspace*{-20mm}  
\F^2 \wedge \Hm^2  & = & [ - (e_1 d_1+ e_2 d_2 + e_3  d_3) + ( b_1 h_1+ b_2 h_2 
	+ b_3  h_3)]\ \dd t \wedge  \dd x^1 \wedge \dd x^2 \wedge \dd x^3,   	\nonumber \\	
\hspace*{-20mm}   &  = &  [ - (\be, \bd) + (\bfb, \bh ) ] \ \V^4.     			\label{A-F2H2} \\
\hspace*{-20mm}  && (\be,\bd)= e_1 d_1+ e_2 d_2 + e_3 d_3, \hskip6mm (\bfb,\bh)= b_1 h_1+ b_2 h_2 + b_3 h_3. \label{scalar}
\end{eqnarray}
This is  analogous to the Lagrangian 4-form of the electromagnetic field  [Hehl \& Obukhov (2003), \S B.2.3); 
Landau \& Lifshitz (1975) \S27 ]. Understanding of this remarkable property may be deepened by considering 
the Hodge star operation $*$.

Consider an $n$-dimensional vector space with a metric $g$. The Hodge star  $*$ is defined as 
a linear map, with associating  each $p$-form such as $\om^p, \eta^p$,  $\cdots$ with an $(n-p)$-form 
$*\om^p,*\eta^p$,  $\cdots$ respectively, such that
\[	\om \wedge *\eta = (\om,\,\eta)\, \V^n		\]
where  $(\om,\,\eta)$ is the $p$-dimensional scalar product (like (\ref{scalar}) of 3D) between two forms 
$\om^p$ and $\eta^p$, and $\V^n$ is an $n$-dimensional volume form induced by $g$.  Needless to say, the scalar 
product $( \cdot,\,\cdot)$ is {\it positive definite} and symmetric.  

Taking an example, let us consider the  Hodge star dual of $\F^2$ of (\ref{F2-xeb}), defined by 
$*\F^2= -*(E^1 \wedge \dd t) - *\Bb^2$ from (\ref{F2-xeb}). In view of $E^1$ and  $\Bb^2$ defined by (\ref{A-EB}), 
each term of the Hodge dual $*\F^2$ is given respectively by
\begin{eqnarray} 
*( E^1 \wedge \dd t) \ & \equiv & \ \EE^2  = e_1\,\dd x^2\wedge \dd x^3 + e_2\,\dd x^3\wedge \dd x^1 
		+ e_3 \,\dd x^1\wedge \dd x^2 .				\label{*E1}  \\
\ \hspace{5mm} *\Bb^2 \ & \equiv & \ \dd t \wedge B^1,  \hskip10mm B^1  = b_1 \dd x^1 + b_2 \dd x^2 + b_3 \dd x^3.
\label{*B2}
\end{eqnarray}
Thus, we obtain positive definite expressions:
\[ \hspace*{-5mm} (E^1 \wedge \dd t) \wedge \, *( E^1 \wedge \dd t) = (\be, \be)\,\V^4,  
	\hskip10mm \Bb^2 \wedge \, *\Bb^2 = (\bfb, \bfb)\,\V^4.					\]
In order to represent $\Hm^2$ of (\ref{A-H2-hd}), it is proposed  by analogy with electromagnetism that 
the excitation $\Hm^2$ is related to the field strength  $\F^2$ by a constitutive relation 
$\Hm^2= C_{(4)}\,^*\F^2$, according to {\it Theorem 1} of Scofield \& Huq (2010, 2014), where $C_{(4)}$ 
is a parameter matrix of fourth order in general. Here, assuming that the matter under consideration is 
homogeneous and isotropic in the space, we express $C_{(4)}$ by using two parameters $\mu$ and $\eps$ 
as follows:
\begin{eqnarray} 
\Hm^2  =  H^1 \wedge \dd t - \Dd^2 = C_{(4)}\, ^*\F^2  & = &  \mu^{-1} B^1 \wedge \dd t  - \eps\,\EE^2,	\label{*F-xBE}	\\
&&	\hspace*{-40mm}  \bh =  \mu^{-1} \bfb,  \hspace{20mm} \bd = \eps \be,			\label{bd-bh-a3}
\end{eqnarray}
where the expressions on the second line are obtained by comparing the second and fourth expressions on the 
first line, where $\eps$ is analogous 
to the electric permittivity  and $\mu$ to the magnetic  permeability of the electromagnetism, respectively.

\subsection{Transvers waves supported by vorticity field}  	\label{a4}
It is understood that the formulation of \S\ref{s22} is made for the total current $\bj=\bj_c+\bj_d$
of (\ref{TW-S-34}), where $\bj_c = \rho \bv$ and $\bv=\bU+\bu$. Since the equations (\ref{TW-a}) 
and (\ref{TW-bf}) are linear with respect to the TW-field, the flux of convection  current $\bj_c=\rho \bv$  
gives rise to a  contribution to the field, separately from that due to $\bj_d$.  Here, we examine  the effect 
of $\bj_c$ only, and field variables are denoted with a suffix $c$.  It is remarkable to find that the 
vorticity field $\bom(\bx,t)$ supports {\it transverse waves} under the constraint of the continuity equation. 
The effect of $\bj_d$ is considered in the main text. 

For the fields associated with $\bj_c=\rho \bv$, the equations (\ref{TW-a}) and (\ref{TW-bf}) are given by 
\begin{equation}
	\pl_t \bfb_c + \curl \,\be_c =0, \hskip15mm \divg \bfb_c=0, 		\label{TWc-a}
\end{equation}
\begin{equation}
- \pl_t \bd_c + \curl \,\bh_c = \bj_c, \hskip15mm  \divg \bd_c = \rho_c. 	\label{TWc-b}
\end{equation}
respectively.  The two equations on the second line  state just the current conservation: 
$\pl_t \rho_c+ \divg \bj_c=0$, and the vector fields $\bd_c$ and $\bh_c$ are regarded as a pair of 
vector potentials for the conserved 4-current  $J_c =(\rho_c, \bj_c)$. 

Regarding the first line, one can introduce a 4-vector potential $(A^\mu) =(\phi_c, v_i)$ where 
$\bv=(v_i)$ is a velocity vector, and represent the vectors $\be_c$ and $\bfb_c$ in the two equations 
of (\ref{TWc-a}) as
\begin{equation}
\be_c=  - \pl_t \bv - \nabla \phi_c, \hskip10mm \bfb_c =  \nabla\times \bv = \bom,  	\label{TWc-eb}  
\end{equation}
By these defining equations, the  set of equations of (\ref{TWc-a}) is satisfied identically. 

Here we consider a flow field of an inviscid fluid of uniform entropy, in which $\tau^{(vis)}=0$ and
$\nabla s=0$. 
If we define $\phi_c$ by $\phi_c= h +\half |\bv|^2$, from the equation of motion of (\ref{EqM-S21}), 
we find
\begin{equation}
  \be_c= \bom \times \bv . 		\label{bec}  
\end{equation}
 Using (\ref{bec}), the first of (\ref{TWc-a}) reduces to
\[   	\pl_t \bom =  -\curl( \bom \times \bv ).  \]
This is nothing but the vorticity equation of (\ref{Vort-S21}) under the assumed conditions.  Kambe (2010) investigated 
this aspect for fluids both with and without viscosity, clarifying that the role of vector potential of $\be_c$ 
and $\bfb_c$ is played by the velocity vector $\bv$, and that the fluid electric field $\be_c$ is given by (\ref{bec}).

Thus it is found that the equations (\ref{TWc-a}) and (\ref{TWc-b}) are consistent with the vorticity 
equation (\ref{Vort-S21}) and  the continuity equation (\ref{FF-C}) of FF-field under the equation of 
motion  (\ref{EqM-S21}) for an inviscid fluid of uniform entropy in the absence of TW field.
One of the merits to introduce the Maxwell-type equations (\ref{TWc-a}) and (\ref{TWc-b}) is as follows.

One can introduce a system of Maxwell equations (\ref{TWc-a}) and (\ref{TWc-b}) for  turbulence. We 
assume that it is characterized by {\it field-material} parameters $\eps$ and $\mu$ such 
that $\bd_c=\eps \be_c$ and $\bh_c = \mu^{-1} \bfb_c$ (see Appendix A.3 and A.4).  Then the system 
of Maxwell equations supports transverse waves of phase velocity $c_t=1/\sqrt{\eps\mu}$, where 
$\bh_c = \mu^{-1} \bom$ and $\bd_c = \eps\,\bom \times \bv$. Thus, it is seen that the vorticity field 
$\bom(\bx,t)$ supports transverse waves  described by (\ref{TWc-a}) and  (\ref{TWc-b}).

\section{Equations of energy, momentum, entropy and transverse waves}  \label{bb}

\subsection{Equations of  FF-field}    			\label{b1} 
Basic equations of current theory of fluid mechanics are reviewed, particularly in  ($a$)  the definition of the 
viscous stress $\tau^{(vis)}_{ij}$  and in ($b$) how the energy equation is derived. The latter is 
a preparation for the derivation of energy equation of the combined system in \S\ref{b3} ($b$). \\
\noindent ($a$) {\it Momentum equation}:\, Momentum equation of the FF-field is given by 
\begin{equation}
\pl_t (\rho \bv)_i + \pl_j \Pi_{ij} = 0.   \hskip8mm 
\Pi_{ij} \equiv \rho v_i v_j + p\,\del_{ij} - \tau^{(vis)}_{ij}		 	\label{B-FFmom}
\end{equation} 
where $\bv$ is the fluid velocity, and $\tau^{(vis)}_{ij}$ is the viscous stress tensor defined by
\begin{equation}
\tau^{(vis)}_{ij} = \eta \,(\pl_i v_j + \pl_j v_i - \twothird \delta_{ij}\, \pl_k v_k) 
		    + \zeta\, \delta_{ij}\, \pl_k v_k, 				\label{vis-str}
\end{equation}
where $\eta$ and $\zeta$ are coefficients of viscosity (assumed constant). From this, we obtain 
\begin{equation}
\nabla \cdot \tau^{(vis)} = \eta \nabla^2 \bv + (\zeta + \third \eta)\,\nabla (\divg \bv), \qquad 
\mbox{where} \quad   (\nabla \cdot \tau^{(vis)})_i = \pl_j \tau^{(vis)}_{ij}.  			\label{vis-for}
\end{equation}
Using the mass conservation equation, 
\begin{equation}    
\pl_t\rho  + \divg \rho\bv = 0,					\label{FF-C-b1} 
\end{equation}
 and a thermodynamic equation $(1/\rho) \dd p= \dd h -T \dd s$ ($s$: specific entropy), the momentum equation 
 (\ref{B-FFmom}) is transformed to an equation of motion of a viscous  fluid: 
\begin{equation}  
\rho \,\pl_t \bv = - \half \rho\nabla( |\bv|^2) +\rho \bv\times \bom -\nabla\,p + \nabla\cdot \tau^{(vis)}, \label{EqM-b1}  
\end{equation}

\noindent ($b$) {\it Energy equation}:\,  \label{b12} 
The energy equation of  FF-field (only) is given by Landau \& Lifshitz (1987) [\S49]. Taking scalar product of  $\bv$ 
with  (\ref{EqM-b1})  on both sides, we obtain
\begin{equation}
 \pl_t \big[ \rho(\frac{1}{2} v^2 + \ine) \big] + \divg(\bq_{\rm f} ) = \rho T (\D s/\D t) 
	- Q_{vis}  - Q_T =0,		 	\label{FF-E-b1}
\end{equation}
where $\ine$ is the internal energy.  Owing to the two relations  
\[  \dd p= \rho \dd h - \rho T \dd s, \qquad  \rho \pl_t \ine = \rho T \pl_t s  + (p/\rho) \pl_t \rho 
			= \rho T \pl_t s - (p/\rho) \divg(\rho \bv), 		\]
the {\it FF}-energy-flux $\bq_{\rm f}$ is defined  by the following, 
\begin{equation}
\bq_{\rm f} = \rho \bv (\frac{1}{2} v^2+ h)  - \bv\cdot \tau^{(vis)} - k_T\,\nabla T,	\label{q-FF-b1}
\end{equation}
and the right hand side of (\ref{FF-E-b1}) vanishes owing to the entropy equation:
\begin{equation}
\rho T \frac{\D}{\D t} s = Q_{\rm vis} + Q_T ,	\qquad Q_{\rm vis} \equiv \mbox{$\sum_{i,j}$} \,
	\pl_jv_i\ \tau^{(vis)}_{ij}, \quad Q_T \equiv \divg (k_T \nabla T),  			\label{FF-entropy-b1}  
\end{equation}
where $h = \ine + p/\rho$ is the enthalpy, and $k_T$  the thermal diffusivity. 

\subsection{Equations of transverse wave}   \label{b2} 
From the mathematical analyses in the Appendix \ref{aa}, we have now a set of Maxwell equations 
for a turbulent flow characterized by the {\it field} parameters $\eps$ and $\mu$. The source of 
the transverse wave field (TW field) is a conserved 4-current $J =(\rho, \bj)$, satisfying  
$\pl_t \rho + \divg \bj=0$,  From the equations (\ref{A-db}), (\ref{A-dtb-de}), (\ref{A-dd}) and 
(\ref{A-dtd-dh}), we have the following set of TW equations:
\begin{equation}
	\pl_t \bfb + \curl \,\be =0, \hskip15mm \divg \bfb=0, 		\label{A-TW-a}
\end{equation}
\begin{equation}
- \pl_t \bd + \curl \,\bh = \bj, \hskip15mm  \divg \bd = \rho, 		\label{A-TW-b}
\end{equation}
where $\bd=\eps \be$ and $\bh= \mu^{-1} \bfb$.   Transverse waves are naturally  supported by(\ref{A-TW-a}) 
and (\ref{A-TW-b}). In fact, wave equations are derived  for  $\be$ and $\bh$ from the above four equations:
\begin{eqnarray} 
\Bigl[\nabla^2 - c_t^{-2} \pl_t^{\,2} \Bigr]\,\be & = & \mu\,\pl_t \bj +\eps^{-1}\nabla \rho,  \label{WE-be-b2} \\
\Bigl[ \nabla^2 - c_t^{-2} \pl_t^{\,2} \Bigr] \,\bh & = & - \nabla \times \bj,      \label{WE-bfb-b2}  	
\end{eqnarray}
where $\eps$ and $\mu$ are assumed constant, and $c_t = 1/ \sqrt{\eps \mu}$.

The energy equation and momentum equation of the TW field are derived immediately from the system of 
(\ref{A-TW-a}) and (\ref{A-TW-b}), and are given respectively by
\begin{eqnarray}
\pl_t w  + \divg \bq_{\mbox{\tiny fP}} & = & - \be \cdot \bj,   \label{A-en}  \\
\pl_t \bg + \nabla \cdot M  & = & -\bF_{\mbox{\tiny L}}, \qquad \mbox{where} \quad (\nabla \cdot M)_i = \pl_j M_{ij}.
			\label{A-mom}
\end{eqnarray}
where $w= \frac{1}{2}(\be\cdot\bd + \bh\cdot\bfb)$ is an energy density of  TW-field, and $\bg= 
\bd\times \bfb = \bq/c_t^{\,2}$ is its momentum density, and $\bq_{\mbox{\tiny FP}} \equiv \be \times \bh$ 
is a {\it fluid} Poynting vector. The tensor $M=(M_{ij})$ is a {\it fluid} Maxwell-stress, the vector 
$\bF_{\mbox{\tiny L}}$ is a {\it fluid} Lorentz force, and $c_t$ (written as $c$ in \S\ref{a4}) is the phase speed 
of transverse waves in turbulence, each of which is defined respectively by
\begin{eqnarray}
\bF_{\mbox{\tiny L}} & = & \rho\,\be +\bj\times \bfb, \hskip10mm 
		c_t= 1/\sqrt{\eps \mu}, 					\label{A-F_Lor} \\
M_{ij} & = & -(e_id_j + h_ib_j) + \half (\be\cdot \bd + \bh \cdot \bfb)\, \delta_{ij}. 	\label{M-stress}
\end{eqnarray}

\subsection{Whole field}  \label{b3} 
The whole system consists of the {\it TW}-field and  {\it FF}-field  (Scofield \& Huq, 2014). \\
\noindent ($a$) {\it Momentum equation}:\, \label{b321} 
The momentum equation of the FF-field is given by 
\begin{equation}
\pl_t (\rho \bv)_i + \pl_j \Pi_{ij} =  (\bF_{\mbox{\tiny L}})_i.  	\label{FF-mom-b3}	 	
\end{equation} 
whereas the momentum equation of the TW-field is given by (\ref{A-mom}):
\begin{equation}
 \pl_t (\bg)_i + \pl_j M_{ij} = -  (\bF_{\mbox{\tiny L}})_i.		\label{TW-mom-b3}	 
\end{equation} 
Adding (\ref{FF-mom-b3}) and (\ref{TW-mom-b3}) side by side results in the momentum equation of the whole field: 
\begin{equation}
\pl_t( \rho \bv +\bg)_i +\pl_j(\Pi_{ij}+ M_{ij})=0. 			\label{A-tot-mom}
\end{equation}
It is seen that both fields  interact each other by exchanging rate of change of momentum  $\bF_{\mbox{\tiny L}}$. 
\\[-3mm]

\noindent ($b$) {\it Energy equation}:\,  \label{b32} 
Energy equation the FF-field is given by 
\begin{equation}
 \pl_t \big[ \rho(\frac{1}{2} v^2 + \ine) \big] + \divg(\bq_{\rm f} ) = \be \cdot \bj, 		\label{FF-E-b32}
\end{equation}
See \S\ref{s23} in the main text how the term $\be \cdot \bj$ appears on the RHS.
The energy equation of the TW-field is given by (\ref{A-mom}):
\begin{equation}
 \pl_t w  + \divg \bq_{\mbox{\tiny fP}} = - \be \cdot \bj,    		\label{TW-E-b32}	 
\end{equation} 
Thus, adding (\ref{FF-E-b32}) and (\ref{TW-E-b32}) side by side yields the equation of total energy:
\[ \pl_t \big[ \rho(\frac{1}{2} v^2 + \ine) + w \big] + \divg( \bq_{\rm f} 
	+ \bq_{\mbox{\tiny fP}} ) =0, 					\label{A-tot-en}  \]  
where $\ine$ is the internal energy of fluid, and  $\bq_{\rm f}$ is the FF-energy flux given by 
\begin{equation}
\bq_{\rm f} =\rho \bv (\frac{1}{2} v^2+ h) -\bv\cdot \tau^{(vis)}
	- k_T\, \nabla T,				\label{A-q_FF}
\end{equation}
(Landau, L.D. and Lifshitz (1987), \S49).
The entropy equation is modified in this combined case and a new term $Q_{\mbox{\scriptsize D}} $ is 
added to (\ref{FF-entropy-b1}):
\begin{equation}
\rho T \frac{\D}{\D t} s = Q_{\mbox{\scriptsize D}}  + Q_{vis} + Q_T,				\label{entropy-s32}
\end{equation}
where  $Q_{\mbox{\scriptsize D}} =\be \cdot \bj_d$, which is a new term of  rate of heating due to 
{\it turbulence}-Darcy effect.

\subsection{Energy-momentum tensor: \ free FF-field  }  	\label{b4}
Having in mind  application to the whole combined field of {\it FF}-field and {\it TW}-field (\S\ref{s23} in the  main
text), general formalism of {\it theoretical physics} is presented here for the free {\it FF}-field (\ie in the
absence of TW-field)  on the basis of the Lagrangian density $\Lam_{\rm f}$ and hence the variational principle.  
In this section, we  derive the same equations (\ref{FF-M0}) and (\ref{FF-E})   from the general principle.

Field equations are derived in accordance with the principle of least action in four-dimensional space-time \ 
$x^\al =(x^0, \bx)$ (with $x^0=ct$, $\bx=(x^1, x^2, x^3)$ with $c$ the light velocity). In a general form, 
the Lagrangian density is a certain functional of the quantities $q_\gam(x^\al)$ describing the state of the 
system, where included in $q_\gam(x^\al)$ are three  components of velocity field and two thermodynamic variables, 
\etc.  The action  $S_{\rm f}$ for the fluid flow is defined in the form, $S_{\rm f} = 
\int \Lam_{\rm f}(\,q_\gam(x^\al)\,)\,\dd \Om$, where $\dd \Om= \dd x^0 x^1 x^2 x^3$. The governing equations 
of motion are derived as the Lagrange's equation in general with taking variation of the Lagrangian density 
$\Lam_{\rm f}$ by varying $q_\gam$. 

However, we are interested here in deriving the conservation equations of energy and momentum, which are 
represented by
\begin{equation}	
	\pl_\al T^{\al\beta}_{\rm f} = 0,		\label{T-cons-b1}
\end{equation}
where $T^{\al\beta}_{\rm f} $ is the {\it energy-momentum} tensor (or stress tensor) of fluid flow, defined by
\begin{eqnarray} 
T^{\al\beta}_{\rm f}  & = & \frac{\pl \Lam_{\rm f}}{\pl(\pl_\al a^\lam)}\,\pl^\beta a^\lam 
	- g^{\al\beta} \Lam_{\rm f}, 
\hspace*{10mm} \pl_{\al} \equiv (\pl_\tau, \nabla), \quad  \pl^{\beta}=g^{\beta\al}\pl_\al
  	= (\pl_\tau, -\nabla), 							\label{EM-tensor-b1}
\end{eqnarray}
with the metric tensors given by $ g_{\mu\nu} = g^{\mu\nu} = {\rm diag}(1, -1, -1, -1) $.\footnote{In the space-time 
representation, greek letters such as $\al,\,\beta,\, \mu,\, \nu, \,\lam$ denote $(0,1,2,3)$ and roman letters such 
as $i, \,k$ denote $(1,2,3)$.}

The {\it energy-momentum} tensors are considered in the text {\it Fluid Mechanics} (Landau \& Lifshitz 1987) at 
Chap. XV  "{\it Relativistic Fluid Dynamics}" where relativistic energy-momentum tensors are given, together with their
 non-relativistic  limits as the flow velocity $v$ is much less than the light velocity $c$. Our study corresponds 
 to the latter case. From the section \S133, we find the following expressions of the  non-relativistic 
$T^{\al\beta}_{\rm f}$. To make it clear, we show it in the following matric form:

 {\small 
\begin{equation}	
T^{\al\beta}_{\rm f}  =   \left( \begin{array}{cccc}  
	\overline{T}^{00}  &  \overline{T}^{01}  & \overline{T}^{02}  & \overline{T}^{03} 	\\	 
        \overline{T}^{10}  &  \Pi^{11}           & \Pi^{12}           & \Pi^{13}  		\\	
	\overline{T}^{20}  &  \Pi^{21}           & \Pi^{22}           & \Pi^{23}  		\\	
	\overline{T}^{30}  &  \Pi^{31}           & \Pi^{32}           & \Pi^{33}  		
        \end{array}  \right), 					
\qquad   \left.   \begin{array}{lcl}  
	  \overline{T}^{00} & = & \half \rho v^2 + \rho \ine \ \ (+ \ \rho c^2),	     \\
          \overline{T}^{k0} & = &  c^{-1} \rho v^k (\half v^2 + h ) \ \ (+ \ c \rho v^k),  \\
          \overline{T}^{0k} & = &  c \rho v^k + c^{-1}  \rho v^k (\half v^2 + h ), \\
          \Pi^{ik}  	    & = &  \rho v^i v^k + p\,\del_{ik} = \Pi^{ki}
        \end{array}   \quad  \right\}    		      			\label{T-f-b1}  
\end{equation}   }
where $i,\, k = 1, 2, 3$ and $h =\ine + p/\rho$ is the enthalpy.

The {\it time} component ($\beta=0$) of Eq.(\ref{T-cons-b1}) is  given by $\pl_\al T^{\al 0}_{\rm f} = 0$; 
namely, we have 
\begin{equation}
 \pl_\tau \overline{T}^{00} + \pl_k \overline{T}^{k0} = \frac{1}{c}\,\Big( \pl_t \big[ \rho ( \half  v^2 
 	+ \ine )\big]  + \divg \big[ \rho \bv (\half v^2 + h ) \big] \Big) = 0,	 \label{e-cons-b1}
\end{equation}
where the expressions of (\ref{T-f_ab}) are used with excluding the rest energy term $\rho c^2$ in the 
parenthesis of the energy density  $\overline{T}^{00}$ and also the term $ c \rho v^k$ in the parenthesis of 
the energy flux density  $\overline{T}^{k0}$. This equation is equivalent to the energy  equation (\ref{FF-E}) 
or (\ref{FF-E-b1}) with the energy flux vector $\bq_{\rm f}$ replaced by $\bq_{\rm f}^{(0)}$  without viscosity 
effect and thermal conduction effect.

If the term $c \rho v^k$ in the parenthesis of   $\overline{T}^{k0}$ is included,  the tensor 
$T^{\al\beta}_{\rm f}$ becomes symmetric with respect to $\al$ and $\beta$.  Then, this gives an additional
term  $\rho c^2\, \bv$  within $\divg [ \cdots ]$ of (\ref{e-cons-s21}), which denotes flux of the rest energy.
However, the rest energy is not taken into account in the  non-relativistic fluid dynamics. Thus, the 
{\it energy-momentum} tensor becomes non-symmetric.

The {\it space} component ($\beta= k$ with $k =1,,2,3$) is  given by $\pl_\al T^{\al k}_{\rm f}= 0$; namely, we have 
\begin{equation}  \hspace*{-20mm}  
 \pl_\tau \overline{T}^{0 k} + \pl_i \Pi^{ik} = \Big[ \pl_t \big( \rho v^k ) 
 	+ \pl_i \big( \rho v^i v^k + p\,\del_{ik} \big) \,\Big]  + O( (v/c)^2 ) = 0,  
 		\quad (k =1, 2, 3),	\label{mom-cons-b1}  
\end{equation}
 This  is equivalent to the momentum  equation (\ref{FF-M0}) if terms of O($(v/c)^2$) are neglected.
 
Thus, it is shown that the conservation equations of energy and momentum in the current theory are interpreted by 
the general formalism of theoretical physics.

\subsection{Energy-momentum tensor: \ free TW-field }  	\label{b5}

The Lagrangian functional $\Lam_0$ of the {\it TW}-system free from external excitation is defined by
using (\ref{A-F2H2}) and (\ref{bd-bh}) as follows: 
\begin{eqnarray} \ \hspace*{-10mm}	
\Lam_0 \,\V^4 \equiv - \Half \F^2 \wedge \Hm^2 & = & \Half [(\be, \bd)-(\bfb, \bh)] \, \V^4  	
		= \Half [\,\eps (\be, \be) - \mu^{-1}  (\bfb, \bfb )]  \, \V^4,	   	\label{L-Lag1}  \\
\Lam_0  & = & \Half [\,\eps (\be, \be) - \mu^{-1}  (\bfb, \bfb )] 
	 = \frac{1}{2\mu} [\,(\overline{\be},\, \overline{\be}) -  (\bfb, \bfb )] 	\label{LD-eb}	\\
\overline{\be} & \equiv & \be/c , \qquad \overline{e}_{\,\lam} 
		= - \pl_\tau a^\lam - \pl_\lam \overline{\phi} .   \label{ec-a4}
\end{eqnarray}
where  $c= 1/\sqrt{\eps\mu}$ (phase velocity of {\it TW}-waves), \ $x^0 \equiv \tau=c\,t$, and  $\overline{\phi}
= \phi_a/c$.   General  representation of Lagrangian  has an interaction term $- j_\al a^\al \,\V^4$, 
which is not included because we consider the energy-momentum tensor of the TW-field in free state. 

The {\it new} variable $x^\nu =(\tau, x^i)$ is used in this section, where $t$ is replaced by $\tau=c\,t$ with 
the other three $x^k$ unchanged. Correspondingly, we use here the new field $\overline{\be}= \pl_\tau \ba 
- \nabla \overline{\phi}$ in stead of $\be$,  and use the new 4-vector $a^\nu = (\overline{\phi}, a^k)$ 
and co-vector $a_\nu  = (\overline{\phi}, -a^k)$ for $k=1,2,3$.  The new field strength tensors are defined by 
\begin{equation} 
\overline{F}_{\nu\lam}  = \pl_\nu a_\lam - \pl_\lam a_\nu, \qquad \overline{F}^{\nu\lam}  
=  \pl^\nu a^\lam - \pl^\lam a^\nu.
\end{equation}
In matrix form, the new  $\overline{F}_{\mu\nu}$ and $\overline{F}^{\mu\nu}$ are expressed by 
 {\small 
\begin{equation}	\hspace*{-10mm}  
\overline{F}_{\nu\lam} =  \left( \begin{array}{cccc}  
		0   &   \overline{e}_1  &   \overline{e}_2 &   \overline{e}_3	\\	 
        	-\overline{e}_1 & 0   & -b_3  & \ b_2  		\\	
		-\overline{e}_2 &  \ b_3 & 0    &  -b_1 	\\	 
        	-\overline{e}_3 & \,-b_2 & \ b_1 &  0  	
	\end{array}  \right),   \qquad 
\overline{F}^{\nu\lam} =   \left( \begin{array}{cccc}  
	0   &  - \overline{e}_1  &  - \overline{e}_2 &  - \overline{e}_3	\\	 
        \overline{e}_1 & 0   & -b_3  & \ b_2  \\	
	\overline{e}_2 &  \ b_3 & 0    &  -b_1 	\\	 \overline{e}_3 & \,-b_2 & \ b_1 &  0  	
	\end{array}  \right), 							     \label{F_munu}  
\end{equation}   }
The Lagrangian $\Lam_0$ of (\ref{LD-eb}) can be represented in terms of the field strength tensors as 
\begin{equation} 
\Lam_0 = - \frac{1}{4\mu}\,\overline{F}_{\nu\lam}  \overline{F}^{\nu\lam} 
	 = \frac{1}{2\mu} [\,(\overline{\be},\, \overline{\be}) -  (\bfb, \bfb )] .		\label{LD-Fab}
\end{equation}
The energy-momentum tensor of this system is defined by 
\begin{eqnarray} 
T_{\rm w}^{\al\beta} & = & \frac{\pl \Lam_0}{\pl(\pl_\al a^\lam)}\,\pl^\beta a^\lam - g^{\al\beta} \Lam_0,
			\label{EM-tensor-b2} \\	
  &&  \hspace*{15mm} \pl_{\al} \equiv (\pl_\tau, \nabla), \qquad  \pl^{\beta}=g^{\beta\al}\pl_\al
  	= (\pl_\tau, -\nabla), \nonumber
\end{eqnarray}
(Jackson 1999;  Landau \& Lifshitz 1975),  where $g^{\al\beta} =\mbox{diag}(1, -1,-1,-1)=g_{\al\beta}$, and ($\al$, 
$\beta$) take one of $0,1,2,3$, and $x^0 =\tau \equiv c\, t$.  The energy-momentum tensor is fundamentally 
important because the conservation law in free space is represented by 
\begin{equation}	
	\pl_\al T_{\rm w}^{\al\beta} = 0.		\label{T-con-law-b2}
\end{equation}
Carrying out straightforward but non-trivial calculus of (\ref{EM-tensor-b2}), the tensor $T_{\rm w}^{\al\beta}$ can 
be transformed to the following forms (see Jackson (1999) [\S12.7, 12.10]):
\begin{eqnarray} 
T_{\rm w}^{\al\beta} & = & - \frac{1}{\mu} g^{\al\nu}\overline{F}_{\nu\lam} \pl^\beta a^\lam - g^{\al\beta} \Lam_0 	
			=  \Theta^{\al\beta}  + T^{\al\beta}_{\,D},			\label{T-ab-b2}	\\
\Theta^{\al\beta}   & = & \frac{1}{\mu} \, g^{\al\nu}\overline{F}_{\nu\lam}\overline{F}^{\lam\beta} 
			+ \frac{1}{4\mu} \,g^{\al\beta} \overline{F}_{\nu\lam} \overline{F}^{\nu\lam} 
                        = \Theta^{\beta\al},  \\
T^{\al\beta}_{\,D}  & = & - \frac{1}{\mu} g^{\al\nu}\,\overline{F}_{\nu\lam} \,\pl^\lam a^\beta
			= \frac{1}{\mu}\,\pl_\lam (\overline{F}^{\lam\al} a^{\beta}).
\end{eqnarray}
The expression (\ref{T-ab-b2}) expresses decomposition of the tensor $T_{\rm w}^{\al\beta}$ into a symmetric 
component $\Theta^{\al\beta}$ and a remaining part $T^{\al\beta}_{\,D}$. This decomposition is important 
because one can show immediately the vanishing of $\pl_\al T^{\al\beta}_{\,D}$. Namely, 
$\pl_\al T^{\al\beta}_{\,D} = \mu^{-1}\, \pl_\al \pl_\lam (\overline{F}^{\lam\al} a^{\beta})  = 0$.
This is because, exchanging the parameters $\al$ and $\lam$, we have $ \pl_\al \pl_\lam = \pl_\lam \pl_\al$,
but $\overline{F}^{\lam\al} = - \overline{F}^{\al\lam}$.  Therefore, the conservation equation 
(\ref{T-con-law-b2}) is replaced by 
\begin{equation}	
	\pl_\al \Theta^{\al\beta} = 0.		\label{Theta-con-law}
\end{equation}
($a$) \ Its {\it time} component ($\beta=0$) is  given by $\pl_\al \Theta^{\al 0} = 0$; namely, we have 
\begin{equation}
 \pl_\tau \Theta^{0 0} + \pl_i \Theta^{0 i}    = 0,  \qquad (i=1, 2, 3),	\label{e-cons-a4}
\end{equation}
where, 
\begin{eqnarray}
\Theta^{0 0}  & = & \frac{1}{\mu} \, g^{0 \nu}\overline{F}_{\nu\lam}\overline{F}^{\lam 0} 
	        + \frac{1}{4\mu} \,g^{00} \overline{F}_{\nu\lam} \overline{F}^{\nu\lam} = \,w,	 \label{T-00}	\\
\Theta^{0 i}  & = & \frac{1}{\mu} \, g^{0 \nu}\overline{F}_{\nu\lam}\overline{F}^{\lam i}  = \frac{1}{\mu} \,
	(\overline{\be} \times \bfb)_i = c^{-1} (\be \times \bh)_i = c^{-1} (\bq_{\mbox{\tiny fP}})_i,  \label{T-0i}
\end{eqnarray}
\begin{eqnarray}
 w   & \equiv & \frac{1}{2\mu} [\,(\overline{\be},\, \overline{\be}) + (\bfb, \bfb )]
 	=   \frac{1}{2} [\,\eps (\be,\, \be) + \mu^{-1} (\bfb, \bfb )]			 \label{energy-D}  \\
 \bq_{\mbox{\tiny fP}} & \equiv  & \be \times \bh				           \label{energy-F}
\end{eqnarray}
Substituting (\ref{T-00}) and (\ref{T-0i}) into (\ref{e-cons-a4}) and noting $ \pl_\tau =c^{-1} \pl_t$, we obtain 
\begin{equation}
\pl_t w  + \divg \bq_{\mbox{\tiny fP}} = 0,   	\label{TW-E-a4}  
\end{equation}
where $w$ is the energy density, and $\bq_{\mbox{\tiny fP}}$ is the energy flux density, called 
{\it fluid}-Poynting vector corresponding to the Poynting vector in electromagnetism. Thus it is seen that 
the equation (\ref{TW-E-a4}) describes {\it  energy conservation} in the system free from external excitation.

\noindent
($b$) \ The {\it space} component ($\beta=k$ with $k=1,,2,3$) is  given by $\pl_\al \Theta^{\al k} = 0$; namely, 
we have 
\begin{equation}
 \pl_\tau \Theta^{0 k} + \pl_i \Theta^{ik}    = 0,  \qquad (k =1, 2, 3),	\label{mom-cons-a4}
\end{equation}
The term $\Theta^{0 k}$ is given by (\ref{T-0i}), whereas the term $\Theta^{ik} $ is given by
\begin{eqnarray}
\Theta^{0 k} & = & c^{-1} (\be \times \bh)_i = c ((\eps \be) \times (\mu \bh)) = c \,\bg \ 
      (= \frac{1}{c} \,\bq_{\mbox{\tiny fP}}\,), \qquad  \bg = (g_i) \equiv \bd \times \bfb,	\label{T-0k}	\\
\Theta^{ik} & = & \frac{1}{\mu} \, g^{i \nu}\overline{F}_{\nu\lam}\overline{F}^{\lam k} 
	        	+ \frac{1}{4\mu} \,g^{ik} \overline{F}_{\nu\lam} \overline{F}^{\nu\lam}	
              = -  \frac{1}{\mu} \, \overline{F}_{i\lam}\overline{F}^{\lam k} 		
	      -  \frac{1}{2\mu} g^{ik} [\,(\overline{\be},\, \overline{\be}) - (\bfb, \bfb )]  \nonumber \\
   & = & M_{ik},  									\label{T-ik}	 \\
M_{ik} & \equiv  & - (\eps \,e_i e_k + \mu^{-1}\, b_ib_k) 
				+ \half \del_{ik}\,[\,\eps (\be,\, \be) + \mu^{-1} (\bfb, \bfb )].  \label{M-ik}
\end{eqnarray}
Substituting (\ref{T-0k}) and (\ref{T-ik}) into (\ref{mom-cons-a4})  and noting $ \pl_\tau =c^{-1} \pl_t$,
we obtain an equation of {\it momentum conservation} in the system free from external excitation:
\begin{equation}
\pl_t\,g_k  + \pl_i M_{ik} = 0,   	\label{TW-M-a4}  
\end{equation}
The field  $\bg = \bd \times \bfb $ is interpreted as   {\it momentum density} of the field, and $M=(M_{ik})$
is a {\it fluid} Maxwell-stress. See (\ref{A-en}) and (\ref{A-mom}) for the conservation equations of energy
and momentum  in the presence of external excitation.

\subsection{In what circumstances the viscous stress is valid ?} \label{b6}
Traditionally, it is understood that dissipation of kinetic energy of turbulence is caused by the viscous 
stress only.  The viscous stress is assumed to depend on the rate of distortion of a fluid-element  
during motion, and its mathematical expression is derived by assuming linear dependence on the 
rate-of-strain tensor, which is {\it of a purely tentative character}.  This is mentioned in the 
classical textbook {\it Hydrodynamics} by Horace Lamb (1932, Art.326),  describing moreover as
follows: "{\it Although there is considerable {\it $\grave{a}$ priori} probability that it will 
represent the facts accurately in the case of infinitely small motions, we have so far no assurance 
that it will hold generally}."

In addition, the viscous force is a surface traction, expressed as a force acting on an O($l^2$)-surface 
 of a fluid particle of infinitesimally small length-scale $l$. This is because the force acting on its mass is 
O$(l^3)$ which is higher order than O($l^2$) for an infinitesimal $l$, and also because the internal 
stresses are due to molecular forces which are {\it near-action}. By the same reason, a reaction to an 
{\it inertial} force due to the particle acceleration brought about by its surrounding is 
of O($l^3$) and omitted (Lamb, 1932).

Now, consider a tiny fluid particle convected around by violently turbulent flow at very high Reynolds 
numbers. Local motion of the particle in the turbulence is forced by strong linear acceleration and 
strained by intense shear of rotational motion of such flows.  Namely the frame of reference moving 
with the fluid particle is not only {\it non-inertial} but also under rapidly straining deformation.  
Omission of this influence is not justified for such {\it soft} particles exposed to intense turbulent motion.

\setcounter{footnote}{0}
\section{\small Perturbation  and production term  in the absence of TW-field}  \label{cc}

\vspace{-2mm}
\subsection{An action of a perturbation field}    				\label{c1}

In the absence of TW-field (current theory),  let us define the 
Navier-Stokes operator NS$[\bv]$ for a velocity vector field $\bv$ satisfying $\nabla\cdot\bv=0$ by
\begin{eqnarray}
{\rm NS}[\bv]  & \equiv & \pl_t \bv + (\bv\cdot\nabla) \bv +\nabla P_v - \nu \nabla^2\bv, \label{NS-a-c} \\
& \equiv & \pl_t \bv + \bom_v \times \bv +\nabla(P_v +\half |\bv|^2) - \nu \nabla^2\bv,   \label{NS-b-c}  
\end{eqnarray}  
where $\bom_v=\nabla\times\bv$ is the vorticity,\footnote{The following identity is used: ($i$) 
$(\bv\cdot\nabla)\bv = \bom \times \bv + \nabla(\half |\bv|^2)$.   \\  \hspace*{5mm} The vorticity $\nabla\times\bv$ 
is written as $\bom_v$, or as $\bfb_v$ like (\ref{bf_L}). \ Another identity (for a vector $\be$)   \\ \hspace*{5mm}
($ii$) $\nabla \times (\nabla\times \be) = \nabla(\nabla\cdot \be) - \nabla^2 \be$ \ is used  in the main text.} 
$\pl_t=\pl/\pl t$, $P_v = p_{\bv}/\rho$ with $p_{\bv}$ the pressure associated with the $\bv$-field, 
$\rho$ the density (constant) and $\nu$ the kinematic viscosity (constant). Now consider two neighboring vector 
fields $\bv_0=\bU$ and $\bv=\bU+\bu$ having coherent structures similar to each other, or similar 
vortex structures. As an  example of the basic flow $\bU$, one may think of a streaky boundary layer flow.

Let us consider the dynamics of $\bv$ satisfying ${\rm NS}[\bv] =0$, namely 
\begin{equation}
 \pl_t \bv + (\bv\cdot\nabla) \bv = - \nabla P_v + \nu \nabla^2\bv.  \label{NS-v} 
\end{equation}   
Difference of NS$[\bv_0]$ from NS$[\bv]$ (here, NS$[\bv_0]$ does not necessarily vanish) is given by
\begin{eqnarray}
&& \hspace*{-18mm}  {\rm NS}[\bv_0] - {\rm NS}[\bv] = \bff_{\rm L}[\bu] + \bu\times \bOm + \nu  \nabla^2\bu,
			\qquad 	\bOm =\nabla \times \bU, \qquad \bv= \bU+\bu, 	\label{dNS} \\
&& \hspace*{-3mm}  
\bff_{\rm L}[\bu]  \equiv \be_u + \bv\times \bfb_u,  \qquad \be_u \equiv -\pl_t\bu -\nabla \phi_u, 
	\qquad \bfb_u \equiv \nabla\times \bu, \ = \bom_u,				\label{bf_L}
\end{eqnarray}
where $\phi_u= (P_{\bv} - P_{\mbox{\tiny $\bU$}}) + \bu\cdot\bU + (1/2)|\bu|^2$.  If\, 
$\bu \parallel \bOm$ so that $\bu\times \bOm=0$, the  condition of ${\rm NS}[\bv]=0$ results in the 
equation ${\rm NS}[\bU] = \bff_{\rm L}[\bu]+ \nu  \nabla^2\bu$.  

Thus, it is found that the dynamics of 
$\bv=\bU+\bu$ is equivalent to that of  $\bU$ under the actions of {\it Lorentz}-like acceleration 
$\bff_{\rm L}[\bu]$ and the viscous retardation $\nu  \nabla^2\bu$. It is seen that the deviation-velocity
$\bu$ takes the part of a vector potential in the definition equation (\ref{bf_L}) of $\be_u$ and 
$\bfb_u$ just like the theory of Electromagnetism (\eg Jackson (1999) Ch.6).

\vspace{-4mm}
\subsection{Production term of turbulence (the current theory)}  \label{c2}
\vspace{-1mm}
Let us consider the equation (\ref{dNS}) under ${\rm NS}[\bv]=0$ and $\nabla\cdot \bv=0$ with $\bv_0=\bU$ 
and $\bv=\bU+\bu$. Namely, the equation ${\rm NS}[\bv]=0$ is rewritten as 
\begin{eqnarray}  
{\rm NS}[\bU]  & = & 	\bff_{\rm L}[\bu] + \bu\times \bOm + \nu  \nabla^2\bu  \label{NS-U1}  \\
 	& = &   -\pl_t\bu -(\bU \cdot\nabla) \bu - (\bu \cdot\nabla) \bU - (\bu \cdot\nabla) \bu
 		- \nabla \phi_p  + \nu  \nabla^2\bu,				\label{NS-U2} 
\end{eqnarray}
where  $\bff_{\rm L}[\bu] = \be_u + \bU\times \bom_u + \bu \times \bom_u$ and 
$\phi_p = (p_{\bv}-p_{\mbox{\tiny $\bU$}})/\rho $. It is assumed that the vector  $\bU$ describes a slowly 
varying velocity field with respect to Cartesian space coordinates $\bx=(x, y, z)$ and time  $t$, and it is 
supposed that this field is characterized with a coherent structure expressed by the vorticity field $\bOm$, 
while the vector $\bu$ describes a rapidly fluctuating velocity field with respect to $\bx$ and $t$. 

A typical example of the flow $\bU$ 
may be a streaky steady boundary layer flow and $\bu$ a fluctuating perturbation  of a time scale $\tau_u$. 
We consider   statistical time-average over a time span smaller than the time scale $\tau_U$ of $\bU$, 
on the assumption that the time scales of both fields are separated significantly, \ie $\tau_u \ll \tau_U$.
We denote the average over a time span of the order of $\tau_u$ as $\overline{\bu}$ and assume that 
$\overline{\bu}=0$ like turbulent fluctuations. The average $\overline{\bU}$  may be time-dependent and
is written as $\bU$ for simplicity.

Let us take a scalar product of $\bU=(U_k)$ with the equation (\ref{NS-U1}) (or (\ref{NS-U2})) 
written as (LHS34)$_k =$(RHS34)$_k$, and consider time average of the scalar product. Using the 
definition (\ref{NS-a}) of ${\rm NS}$-operator, the left hand side (LHS) is given by 
\begin{eqnarray} 
\overline{U_k\,({\rm LHS34})_k} 	& = & \overline{U_k\,{\rm NS}[\bU]_k} 
		= \pl_t(\half U^2) + \nabla \cdot \bT + D_\nu,  \label{U-NSL} 		\\ 
&&  \bT =\half U^2 \bU + P\bU - \nu\nabla(\half U^2), \quad D_\nu = \nu(\pl_iU_k)^2 (\ge 0). \nonumber
\end{eqnarray}
Using $\overline{\bu}=0$,   average of the product of $U_k$(RHS34)$_k$, is
\begin{equation}   
\overline{U_k\,({\rm RHS34})_k} = \overline{U_k\,(\bff_{\rm L}[\bu])_k} 
	= - U_k\,\pl_i (\overline{u_iu_k}) - \pl_k (U_k \overline{\phi_p}). 	  \label{U-NSR}
\end{equation}
Equating  the RHS's of (\ref{U-NSL}) and (\ref{U-NSR}), we integrate the resulting equation over 
a simply connected 3-dimensional volume $V_3$ in the fluid, enclosed by 2-dimensional boundary 
surface $\pl V_3$, chosen arbitrarily in the space of the coordinates $\bx=(x_i)=(x,y,z)$. 
Thus, we obtain
\begin{eqnarray}
\frac{\dd}{\dd t} K_U + D_{vis} & = & W_f[\bU, \bu] + \ I_{\pl V_3}, \quad = -P  \ \ (+\, I'_{\,\pl V_3}), 
				\label{dE-eq-c}	\\
&& \hspace*{-11mm} K_U =\int_{V_3} \half U^2\,\dd^3\bx, \hskip17mm D_{vis} = \int_{V_3} D_\nu\,\dd^3\bx, \nonumber \\
 && \hspace*{-21mm} W_f[\bU, \bu] \equiv \int_{V_3} \overline{U_k\,(\bff_{\rm L}[\bu])_k}\,\dd^3\bx,  
	\qquad  P = - \int_{V_3} \overline{u_iu_k}\ \pl_iU_k\,\dd^3\bx,   		\label{dE-def-c}
\end{eqnarray}
where the term $I_{\pl V_3}$ denotes integration  over the surface $\pl V_3$, obtained from 
integrating the term like $\pl_k(\cdot)$ over $V_3$, but those surface integrals  are neglected here and 
below. The integral $K_U$ is the total kinetic energy of the flow field $\bU$ in $V_3$, and the term 
$P$ is an integral of the so-called the {\it production term}, \ie rate of turbulence production by the 
action of the Reynolds stress $\overline{u_iu_k}$. This is because the fluctuation energy $k_u=
\int \half |\bu|^2 \,\dd^3\bx$ increases if $P>0$ (see \eg Pope (2000) Ch.\,5). The term $D_{vis}$ 
denotes the bulk rate of viscous dissipation of energy $K_U$ in $V_3$.  Note that
\begin{equation}  \hspace*{-15mm} 
W_f[\bU, \bu] = \int_{V_3} \overline{U_k\,(\bff_{\rm L}[\bu])_k} \,\dd^3\bx
	= \int_{V_3} \overline{u_iu_k}\ \pl_iU_k\,\dd^3\bx \ (+\, I''_{\,\pl V_3}) 
        = -P  \ (+\, I''_{\,\pl V_3}).				\label{W-fL}
\end{equation}
It is interesting to find the following. If $P>0$, the energy $K_U$ decreases (neglecting the contribution 
from $I'_{\,\pl V_3}$), equivalently $W_f[\bU,\bu]$ would be negative (neglecting $I''_{\,\pl V_3}$). Namely, 
the $\bU$-field do work against the force $\bff_{\rm L}$, and the energy of $\bU$-field is extracted by the 
force.  This is an important mechanism and is extended to application  to the combined system ${\rm NS}[\bv] 
= \bff_{\rm L}[\ba]$ of (\ref{NS-fLa}) in the main text (\S3.4), in which  $W_f[\bU, \bu]$ is replaced by 
$W_f[\bv,\ba]$.  Thus, if $W_f[\bv, \ba] < 0$, then the $\ba$-field would be produced.

\section{Axial components $e_x$ and $h_x$ determine other components} \label{dd}

The problem of transverse waves traveling along the axis ($x$-axis) of a straight pipe studied in
\S\ref{s5} and \ref{s6}, the system of TW equations is reduced to two equations of (\ref{TW-eh}). 
If $\pl_t$ is replaced with $-i\om$, those are reduced to
\[  
\nabla \times \be = ip \bh, \hskip10mm \nabla \times \bh = (\sg - iq)\be +\bj_c,  \label{B-TW-eh1}  
\]  
where $p=\om \mu$ and $q=\om\eps$. Taking derivative with respect to $x$, \ie operating $\pl_x$,
is equivalent to multiplication by $ik$. Then, the $y$ and $z$ components of the first equation in 
the cross-sectional plane are 
\[ 
ik \pl_z e_x -(ik)^2 e_z = -pk \,h_y, \hskip10mm (ik)^2 e_y - ik \pl_y e_x  = -pk \,h_z.  \label{B-TW-h_yz}  
\] 
Similarly, the $y$ and $z$ components of the second equation are 
\[ 
ik \pl_z h_x + k^2 h_z = ik (\sg - iq)e_y + \pl_x j_{c,y}, \hskip5mm 
- k^2 h_y - ik \pl_y h_x = ik (\sg - iq)e_z + \pl_x j_{c,z}.  \label{B-TW-e_yz}  
\] 
From these, we obtain    {\small 
\begin{eqnarray}   \hspace*{-20mm}  
 e_y = \frac{1}{K^2} (ik \,\pl_y e_x + ip\, \pl_z h_x - c_t\mu \,\pl_x j_{c,y}), \hspace{3mm} & \quad &  
 	e_z = \frac{1}{K^2} (ik\, \pl_z e_x - ip \,\pl_y h_x - c_t\mu\, \pl_x j_{c,z}), \label{B-e_yz}  \\
 \hspace*{-20mm}  
 h_y = \frac{1}{K^2} (ik\, \pl_y h_x + (\sg-iq) \pl_z e_x + \pl_x j_{c,z}),   & \quad &  
	\hspace{0mm}  h_z =\frac{1}{K^2} (ik \,\pl_z h_x - (\sg-iq) \pl_y e_x - \pl_x j_{c,y}),    \label{B-h_yz} 
\end{eqnarray}   }
where $K^2=k_0^{\,2}-k^2+ip\sg$ and $pq=\om^2 \eps\mu=(\om/c_t)^2=k_0^{\,2}$. 

Suppose that the convection current $\bj_c$ is given. Then the equations (\ref{B-e_yz}) and (\ref{B-h_yz}) 
state that the $y$ and $z$ components of $\be$ and $\bh$ in the cross-sectional plane are determined
once the axial components $e_x$ and $h_x$ are known.


\vspace{5mm}

\section*{References}
\begin{harvard}

\item[]	 
Adrian, R.J., Hairpin vortex organization in wall turbulence.  
{\itshape Phys. Fluids} 2007, {\bfseries 19}, 041301.

\item[]	 
Bailey, S.C.C., Hultmark, M., Smits, A.J. and Schultz,M., Azimuthal structure of turbulence 
in high Reynolds number pipe flow. {\itshape J. Fluid Mech.} 2008, {\bfseries 615}, 121-138.

\item[]	 
Bailey, S.C.C. and  Smits, A.J., Experimental investigation of the structure of large- and
very-large-scale motions in turbulent pipe flow. {\itshape J. Fluid Mech.} 2010, 
{\bfseries 651}, 339-356.

\item[]	 
Boberg, L. and  Brosa, U., Onset of turbulence in a pipe.
{\itshape Z. Naturforsch.} 1988, {\bfseries 43a}, 697--726.

\item[]	 
 Brosa, U., Disturbances in pipe flow excited by magnetic fields.
{\itshape Z. Naturforsch.} 1991, {\bfseries 46a}, 473--480.

\item[]	 
Bullock, K. J., Cooper, R. E. and Abernath, F. H., Structural similarity in radial 
correlations and spectra of longitudinal velocity fluctuations in pipe flow. 
{\itshape J. Fluid Mech.} 1978, {\bfseries 88}, 585--608. 

\item[]  
Butler, K.M. and Farrell, B.F., Three-dimensional optimal perturbations in viscous shear flow.
{\itshape Phys. Fluids A} 1992, {\bfseries 4}, 1637--1650.

\item[]	 
Cess, R.D., A survey of the literature on heat transfer in turbulent pipe flow.
{\itshape Rep.}\ 8-0529-R24 (Westinghouse Research),  1958.

\item[]	 
Cossu, C. and Brandt, L., On Tollmien-Schlichting waves in streaky boundary layers. 
{\itshape Europ. J. of Mech. B/Fluids} 2004, {\bfseries 23}, 815--833. 

\item[]	 
Del \'{A}lamo, J. C. and Jim\'{e}nez, J.,  Linear energy amplification in turbulent channels.
{\itshape J. Fluid Mech.} 2006, {\bfseries 559}, 205-213. 

\item[]	 
Faisst, H. and  Eckhardt, B., Traveling waves in pipe flow. 
{\itshape Phys. Rev. Lett.} 2003, {\bfseries 91}, 224502.

\item[]	 
Frankel, T., {\itshape The Geometry of Physics -- An Introduction}, 1997 (Cmbridge University Press).

\item[]	 
Fransson, J.H.M., Talamelli, A., Brandt, L. and Cossu, C., Delaying transition to turbulence by
a passive mechanism. {\itshape Phys. Rev. Lett.} 2006, {\bfseries 96}, 064501.

\item[]	 
Frisch, U., \textit{Turbulence}, 1995 (Cmbridge University Press).

\item[]	 
Guala, M.,  Hommema, S.E. and  Adrian, R.J., Large-scale and very-large-scale motions 
in turbulent pipe flow. {\itshape J. Fluid Mech.} 2006, {\bfseries 554}, 521--542.

\item[]   
Gustavsson, L. H., Energy growth of three-dimensional disturbances in plan Poiseuille flow. 
{\itshape J. Fluid Mech.} 1991, {\bfseries 224}, 241--260.

\item[]	 
Hehl, F. W. and Obukhov, Y. N.,  {\itshape Foundations of Classical Electrodynamics
-- Charge, Flux, and Metric}, 2003 (Birkh\"{a}user, Boston).

\item[]   
Henningson, D.S., Lundbladh, A. and Johansson, A.V., A mechanism for bypass transition from 
localized disturbances in wall-bounded shear flows, 
{\itshape J. Fluid Mech.} 1993, {\bfseries 250}, 169--207.

\item[]	 
Hof, B.,  van Doorne, C.W.H., Westerweel, J.,  Nieustadt, F.T.M., Faisst, H.  Eckhardt, B. 
Wedin, H.,   Kerswell R.R. and Waleffe, F.,  Experimental observation of nonlinear traveling 
waves in turbulent pipe flow. {\itshape Science} 2004, {\bfseries 305}, 1594--1598.

\item[]	 
Hussain, A.K.M.F. and Reynolds, W.C.,  The mechanics of an organized wave in turbulent shear flow. 
{\itshape J. Fluid Mech.} 1970, {\bfseries 41}, 241 - 258.

\item[]	 
Hussain, A.K.M.F. and Reynolds, W.C.,  The mechanics of an organized wave in turbulent shear flow. Part 2.
Experimental results. {\itshape J. Fluid Mech.} 1972, {\bfseries 54}, 241 - 261.

\item[]	 
Hussain, A.K.M.F. and Reynolds, W.C.,  Measurement in fully developed turbulent channel flow. 
 {\itshape J. Fluids Eng.} 1975, {\bfseries 97}, 568 - 578.

\item[]	 
Hutchins, N. and Marusic, I., Evidence of very long meandering features in the logarithmic 
region of turbulent boundary layers. {\itshape J. Fluid Mech.} 2007, {\bfseries 579}, 1-28.

\item[]	 
Jackson, J. D.,  \textit{Classical Electrodynamics} (3rd edn),  1999 (Wiley: New York).

\item[]	 
Jim\'{e}nez, J., The largest scales of turbulent wall flows.  In \textit{Ann. Res. Briefs}  
(Center for Turbulence Research, Stanford University) 1998, pp.137--154.

\item[]	 
Kambe, T. and Takao, T., Motion of distorted vortex rings, {\itshape J. Phys. Soc. Jpn.} 
1971, {\bfseries 31}, 591--599. 

\item[]	 
Kambe, T., \textit{Elementary Fluid Mechanics}, 2007 (World Scientific).

\item[]	 
Kambe, T., A new formulation of equations of compressible fluids by analogy with Maxwell's 
equation, {\itshape Fluid Dyn. Res.} 2010, {\bfseries 42}, 055502 (18pp).

\item[]	 Kambe, T., A new scenario of turbulence theory and application to pipe turbulence.
	{\it Proceedings} of ETC15 (Delft), 2015.

\item[]	 Kambe, T., New scenario of turbulence theory and wall turbulence.  ICTAM16\_Kambe\_126638\_2PagePDF,
        in {\it Proceedings} of ICTAM-2016 (Montreal), 2016.

\item[]	 
Kim, K.C. and Adrian, R.J., Very large-scale motion in the outer layer. 
{\itshape Phys. Fluids} 1999, {\bfseries 11}, 417--422. 

\item[]	 
Lamb, H., \textit{Hydrodynamics} (6th edn), 1932 (Cambridge University Press).

\item[]	 
Landau, L.D. and Lifshitz, E.M.,  \textit{The Classical Theory of Fields} (4th edn), 1975 (Pergamon). 
       
\item[]	 
Landau, L.D. and Lifshitz, E.M.,  \textit{Fluid Mechanics} (2nd edn), 1987 (Pergamon). 
 
\item[]	 
Lighthill, J., \textit{Waves in Fluids} (6th edn), 1978 (Cambridge University Press).

\item[]	 
Marmanis, H., Analogy between the Navier-Stokes equations and Maxwell's equations: Application 
to turbulence. {\itshape Phys. Fluids} 1998, {\bfseries 10}, 1428--1437.

\item[]	 
Monty, J.P.,  Stewart, J.A.,  Williams, R.C. and Chong, M.S.,  Large-scale features in 
turbulent pipe and channel flow. {\itshape J. Fluid Mech.} 2007, {\bfseries 589}, 147--156.

\item[]	 
Nishi, M., \"{U}nsal, B.,  Durst, F. and Biswas, G., Laminar-to-turbulent transition of pipe 
flows through puffs and slugs.  {\itshape J. Fluid Mech.} 2008, {\bfseries 614}, 425--446.

\item[]	 
Perry, A. E., Henbest, S. M. and Chong, M.S., A theoretical and experimental study of wall 
turbulence {\itshape J. Fluid Mech.} 1986, {\bfseries 165}, 163--199.

\item[]	 
Pope, S. B.,  {\itshape Turbulent Flows},  2000 (Cambridge University Press).

\item[]	 
Priymak, V. G. and  Miyazaki, T., Long-wave motions in turbulent shear flows.
{\itshape Phys. Fluids} 1994, {\bfseries 6},  3454--3464. 

\item[]	 
Reddy, S.C. and Henningson, D.S., Energy growth in viscous channel flows.
	{\it J. Fluid Mech.} 1993, {\bf 252}, 209--238.

\item[]	 
Reynolds O., An experimental investigation of the circumstances which determine whether the motion
of water shall be direct or sinuous, and the law of resistance in parallel channels.
{\itshape Philos. Trans. R. Soc. London A} 1883, {\bfseries 174}, 935--982. 

\item[]	 
 Reynolds, W.C. and Hussain, A.K.M.F.,  The mechanics of an organized wave in turbulent shear flow. Part 3.
Theoretical models and comparisons with experiments. {\itshape J. Fluid Mech.} 1972, {\bfseries 54}, 263 - 288.

\item[]	 
Rosenberg, Hultmark,  Vallikivi, Bailey \& Smits. Turbulence spectra in smooth- and 
rough-wall pipe flow at extreme Reynolds numbers. {\itshape J. Fluid Mech.} 2013, 
{\bfseries 731}, 46-63.

\item[]	 Schoppa W. and Hussain F., Coherent structure generation in near-wall turbulence.
	{\it J. Fluid Mech.} 2002, {\bf 453}, 57-108.

\item[]	 
Scofield D. F. and  Pablo Huq, Concordances among electromagnetic, fluid dynamical, and 
gravitational field theories. {\itshape Phys. Lett. A} 2010, {\bfseries 374}, 3476--3482.

\item[]	 
Scofield D. F. and  Pablo Huq, Fluid dynamical Lorentz force law and Poynting theorem -- 
derivation and implications. {\itshape  Fluid Dyn. Res.} 2014, {\bfseries 46}, 055514 (22pp).
        
\item[]	 
Singer B.A. and  Joslin R.D., Metamorphosis of a hairpin vortex into a young turbulent spot.
{\itshape Phys. Fluids} 1994, {\bfseries 6},  3724-3736.
        
\item[]	 
Smits, A. J., McKeon B. J. \& Marusic, I., High-Reynolds number wall turbulence. 
{\itshape Annu. Rev. Fluid Mech.} 2011, {\bfseries 43}, 353-375.
        
\item[]	 
Townsend A. A., {\itshape The Structure of Turbulent Shear Flow}  (2nd ed.), 1976 (Cambridge
University Press).

\item[]	 
Trefethen, L.N., Trefethen, A.E., Reddy, S.C. and Driscoll, T.A.. Hydrodynamics stability
without eigenvalues.  {\itshape SCIENCE} 1993, {\bfseries 261}, 578--584..

\item[]	 
Tsinober, A., {\itshape An Informal Conceptual Introduction to Turbulence} (2nd ed.), 
FMIA 92. 2009 (Springer). 

\item[]	 
Waleffe, F., On a self-sustaining process in shear flows. {\itshape Phys. Fluids} 
1997, {\bfseries 9}, 883-900.

\item[]	 
Waleffe, F., Three-dimensional coherent states in plane shear flows.
{\itshape Phys. Rev. Lett.} 1998, {\bfseries 81}, 4140-4143.

\item[]	 
Waleffe, F., Homotopy of exact coherent structures in plane shear flows. {\itshape Phys. Fluids} 
2003, {\bfseries 15}, 1517-1534.

\item[]	 
Wedin, H. and Kerswell, R.R.,  Exact coherent structures in pipe flow: travelling wave solutions.  
{\itshape J. Fluid Mech.} 2004, {\bfseries 508}, 333-371.

\item[]	 
Zhou, J., Adrian, R.J. \& Balachandar, S., Autogeneration of near-wall vortical structures in 
channel flow.  {\itshape Phys. Fluids} 1996, {\bfseries 8} (1), 288-290.   \\ 

\end{harvard}

\end{document}